\documentclass[%
reprint,
amsmath,amssymb,
aps,
pra,
floatfix,
superscriptaddress]{revtex4-1}

\usepackage{graphicx}
\usepackage{dcolumn}
\usepackage{bm}

\usepackage[T1]{fontenc}
\usepackage{amsmath}
\usepackage{breqn}
\usepackage[bbgreekl]{mathbbol}
\usepackage{bbm}
\usepackage{mathrsfs}
\usepackage{xcolor}
\usepackage{hyperref}

\newcommand{\be}{\begin{equation}}
\newcommand{\ee}{\end{equation}}
\newcommand{\bs}{\begin{split}}
\newcommand{\es}{\end{split}}

\newcommand{\XL}[1]{\textcolor{blue}{#1}}
\newcommand{\equ}[1]{\begin{equation}#1\end{equation}}

\newcommand{\YM}[1]{\textcolor{red}{#1}}

\makeatletter
\let\cat@comma@active\@empty
\makeatother

\begin{document}
\preprint{APS/123-QED}
\title{Coherent coupling completes an unambiguous optomechanical classification framework} 
\author{Xiang Li}
\affiliation{Theoretical Astrophysics 350-17, California Institute of Technology, Pasadena, California 91125, USA}
\author{Mikhail Korobko}
\affiliation{Institut f\"ur Laserphysik und Zentrum f\"ur Optische Quantentechnologien, Universit\"at Hamburg, Luruper Chaussee 149, 22761 Hamburg, Germany}
\author{Yiqiu Ma}
\affiliation{Theoretical Astrophysics 350-17, California Institute of Technology, Pasadena, California 91125, USA}
\author{Roman Schnabel}
\affiliation{Institut f\"ur Laserphysik und Zentrum f\"ur Optische Quantentechnologien, Universit\"at Hamburg, Luruper Chaussee 149, 22761 Hamburg, Germany}
\author{Yanbei Chen}
\affiliation{Theoretical Astrophysics 350-17, California Institute of Technology, Pasadena, California 91125, USA}

\begin{abstract}
In most optomechanical systems a movable mirror is a part of an optical cavity, and its oscillation modulates either the resonance frequency of the cavity, or its coupling to the environment. There exists the third option -- which we call a "coherent coupling" -- when the mechanical oscillation couples several non-degenerate optical modes supported by the cavity. Identifying the nature of the coupling can be an important step in designing the setup for a specific application. In order to unambiguously distinguish between different optomechanical couplings, we develop a general framework based on the Hamiltonian of the system. Using this framework we give examples of different couplings, and discuss in details one particular case of a purely coherent coupling in a ring cavity with a movable mirror inside. We demonstrate that in certain cases coherent coupling can be beneficial for cooling the motion of the mechanical oscillator. Our general framework allows to approach the design of optomechanical experiments in a methodological way, for precise exploitation of the strengths of particular optomechanical couplings.
\end{abstract}

\maketitle

\section{\label{sec:intro}Introduction}
Cavity optomechanics\,\cite{kippenberg2008cavity,favero2009optomechanics,aspelmeyer2014cavity,bowen2015quantum} studies the interaction between light and mechanical systems embedded into optical resonators. The precision at which modern optomechanical experiments operate allows to study the quantum properties of light and matter,
including the cooling of macroscopic oscillators to their quantum ground state\,\cite{bhattacharya2007trapping,yong2013review,sawadsky2015observation}, optomechanical squeezing of quantum fluctuations in light\,\cite{purdy2013strong,kronwald2014dissipative,aggarwal2018room,schnabel2017squeezedSec56}, quantum entanglement between optical and mechanical degrees of freedom\,\cite{bose1997preparation,bose1999scheme,vitali2007optomechanical,miao2010universal} as well as between space-like separated mechanical oscillators\,\cite{bose1999scheme,mancini2002entangling,bhattacharya2008entangling,hartmann2008steady,schnabel2015einstein} and non-classical states of mechanical oscillators\,\cite{bose1997preparation,nation2013nonclassical,schnabel2015einstein,brunelli2018unconditional,davis2018painting}. Optomechanics has become an experimental platform for testing quantum mechanics in the macroscopic world\,\cite{bose1999scheme,mancini2002entangling,chen2013macroscopic,purdy2013observation} and looking for potential paths towards quantum gravity\,\cite{bawaj2015probing,schnabel2015einstein,li2016discriminating,belenchia2017tests}.
Optomechanics has also been established as a toolbox for computational and metrological tasks, such as: frequency-converting microwaves to optical light\,\cite{tian2010optical,hill2012coherent,ockeloen2016low,lecocq2016mechanically}, on-chip signal modulation and processing\,\cite{huang2018dissipative}, nanoscale torque detection \,\cite{wu2014dissipative} and the detection of gravitational waves\,\cite{mcclelland2011advanced,abbott2016observation,abbott2018prospects} with kilometre-scale detectors (Advanced LIGO\,\cite{harry2010advanced,aasi2015advanced}, Advanced Virgo\,\cite{acernese2014advanced,acernese2015advanced}, GEO600\,\cite{luck1997geo600,affeldt2014advanced}, KAGRA\,\cite{aso2013interferometer,somiya2012detector}).

While optomechanical systems vary in scale, frequency and complexity, their theoretical description on the fundamental level can be reduced to simple Hamiltonians. 
Conventionally the coupling between the optical and mechanical degrees of freedom is classified based on intuitive physical picture of the setup.
Most common are the systems with \emph{dispersive} coupling, where the mechanical oscillation modulates the cavity's resonance frequency\,\cite{kippenberg2008cavity}. The simplest case of such systems is a Fabry-P\'{e}rot cavity with a movable end mirror\,\cite{kleckner2006high}. 
Another type of coupling is \emph{dissipative}\,\cite{xuereb2011dissipative}: the oscillation modulates the coupling between the system and the environment. Any system with a movable mirror that simultaneously couples with the cavity modes and the pumping field can be viewed as having dissipative coupling, and the simplest example is a Fabry-P\'{e}rot cavity with a movable front mirror\,\cite{vyatchanin2016quantum}. There exists another type of interaction where the mechanical oscillation modulates the coupling between two or more cavity modes\,\cite{khalili2016generalized}. In practice, a complex optomechanical system might not fit into one single type of interaction presented above, or might be misclassified. Therefore, we want a \emph{mutually exclusive and collectively exhaustive} way of classification.


To illustrate the necessity for such a classification, we show how the coupling could be identified ambiguously when the description of one optomechanical system has different forms depending on the choice of cavity basis modes. 
We consider two optical modes $\hat a_1,\hat a_2$ with frequencies $\omega_1,\omega_2$ coupled via the mechanical oscillation $x$. 
Such system is described by the following intuitive Hamiltonian: 
\equ{\label{eq:nondiag}\hat H_{\rm cav}=\hbar \omega_1\hat a_1^{\dagger}\hat a_1+\hbar \omega_2\hat a_2^{\dagger}\hat a_2+\hbar g_{12}x(\hat{a}_1^\dagger \hat{a}_2+{\rm h.c.}).}
If the system is classified simply based on this Hamiltonian, it could fall into the category of "optical modes coupled by mechanical oscillation".
However, this Hamiltonian would have two different forms based on the parameters of the system.
The first case is when the frequencies of the modes $\hat a_1,\hat a_2$ are equal ($\omega_1=\omega_2$). Then Eq.\,\eqref{eq:nondiag} can be presented in another form:
\equ{\label{eq:Hcase2}\hat H'_{\rm cav}=\hbar(\omega_1-g_{12}x)\hat a'_1{}^{\dagger}\hat a'_1+\hbar(\omega_1+g_{12}x)\hat a'_2{}^{\dagger}\hat a'_2,}
with the following choice of basis modes:
\equ{\label{eq:a1mpa2basis}\hat a'_1=\frac{\hat a_1-\hat a_2}{\sqrt{2}},\ \ \ \hat a'_2=\frac{\hat a_1+\hat a_2}{\sqrt{2}}.}
Such Hamiltonian is a dispersive one: the resonant frequencies of the modes are modulated by the mechanical oscillation. 
The second case is when the eigenfrequencies of the modes $\hat a_{1,2}$ are separated by $\Delta\omega\equiv (\omega_2-\omega_1)/2$. Then we can define a new basis of $x$-dependent modes $\hat a_{1,2}''(x)$:
\equ{\label{eq:a1xa2x}\hat a_1''(x)=\hat a_1-\frac{g_{12}}{2\Delta \omega}x \hat a_2,\ \ \ \hat a_2''(x)=\hat a_2+\frac{g_{12}}{2\Delta \omega}x \hat a_1,}
where we assume mechanical oscillation to be small ($g_{12}x\ll|\Delta\omega|$) and keep only the terms linear in $x$. Under this basis the Hamiltonian in Eq.\,\eqref{eq:nondiag} takes another form:
\equ{\label{eq:Hcase3}\hat H_{\rm cav}''=\hbar \omega_1\hat a''_1{}^{\dagger}(x)\hat a''_1(x)+\hbar \omega_2\hat a''_2{}^{\dagger}(x)\hat a''_2(x),}
where the modes themselves have $x$-dependence.
Such a form of the optomechanical coupling is distinct from either the dispersive or the dissipative coupling. We call it \emph{coherent} coupling and will define rigorously in the next section. 
These three different forms of the Hamiltonian illustrate the ambiguity: Eq.\,\eqref{eq:nondiag} describes the coupling between the two modes via the mechanical oscillation, but in different regimes depending on $\Delta \omega$, it could also be either classified as dispersive coupling in Eq\,\eqref{eq:Hcase2}, or have some new form in Eq.\,\eqref{eq:Hcase3}.
However, one system should have a unique classification, which is determined by the physical properties, not by the choice of basis. Identifying the coupling correctly and uniquely is important for optimizing the design of the experiment. Thus, an unambiguous classification framework is necessary.


In this paper, we establish a general framework for the unambiguous classification of the optomechanical systems. The paper is organized as follows: Sec.\,\ref{sec:classification} provides a step-by-step strategy for expressing the Hamiltonian in a canonical form and discriminating between different $x$-dependence. 
We make emphasis in this section on the coherent coupling, which has not been widely recognized as a separate type of optomechanical coupling. Sec.\,\ref{sec:examples} gives some examples from the literature, including possible ambiguities that could arise in identifying couplings and how our approach helps to resolve them. We provide a further focus on the pure coherent coupling in Sec.\,\ref{sec:ringcavity}, where we investigate an optomechanical ring cavity system\,\cite{Xuereb2011ring,Chesi2015ring,Yilmaz2017ring}. We provide an application example where in certain cases this coupling is beneficial for laser cooling of the mechanical oscillator to its ground state. Sec.\,\ref{sec:discussion} includes a summary of the paper and further discussion.

\section{\label{sec:classification}Classification of optomechanical couplings}
In this section we provide a step-by-step strategy that will lead to a unique classification for each cavity optomechanical system and help to avoid potential amiguity. We start by expressing the total optical Hamiltonian in a canonical form that can describe any optomechanical system with multiple optical and mechanical degrees of freedom:
\equ{\label{eq:generalH}
\hat H({\bf x})=\hbar \hat {\bf{a}}^\dagger({\bf x})\mathbb{\Omega}({\bf x})\hat{\bf{a}}({\bf x})+i\hbar \left(\hat{\bf{a}}^\dagger({\bf x})\mathbb{\Gamma}({\bf x})\hat{\bf{b}} -{\rm h.c.}\right),}
where ${{\bf x}}=\{x_1,x_2,...\}$ are the displacements of mechanical oscillators from their equilibrium positions; 
$\hat {\bf{a}}({\bf x})=(\hat a_1({\bf x}),\hat a_2({\bf x}),...)^{\rm T}$ are the cavity eigenmodes, such that $\mathbb{\Omega}({\bf x})=\mathrm{diag}(\omega_1({\bf x}),\omega_2({\bf x}),...)$ is a diagonal matrix with the corresponding eigenfrequencies; $\hat {\bf{b}}=(\hat b_1,\hat b_2,...)^{\rm T}$ are the external electromagnetic modes, which couple to cavity eigenmodes with coupling rates $\mathbb{\Gamma}({\bf x})=\mathrm{diag}(\sqrt{2\gamma_1({\bf x})},\sqrt{2\gamma_2({\bf x})},...)$ and the optical linewidths are $\gamma_{1,2,...}({\bf x})$. Note that ${\bf x}$ can be treated as quasi-stationary parameters here because the time scale for optical relaxation is much smaller than the mechanical one. For practical calculation, the ${\bf x}$-dependence in $\hat {\bf{a}}({\bf x}),\mathbb{\Omega}({\bf x}),\mathbb{\Gamma}({\bf x})$ can be expanded in series and $\bf x$ can be upgraded to dynamical variables and quantum operators $\hat{\bf x}$ following the canonical formulation\,\cite{law1995interaction,khorasani2017higher,pang2018theoretical}. We further consider a conventional linear regime, where the mechanical oscillation $\bf x$ is much smaller than the optical wavelength $\lambda$, allowing the Hamiltonian to remain only linear $\bf x$ terms for a good approximation. While this approximation is not necessarily applicable to all optomechanical systems \cite{thompson2008strong,karuza2012tunable,xie2017phonon,machado2019quantum}, it covers most of the popular ones. 

One system can be described by different Hamiltonians under different choices of basis, as shown from Eq.\,\eqref{eq:nondiag} to Eq.\,\eqref{eq:Hcase3}, but the canonical form in Eq.\,\eqref{eq:generalH} is always \emph{unique}. This serves as the starting point for establishing an unambiguous classification. Position-dependence in  $\mathbb{\Omega}({\bf x})$ and $\mathbb{\Gamma}({\bf x})$ are intuitive and can be directly understood as dispersive and dissipative couplings separately. We construct a mutually exclusive and collectively exhaustive way of classification by considering the last the last possible $\bf x$-dependence: $\hat {\bf{a}}({\bf x})$, which we call coherent coupling. Such ${\bf x}$-dependent modes $\hat {\bf{a}}({\bf x})$ can be presented as linear combinations of unperturbed modes $\hat {\bf{a}}({\bf 0})$ coupled via the mechanical oscillation $\bf x$, as follows from the linearity of the optical system: 
\equ{\label{eq:axdef}
    \hat a_i({\bf x})=\sum_l f_{il}({\bf x})\hat a_l({\bf 0}),}
where $f_{il}({\bf x})$ are the coupling coefficients and $f_{ii}({\bf x})\equiv1$ (no summation for the repetitive $i$). 



In order to classify an optomechanical system without any ambiguity, we formulate the following steps:
\begin{enumerate}
    \item Write the total Hamiltonian of the optomechanical system including all the optical and mechanical degrees of freedom and the coupling among them in any convenient basis.
    \item Transform the Hamiltonian to the canonical form shown by Eq.\,\eqref{eq:generalH}, where $\mathbb{\Omega}({\bf x})$ and $\mathbb{\Gamma}({\bf x})$ are diagonal and $\hat {\bf{a}}({\bf x})$ is the set of cavity eigenmodes. Environmental modes $\hat{\bf b}$ can be chosen correspondingly.
    \item Classify the type of optomechanical interaction by the ${\bf x}$-dependence feature in $\hat {\bf{a}}({\bf x})$, $\mathbb{\Omega}({\bf x})$ and $\mathbb{\Gamma}({\bf x})$.
\end{enumerate}
We can follow these steps to illustrate the classification of the optomechanical couplings into three types mentioned in the Introduction. We consider a specific example with two cavity modes $\hat{a}_1,\hat{a}_2$ and one mechanical degree of freedom $x$, and expand the $x$-dependence up to a linear order in $x$, where $g_{1,2}$ and $g_{\gamma 1,\gamma 2}$ are the expansion coefficients of diagonal terms in $\mathbb{\Omega}(x)$ and $\mathbb{\Gamma}(x)$ matrices:
\begin{enumerate}
    \item \emph{Dispersive coupling}, where the eigenfrequencies depend on the mechanical oscillation: $\mathbb{\Omega}({\bf x})$, the example reads:
    \equ{
    \mathbb{\Omega}(x)=
    \begin{pmatrix}
    \omega_1-g_1 x & 0 \\
    0 & \omega_2-g_2 x
    \end{pmatrix}.}
    \item \emph{Dissipative coupling}, where the rates of coupling to the external modes depend on the mechanical oscillation: $\mathbb{\Gamma}({\bf x})$, the example reads:
    \equ{
    \mathbb{\Gamma}(x)=
    \begin{pmatrix}
    \sqrt{2\gamma_1}+g_{\gamma 1}x & 0 \\
    0 & \sqrt{2\gamma_2}+g_{\gamma 2}x
    \end{pmatrix}.}
    \item \emph{Coherent coupling}, where the eigenmodes depend on the mechanical oscillation: $\hat {\bf{a}}({\bf x})$, and the derivative of any optical mode $\hat a_i$ with respect to any mechanical displacement $x_j$ includes only the other optical modes (see Eq.\,\eqref{eq:axdef}):
    \equ{\left.\frac{\partial\hat a_i({\bf x})}{\partial x_j}\right|_{{\bf x}={\bf 0}}=\sum_{l\neq i}\left.\frac{\partial f_{il}({\bf x})}{\partial x_j}\right|_{{\bf x}={\bf 0}} \hat a_l({\bf 0}).}
    The simple example can be taken from Eq.\,\eqref{eq:Hcase3} where the original modes $\hat a_{1,2}$ become mixed by $x$ (see Eq.\,\eqref{eq:a1xa2x}) and the corresponding eigenfrequency matrix:
    \equ{\mathbb{\Omega}''(x)=\begin{pmatrix}
    \omega_1 & 0 \\
    0  & \omega_2
    \end{pmatrix},}
    doesn't depend on $x$: $d\mathbb{\Omega}''(x)/dx=0$.
\end{enumerate}
Following our classification strategy, one can clearly distinguish among the different types of interactions, even in cases where several couplings coexist.

\begin{figure}
	\begin{center}
		\includegraphics[scale=0.22]{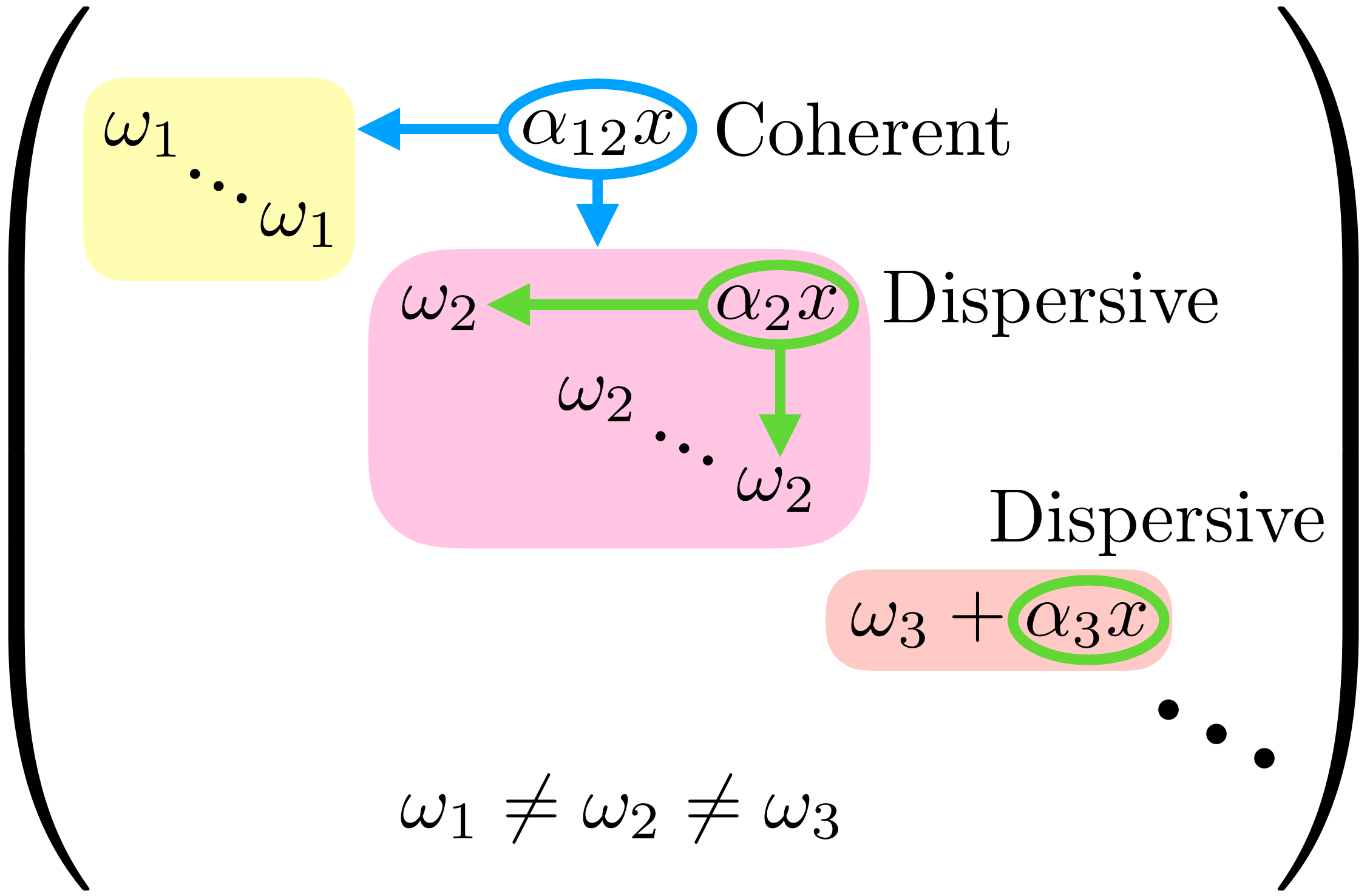}
		\caption{\label{fig:cohvsdisp}The comparison between coherent coupling and dispersive coupling: the $\mathbb{\Omega}({\bf x})$ matrix under the basis of $\hat {\bf a}({\bf 0})$. Each color block represents a frequency degenerate subspace, which can have one or more modes. Different blocks have different frequencies, \textit{e.g.} $\omega_1\neq\omega_2\neq\omega_3$. The influence of mechanical oscillation shows up as $\bf x$-dependent perturbation, which will either directly add on to diagonal terms as dispersive coupling, \textit{e.g.} the $\alpha_3 x$ term, or will show up in off-diagonal terms to couple different modes. The off-diagonal coupling within one color block, \textit{e.g.} the $\alpha_2 x$ term, will open the degeneracy and also cause dispersive coupling. While the coupling between blocks, \textit{e.g.} the $\alpha_{12} x$ term, won't change the eigenfrequencies $\omega_{1,2}$ and will cause coherent coupling. See main text for detailed discussion.}
	\end{center}
\end{figure}

In addition to the definition above, a physical picture of coherent coupling also helps to understand this new concept. The difference between dispersive coupling and coherent coupling can sometimes be not obvious: both of them can be expressed as coupling of optical modes by mechanical motion under some specific cavity basis, as showed the example in Eq.\,\eqref{eq:nondiag}. The distinction between them is illustrated in FIG.\,\ref{fig:cohvsdisp}: when mechanical motion couples different optical modes, these modes are either frequency-degenerate or have different frequencies. When the unperturbed modes are frequency-degenerate, the coupling via the mechanical motion breaks the degeneracy and leads to new $\bf x$-dependent eigenfrequencies, which are the sign of dispersive coupling. On the other hand, when the unperturbed modes have different frequencies, the mechanical displacement leads to a coherent energy transfer between these modes, and such coupling is coherent. Expressed in the canonical Hamiltonian, up to linear order in $\bf x$, the eigenfrequencies remains unchanged but the eigenmodes are the original ones mixed in a $\bf x$-dependent way.


\section{\label{sec:examples}Examples of different couplings}
In this section, we provide some detailed examples of optomechanical coupling of the above three categories. We also cover cases with coexisting couplings.
\subsection{\label{sec:dispersive}Dispersive coupling}
\begin{figure}
	\begin{center}
		\includegraphics[scale=0.25]{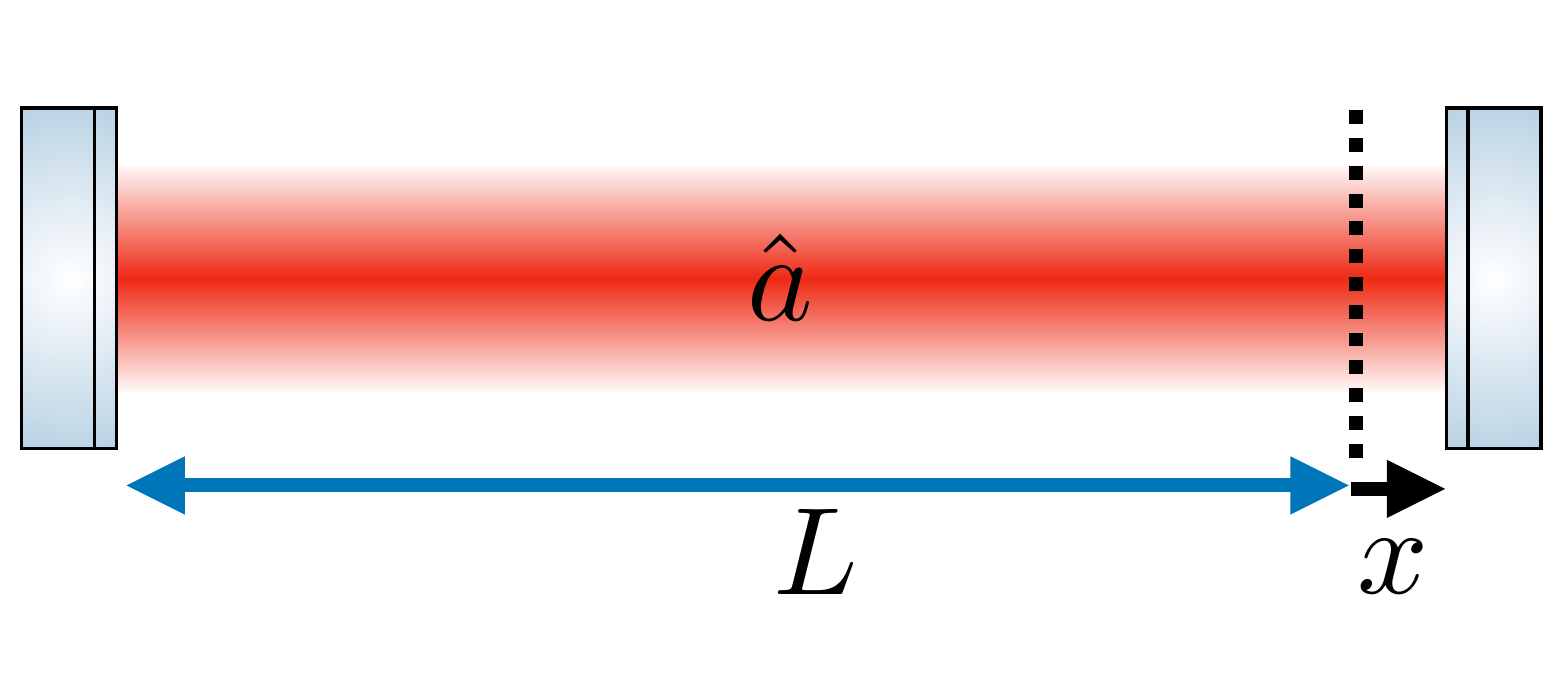}
		\caption{\label{fig:single}Single cavity with a movable end mirror. $L$ is the original cavity length, $x$ is the end mirror displacement from its equilibrium position, $\hat a$ is the cavity optical mode.}
	\end{center}
\end{figure}

Dispersive coupling is the most well-studied type of optomechanical interactions\,\cite{kippenberg2008cavity}. The physical origin of dispersive coupling is the dependence of cavity resonant frequencies on the mechanical oscillation $x$. The Hamiltonian of a single cavity, shown in FIG.\,\ref{fig:single}, reads:
\equ{\label{eq:ax}\hat H_{\rm cav}=\hbar (\omega_a-g_{\omega}x)\hat a^\dagger\hat a,}
where $\omega_a$ is the resonant frequency not affected by the mechanical oscillation, $g_{\omega}=\omega_a/L$ is the dispersive coupling strength, $x$ is the end mirror displacement from its equilibrium position (see detailed derivation in Appendix\,\ref{sec:HamDisper}). 

In this section, we discuss a metrological system that features the dispersive coupling: the Laser Interferometer Gravitational-Wave Observatory (LIGO)\,\cite{aasi2015advanced, khalili2016generalized}. 
This detector takes advantage of two Fabry-P\'{e}rot cavities in the arms of the Michelson interferometer (arm cavities), which sense the gravitational-wave-induced displacement of the test masses. The two arm cavity modes are represented by $\hat a,\hat b$ and their resonance frequencies are by $\omega_0$. These two modes have the same dispersive coupling strength $g$, but they couple to two different displacements $x_1,x_2$. The cavity Hamiltonian can be expressed as:
\equ{\label{eq:ax1bx2}\hat H_{\rm cav} =\hbar(\omega_0-gx_1)\hat a^\dag\hat a+\hbar(\omega_0-gx_2)\hat b^\dag\hat b.} Defining the common and differential mechanical and optical modes as
$x_+=(x_1+x_2)/\sqrt{2},x_-=(x_1-x_2)/\sqrt{2}$; $\hat c_+=(\hat a+\hat b)/\sqrt{2},\hat c_-=(\hat a-\hat b)/\sqrt{2}$, the transformed Hamiltonian takes the form:
\equ{\label{eq:xplusminus}\hat H_{\rm cav} =\hbar(\omega_0-gx_+)\hat{\bf{c}}^\dag\hat{\bf{c}}+\hbar gx_-\hat{\bf{c}}^\dag\hat\sigma_x\hat{\bf{c}},}
where $\hat{\bf{c}}=(\hat c_+,\hat c_-)^{\rm T}$ and $\sigma_x$ is the $x$-component of Pauli matrix. 
Only the differential motion $x_-$ carries the gravitational wave strain signal, so we don't consider the common motion $x_+$. After this operation the transformation from $\hat c_+,\hat c_-$ to $\hat a,\hat b$ is equivalent to the transformation from $\hat a_1,\hat a_2$ to $\hat a_1',\hat a_2'$ in Eq.\,\eqref{eq:a1mpa2basis}. Even though the Hamiltonian can be expressed in different forms in Eqs.\,\eqref{eq:ax1bx2}\eqref{eq:xplusminus}, in our classification strategy, the coupling will \emph{always} be classified as dispersive with eigenmodes $\hat a,\hat b$ and $x_-$-dependent eigenfrequencies: $\omega_{\pm}(x_-)=\omega_0\pm g x_{-}$.

\subsection{\label{sec:diss}Dissipative coupling}
Dissipative coupling happens when the coupling of cavity modes to external modes depends on $x$. For example, for a single cavity mode $\hat a$:
\equ{\label{eq:diss} \hat{H}_{\gamma}=i\hbar\left(\sqrt{2 \gamma}+g_{\gamma}x\right) \left(\hat{a}^{\dagger}\hat{b}-\mathrm{h.c.}\right),}
where $\hat b$ is the external mode and $\sqrt{2\gamma}+g_{\gamma}x$ is the coupling rate, which gives rise to the finite cavity linewidth. The $g_{\gamma}x$ term describes the dependence of the dissipation rate on the mechanical oscillation $x$. The form of dissipative coupling strength $g_{\gamma}$ depends on the specific physical realization.

\begin{figure}
    \begin{center}
        \includegraphics[scale=0.25]{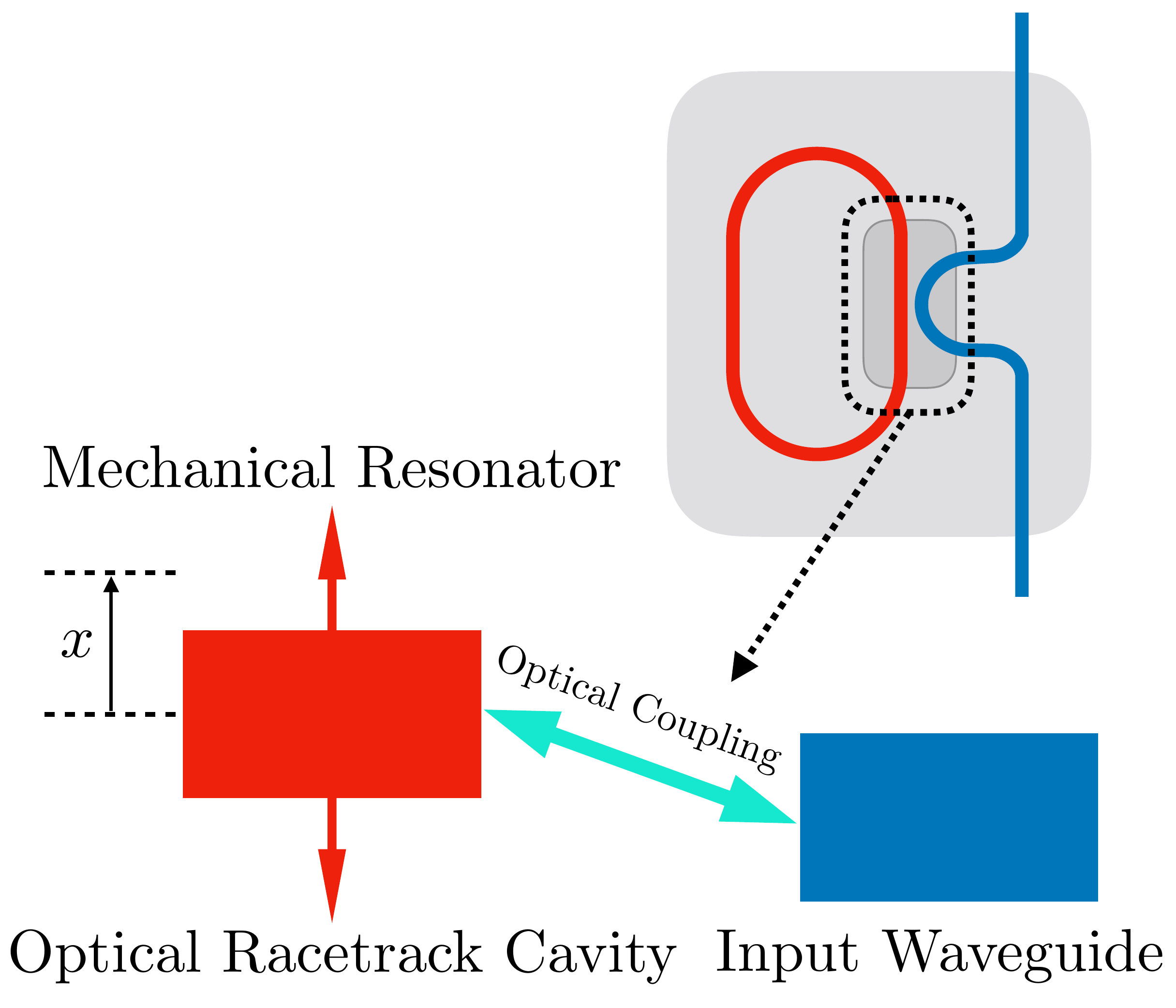}
		\caption{\label{fig:dissi}On-chip Optomechanical coupling between the curved input waveguide and the optical racetrack cavity, adapted from FIG.\,1 in Ref.\,\cite{huang2018dissipative}. The upper right is the top view of the chip, where blue line represents the input waveguide and the red one represents that optical racetrack cavity which is also a mechanical resonator that can have out-of-plane vibrations. The lower left is the schematic of the cross-section inside the dashed area of the upper right. See the main context for discussion.}
	\end{center}
\end{figure}

One recent example is the on-chip dissipative optomechanical resonator\,\cite{huang2018dissipative}. As schematically shown in FIG.\,\ref{fig:dissi}, this system consists of a racetrack optical cavity, which is also a mechanical resonator with out-of-plane vibrations, and a curved input waveguide. Except for the material refractive indices, the optical coupling rate between the racetrack cavity and the input waveguide is determined by the distance between them. The racetrack cavity supports optical mode $\hat a$ and the out-of-plane oscillation expressed by $x$, while the input waveguide carries optical mode $\hat b$. The mechanical oscillation $x$ changes the distance between the racetrack cavity and the input waveguide and thus changes the optical coupling rate between modes $\hat a$ and $\hat b$. The Hamiltonian describing the whole system reads:
\equ{\hat H=\hbar \omega_a \hat a^{\dagger}\hat a+i\hbar\left(\sqrt{2 \gamma}+g_{\gamma}x\right) \left(\hat{a}^{\dagger}\hat{b}-\mathrm{h.c.}\right),}
where neither the cavity mode $\hat a$ nor its resonance frequency $\omega_a$ depends on $x$. There exists only one cavity eigenmode and it already satisfies the canonical form of Eq.\,\eqref{eq:generalH}. Thus, the $x$-dependence in $\hat a,\hat b$ coupling rate shows the feature of dissipative coupling.

\subsection{\label{sec:coherent}Coherent coupling}
The last interaction category to be discussed is coherent coupling where the $x$-dependence appears in the eigenmodes themselves rather than the eigenfrequencies of the optical modes. 

One notable example of coherent coupling is the three-modes optoacoustic interaction\,\cite{miao2009quantum}. It can give rise to important non-linear optomechanical effects such as parametric instability\,\cite{braginsky2001parametric,evans2015observation}, which complicates the operation of gravitational-wave detectors. In a simplified model\,\cite{miao2009quantum}, there are two orthogonal transverse optical-cavity modes $\hat a$ and $\hat b$ with different resonant frequencies $\omega_1$ and $\omega_2$. The acoustic mode has a torsional mode profile and $x$ is its generalized coordinate. The cavity Hamiltonian in this case has the form (see Appendix\,\ref{sec:HamParaInst} for detailed derivation):
\be\label{eq:Hthreemode}\hat H_{\rm cav} =  \hbar\omega_1\hat { a } ^ { \dagger } \hat { a } +  \hbar\omega_2 \hat { b } ^ { \dagger } \hat { b } +  \hbar G _ { 0 } x \left( \hat { a } ^ { \dagger } \hat { b } +\textrm{h.c.} \right).\ee
Note that Eq.\,\eqref{eq:Hthreemode} has the same structure as Eq.\,\eqref{eq:nondiag} and thus follows the same transformation process as in Eq.\,\eqref{eq:a1xa2x}. Up to linear order in $x$, the eigenfrequencies remain the same and the new eigenmodes are the original ones mixed by mechanical oscillation $x$:
\equ{\hat a(x)=\hat a -\frac{G_0}{2\Delta \omega}x\hat b,\ \ \ \hat b(x)=\hat b +\frac{G_0}{2\Delta \omega}x\hat a,}
where $\Delta\omega\equiv (\omega_2-\omega_1)/2$ is the frequency difference. The $x$-dependence in eigenmodes shows the feature of coherent coupling.

\begin{figure}
	\begin{center}
		\includegraphics[scale=0.23]{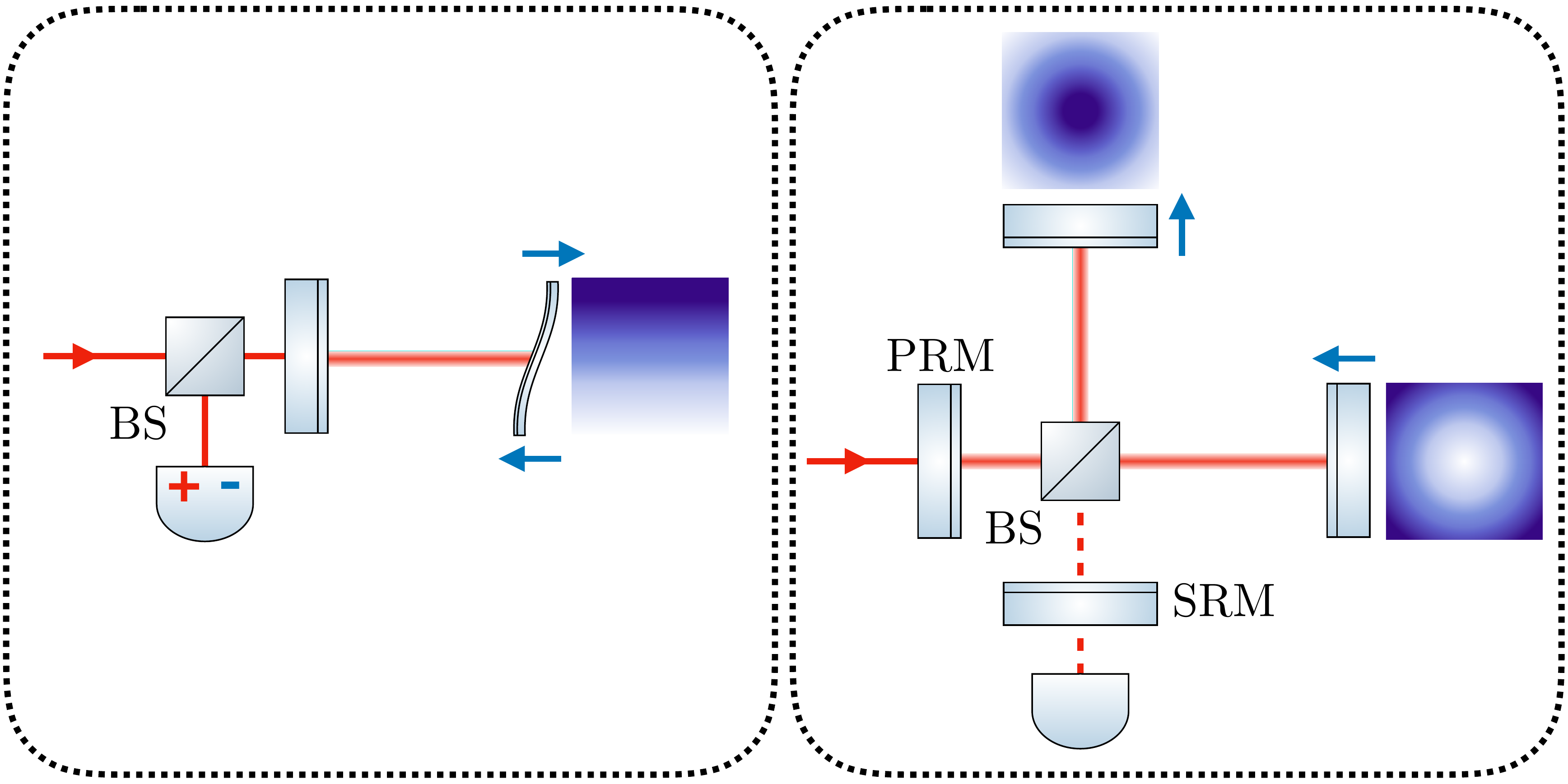}
		\caption{\label{fig:TEM0001}Mapping from three-mode system to a power and signal-recycled interferometer, adapted from FIG.2 in Ref.\,\cite{miao2009quantum}. Although they share a similar three-mode scheme, their physical origins and classification results are different. See main context for detailed discussion.}
	\end{center}
\end{figure}

Note that Ref.\,\cite{miao2009standard} was aware of the $x$-dependence that only happens in eigenmodes, but didn't notice the new coherent coupling category in optomechanics. In Sec.\,\ref{sec:ringcavity} we will investigate a ring cavity system, where the coherent coupling is mediated by the longitudinal oscillation of the mechanical center of mass degree of freedom.

\subsection{\label{sec:Coexisting}Coexisting coupling}
In many cases, different types of optomechanical couplings can coexist. Some optomechanical systems might show different coupling features depending on the parameter regimes that they work in. Following our classification strategy, each type in the coexisting couplings can be clearly distinguished. 

One notable example is the Michelson-Sagnac interferometer\,\cite{xuereb2011dissipative,sawadsky2015observation} with coexisting dispersive and dissipative couplings. 
With careful tuning\,\cite{sawadsky2015observation}, it can become either pure dissipative coupling or pure dispersive coupling. 

\begin{figure}
	\begin{center}
		\includegraphics[scale=0.25]{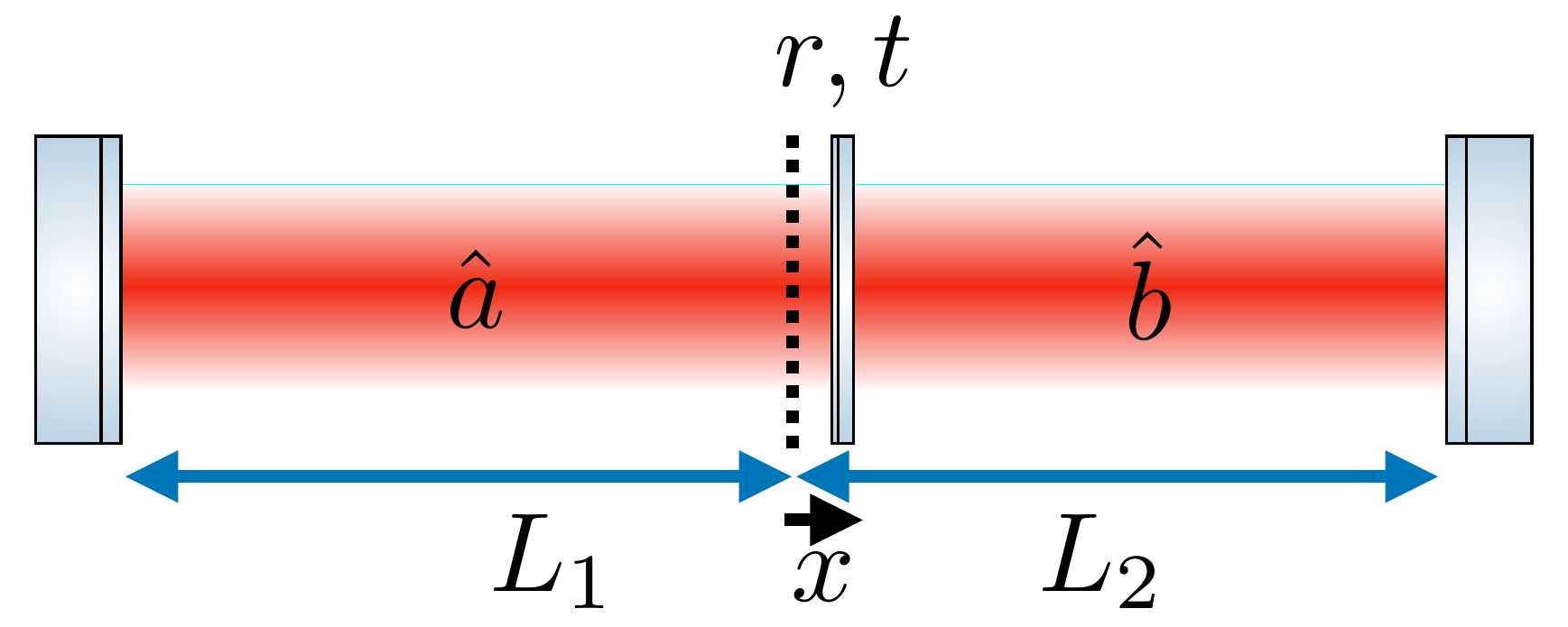}
		\caption{\label{fig:CoupledCavity} The coupled cavity configuration. $L_{1,2}$ are the length of two subcavities, $r$ and $t$ are the amplitude reflectivity and transmittance of the mirror inside and $x$ is membrane oscillation around its equilibrium position. In the main text we consider membrane with low transmittance $t\ll 1$. It is then reasonable to claim that the left and right subcavities can support $\hat a$ and $\hat b$ modes separately. The bare optical frequencies of the two modes are $\omega_{1,2}$ and the optomechanical coupling constants are $g_{1,2}=\omega_{1,2}/L_{1,2}$. Different parameter regime can lead to different classification results. See main text for details.}
	\end{center}
\end{figure}

Another example of a system with coexisting couplings is the system of two coupled cavities separated by a movable mirror, as shown in FIG.\,\ref{fig:CoupledCavity}.
The coupling in such a system can be classified as dispersive, or coherent, or coexisting, depending on the position and optical properties of the central mirror\,\cite{bhattacharya2007trapping,miao2009standard,ma2014narrowing}. 
In the following contents, we give a theoretical description of this system and classify it using our strategy.

When the transmittance of the central mirror is relatively low, $t\ll 1$, we can define two optical modes $\hat{a}, \hat {b}$ for the left and the right subcavities respectively, which are coupled at a characteristic sloshing frequency $\omega_s$.
In terms of these modes the cavity Hamiltonian can be expressed as:
\equ{\label{eq:Hsubab}\begin{split}\hat H_{\rm cav}=&\hbar(\omega_1-g_1 x)\hat a^\dag\hat a+\hbar(\omega_2+g_2 x)\hat b^\dag\hat b\\
&+\hbar\omega_s(\hat a^\dag \hat b+\textrm{h.c.}).\end{split}}
For convenience, we define the average frequency $\omega_0\equiv(\omega_1+\omega_2)/2$ and the frequency difference $\Delta\omega\equiv (\omega_2-\omega_1)/2$. We then convert the Hamiltonian in Eq.\,\eqref{eq:Hsubab} into the canonical form in Eq.\,\eqref{eq:generalH} as required by our classification procedure. 

When the central mirror is perfectly reflective\,\cite{bhattacharya2007trapping}, the sloshing frequency becomes zero ($\omega_s=0$) and Eq.\,\eqref{eq:Hsubab} is already in the canonical form. No optical coupling can happen between $\hat a,\hat b$ modes and they remain to be cavity eigenmodes. The corresponding eigenfrequencies $\omega_1-g_1 x, \omega_2+g_2 x$ are $x$-dependent.
In this case, the system has pure dispersive coupling.

When $\omega_s\neq0$, Eq.\,\eqref{eq:Hsubab} needs to be transformed to the canonical form. The interaction with the mechanical oscillation $x$ couples the original optical eigenmodes"
\begin{align}
    \hat c_{\pm}(0)=(-\Delta \omega\pm \sqrt{\Delta \omega^2+\omega_s^2})\hat a + \omega_s\hat b,
\end{align}
to become:
\begin{align}\label{eq:CoupledCavityModes}
    \hat c_{\pm}(x)=\hat c_{\pm}(0)\pm \frac{g_1+g_2}{4\sqrt{\Delta \omega^2+\omega_s^2}} x \hat c_{\mp}(0),
\end{align}
with the corresponding eigenfrequencies:
\equ{\label{eq:CoupledCavityomega}\begin{split}
\omega_{\pm}(x)&=\omega_0\pm\sqrt{\Delta \omega^2+\omega_s^2}+\\
&+\left(\frac{g_2-g_1}{2}\mp\frac{(g_1+g_2)\Delta\omega}{2\sqrt{\Delta \omega^2+\omega_s^2}}\right)x+\mathcal{O}(x^2).
\end{split}}
Both the eigenmodes and eigenfrequencies depend on $x$, which reveals the coexisting coherent and dispersive couplings. 

The system can have a pure coherent coupling if the central mirror has low transmittance and the two subcavities have the same length $L_1=L_2$\,\cite{miao2009standard,ma2014narrowing}. In this case, the sub cavity frequencies and the corresponding coupling rates in Eq.\,\eqref{eq:Hsubab} become equal: $\omega_1=\omega_2=\omega_0,\Delta\omega=0,g_1=g_2$. The original eigenmodes are $\hat c_{\pm}(0)\propto\hat b \pm \hat a$ and the dispersive feature is absent as the eigenfrequencies of Eq.\,\eqref{eq:CoupledCavityomega} no longer have $x$-dependence: $\omega_{\pm}=\omega_0\pm\omega_s$.

The pure coherent coupling in this coupled cavity example only happens in some specific parameter regimes. In the following section, we will discuss a ring cavity system which always has a pure coherent coupling.

\section{\label{sec:ringcavity}Purely coherent coupling in a ring cavity system}
\begin{figure}[t!]
	\begin{center}
		\includegraphics[scale=0.275]{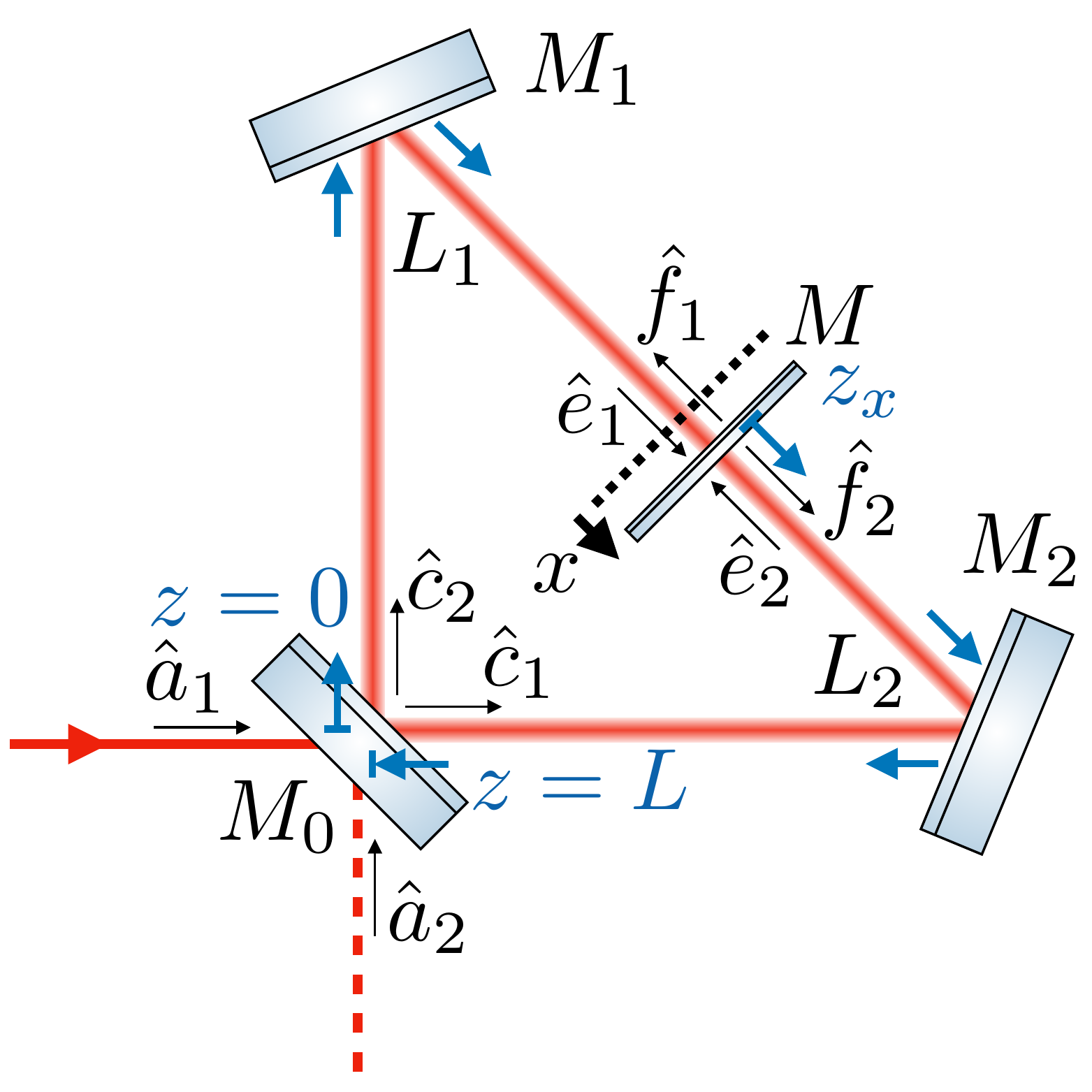}
		\caption{\label{fig:ringconfig}Ring cavity configuration and field labeling. Here $M_0$ is the front mirror with amplitude reflectivity $r_0$ and transmittance $t_0$, $M_{1,2}$ are two fixed totally reflective end mirror, $M$ is the movable membrane with amplitude reflectivity $r$ and transmittance $it$. $L$ is the total cavity length, $L_1=L_2=L/2$ are the distances from the $M_0$ to the equilibrium position of $M$ in clockwise and counterclockwise directions, $x$ is the microscopic displacement of $M$ from its equilibrium. For fields, $\hat c_{1,2}$ are the counterclockwise, clockwise propagating field directly coupled from outside continuum $\hat a_{1,2}$. Defined at the instantaneous position of the membrane, $\hat e_{2,1}$ ($\hat f_{1,2}$) are the propagating fields towards (away from) the membrane in counterclockwise, clockwise direction separately.}
	\end{center}
\end{figure}

In this section we discuss an example of purely coherent coupling in an optomechanical ring cavity system where two resonant modes are coupled via the oscillation of a partially reflective mirror, see FIG.\,\ref{fig:ringconfig}. Similar ring cavity systems with one or multiple scattering objects inside have been studied, including some cases with membranes or mirrors\,\cite{xuereb2011amplified,chesi2015diabolical,yilmaz2017optomechanical} and some other cases with cold atom  clouds\,\cite{nagorny2003collective,kruse2003observation,elsasser2004optical,klinner2006normal,slama2007superradiant,ritsch2013cold,schmidt2014dynamical,mivehvar2018driven}. However, no systematic Hamiltonian construction with a clear definition of optical modes has been done. That motivates our derivation in this section. We also analyze how the coherent coupling helps the laser cooling of mechanical oscillation and compare it with the single cavity dispersive coupling case\,\cite{yong2013review}.

\subsection{\label{sec:ringHMT}Cavity modes and the Hamiltonian}
The detailed derivation of the total Hamiltonian of the ring cavity system can be found in Appendix\,\ref{sec:ringH}. Here we only sketch the key steps of the derivation.

\begin{figure}
	\begin{center}
		\includegraphics[scale=0.37]{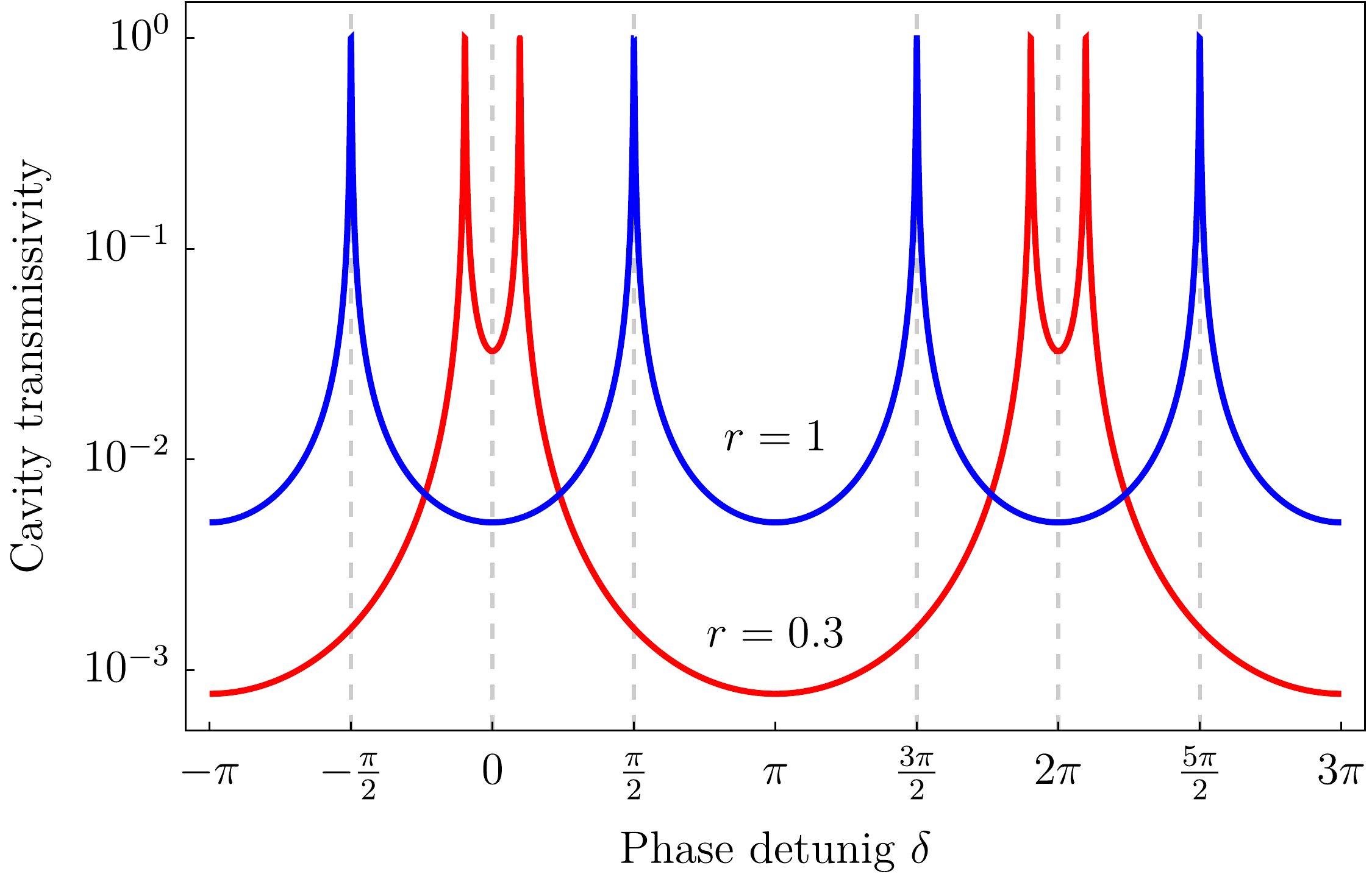}
		\caption{\label{fig:modesplit}Cavity modes splitting caused by different membrane reflectivity $r$. The horizontal axis is a relative phase defined as $\delta\equiv (k_pL+\pi/2)-N2\pi$ and the vertical axis is the ratio of the field amplitude in and out of the system from the input port. The red and blue line correspond to $r=0.3, 1$ respectively. As analyzed in the main context, the cavity resonant frequencies do not depend on $x$ while the electromagnetic mode profiles depend on that. This optomechanical interaction is called coherent coupling. The cavity free spectral range (FSR) is $\Delta\omega_{\rm FSR}=2\pi c/L$ and the linewidth is $\gamma=ct_0^2/2L$.}
	\end{center}
\end{figure}

Without the membrane, the ring cavity can support degenerate clockwise and counterclockwise modes that propagate independently. The membrane reflection couples the two circulating waves and opens the mode degeneracy, as shown in FIG.\,\ref{fig:modesplit}. We first consider resonant cavity modes assuming a perfectly reflective front mirror $M_0$. In this case no outside field can couple into the cavity and the field operator vector $\hat {\bf e}(k)=(\hat e_1(k),\hat e_2(k))^{\rm T}$ obeys the following matrix formula:
\equ{\label{eq:eTcMT}\mathbb{T}_c(k)\hat {\bf e}(k)=\hat 0,}
where $\mathbb{T}_c(k)$ is the closed form transfer matrix:
\equ{\mathbb{T}_c(k)=\begin{pmatrix}
1-ite^{ikL} & -re^{ikL} \\
-re^{ikL} & 1-ite^{ikL}
\end{pmatrix}.}
Solving this equation allows to find wave numbers $k_{\pm}$ of the resonant fields:
\begin{equation}
    k_{\pm} = \frac{1}{iL}\log(\pm r - it)
\end{equation}
The two corresponding resonant frequencies $\omega_{\pm}=c k_{\pm}$ within one free spectral range (FSR) $\Delta\omega_{\rm FSR}=2\pi c/L$ are separated by: 
\equ{\label{eq:omegasMT}\omega_s=\frac{c\arcsin r}{L}.} The two resonant modes have the following feature (see Eq.\,\eqref{eq:espatial}):
\begin{subequations}\label{eq:espatialMT}\begin{align}
\hat e_{1}(k_+)&=\hat e_{2}(k_+),\\
\hat e_{1}(k_-)&=-\hat e_{2}(k_-),
\end{align}\end{subequations}
and the field operators $\hat e_{1,2}(k_{\pm})$ are denoted by $\hat e_{1,2 \pm}$ in the following contents for notational convenience.

We assume that the membrane has a low reflectively ($r\ll 1$), which allows us to work only with two modes that are close to each other within one FSR, \textit{i.e.} $\omega_s\ll\Delta\omega_{\rm FSR}$, as shown approximately by the red line in FIG.\,\ref{fig:modesplit}. In this case, we do not need to consider the other optical resonances out of one FSR. We then assign $\hat{c}_{\pm}$ to represent the annihilation operator of the two cavity modes with optical frequencies $\omega_{\pm}$. The only nonzero commutators between them are:
\equ{[\hat{c}_-, \hat{c}_-^{\dagger}]=1,\ \ \textrm{and}\ \ [\hat{c}_+, \hat{c}_+^{\dagger}]=1.}
According to Eq.\,\eqref{eq:espatialMT}, $\hat c_+$ is named symmetric mode and $\hat c_-$ is named antisymmetric mode. $\hat c_{\pm}$ can be constructed from $\hat e_{1,2\pm}$ fields :
\begin{subequations}\label{}
\begin{align}
\hat c_+\equiv \frac{\hat e_{2+}+\hat e_{1+}}{\sqrt{2}},\\
\hat c_-\equiv \frac{\hat e_{2-}-\hat e_{1-}}{\sqrt{2}},
\end{align}
\end{subequations}
such that $\hat e_{2+}=\hat c_+/\sqrt{2}$ and $e_{2-}=\hat c_-/\sqrt{2}$. 

\begin{figure}
	\begin{center}
		\includegraphics[scale=0.375]{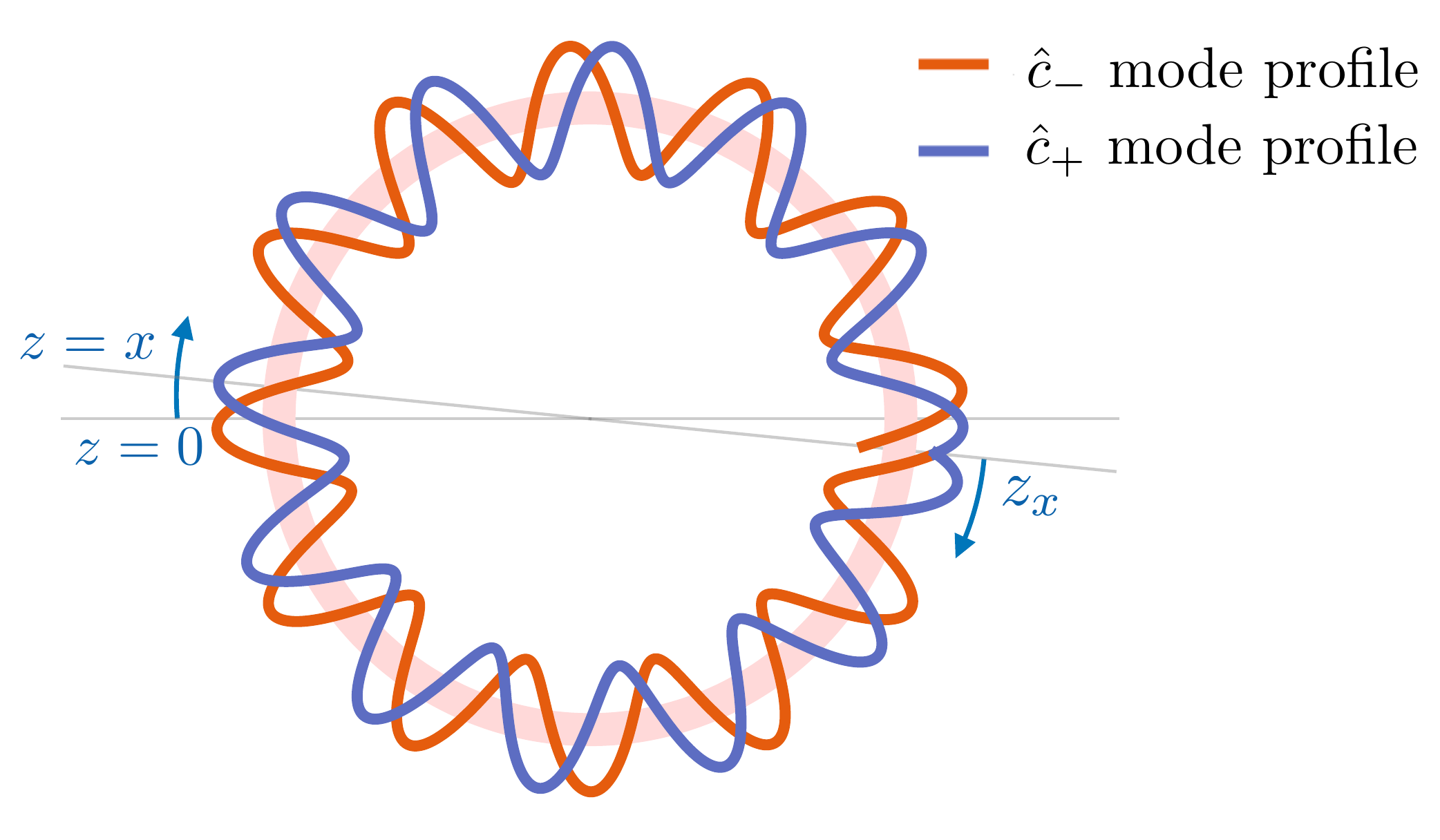}
		\caption{\label{fig:modeprofile}The illustrating plot of $\hat c_{\pm}$ wavefunction $P_{\pm}(z;x)$ in Eq.\,\eqref{eq:PmPp}. We use here a circle instead of a triangle to represent the space inside the ring cavity for plotting convenience. The coordinate system is the same as that in FIG.\,\ref{fig:ringconfig}: $z=0$ is the position of the front mirror $M_0$ and $z_x$ is the position of the membrane. The position of the nodes for both symmetric and antisymmetric modes are shifted with the position of the membrane $z_x$. Starting from $z_x$ and going in two directions, the standing wave amplitude of $\hat c_{+(-)}$ mode remains the same (opposite sign), until it reach maximum (zero) at $z=x$.}
	\end{center}
\end{figure}

To quantitatively describe the electric field distribution, we introduce a coordinate system inside the ring cavity, as shown in FIG.\,\ref{fig:ringconfig}. The origin of this $z$-coordinate is the front mirror $M_0$ and it increases clockwise along the optical axis of the ring cavity. It becomes $z_x= L/2+x$ at the instantaneous position of the membrane and finally becomes $z=L$ when it reaches the front mirror again. The coordinate system here is circular and thus $z=L$ represents the same position as $z=0$. The electric field inside the ring cavity can be represented by the standing wave distribution of two optical modes (see Eq.\,\eqref{eq:EfromcP}):
\equ{\label{eq:EfromcPMT}\hat{E}^+(z;x)=\mathcal{N}(\omega_-)P_-(z;x)\hat c_-+\mathcal{N}(\omega_+)P_+(z;x)\hat c_+,}
where $\mathcal{N}(\omega)=\sqrt{\hbar \omega/2 \mathcal{A} \epsilon_0L}$ is the frequency-dependent normalization factor for a beam with cross-sectional area $\mathcal{A}$ inside the ring cavity, and $P_-(z;x),P_+(z;x)$ are the wavefunctions of $\hat c_-,\hat c_+$ modes along $z$ axis:
\begin{subequations}\label{eq:PmPpMT}
\begin{align}
&P_-(z;x)=\left\{\begin{array}{lr}
2i\sin(k_-(z-x)) &z\in(0,z_x),\\
2i\sin(k_-(z-L-x)) &z\in (z_x,L),\\
\end{array}\right.\\
&P_+(z;x)=\left\{\begin{array}{lr}
2\cos (k_+(z-x)) &z\in(0,z_x),\\
2\cos(k_+(z-L-x)) &z\in (z_x,L).\\
\end{array}\right.
\end{align}\end{subequations}
$P_{\pm}(z;x)$ also represent the electric field standing wave distribution and are qualitatively shown in FIG.\,\ref{fig:modeprofile}. The position of the nodes for both symmetric and antisymmetric modes are shifted with the membrane position $z_x$ and $P_{\pm}(z;x)$ have the following features:
\begin{subequations}\begin{align}\label{eq:PmPpsymmetryMT}
P_{\pm}(z_x-\zeta\ {\rm mod}\ L;x)&=\pm P_{\pm}(z_x+\zeta\ {\rm mod}\ L;x),\\
|P_-(z=x;x)|&=0,\\
|P_+(z=x;x)|&=\max\limits_{z\in(0,L)}|P_+(z;x)|,
\end{align}\end{subequations}
where $\zeta$ represents the distance from an arbitrary point to the membrane. That is, starting from $z_x$ and going in two directions, the standing wave amplitude of $\hat c_{+(-)}$ mode remains the same (opposite sign), until it reaches the maximum (zero) at $z=x$, which is $L/2$ away from $z_x$ both clockwise and counterclockwise. The standing wave feature of $\hat c_{+(-)}$ mode agrees with the naming of (anti)symmetric mode. 

The cavity Hamiltonian can be obtained from the total optical energy inside the ring cavity\,\cite{cheung2011nonadiabatic} and it reads:
\equ{\label{eq:H0MT}\hat{H}_{\rm cav}=\hbar \omega_-\hat{c}_-^\dagger\hat{c}_-+\hbar \omega_+\hat{c}_+^\dagger\hat{c}_+.}
It doesn't have $x$-dependence because the ring cavity is a closed quantum system until now, as shown by the $x$-independent equation Eq.\,\eqref{eq:eTcMT} that we start from.

To obtain the total Hamiltonian and reveal the $x$-dependence, we consider the coupling of the cavity modes to the outside modes by assuming the front mirror to have low transmittance ($t_0\ll 1$). The cavity linewidth $\gamma$ can be obtained from the input-output relation (see Eq.\,\eqref{eq:efroma}):
\equ{\label{eq:gammaMT}\gamma=\frac{ct_0^2}{2L},}
and the ring cavity, as an open passive system, only supports the inside field with the pumping frequency $\omega_p=k_p c$. The extent to which $\hat c_{\pm}$ modes are excited depends on the detuning of the pumping frequency to the resonant ones: $\omega_p-\omega_{\pm}$. In the following contents we will use the wavevector $k_p$ of the pumping field instead of the resonant wavevectors $k_{\pm}$. We use $\hat{c}_{1,2}$ to represent the counterclockwise and clockwise propagating fields that the environment fields $\hat{a}_{1,2}$ directly couples to. Thus the cavity-environment interaction Hamiltonian can be expressed as:
\begin{equation}\label{eq:HacMT}
\hat{H}_{\gamma}=i\hbar\sqrt{2\gamma}(\hat c_1^{\dagger}\hat a_1-{\rm h.c.})+i\hbar\sqrt{2\gamma}(\hat c_2^{\dagger}\hat a_2-{\rm h.c.}).
\end{equation}
The front mirror position $z=0$ is a natural choice of phase reference point for modes $\hat{c}_{1,2}$. 
However, the resonant modes $\hat c_{\pm}$ take the membrane position $z_x$ as the phase reference point, as shown in Eq.\,\eqref{eq:PmPpsymmetryMT}. Thus, the transformation between $\hat{c}_{1,2}$ and $\hat c_{\pm}$ depends on $x$: 
\equ{\label{eq:cmpxparaMT}\hat{c}_{\pm}(x)=\frac{1}{\sqrt{2}}e^{ik_pL/2}(e^{-ik_px}\hat{c}_1\pm e^{ik_px}\hat{c}_2),}
where $\hat{c}_{\pm}(0)$ is the \emph{original} cavity modes that the outside modes $\hat{a}_{1,2}$ directly couple to and they have distribution $P_{\pm}(z;x=0)$. For the outside modes, $\hat{c}_{\pm}(x)$ are the new resonant modes when the membrane is displaced by $x$. It is equivalent to saying that the mechanical oscillation $x$ changes the way of interference between $\hat c_{1,2}$ that leads to the formation of different resonant modes $\hat{c}_{\pm}(x)$. 

The cavity optomechanical Hamiltonian linearized with respect to $x$ (see Eq.\,\eqref{eq:modemixing}) reads:
\equ{\label{eq:ringHoptMT}\hat{H}_{\rm opt}(x)=\hat{H}_0+ \hat{H}_{\rm int}(x),}
where the free part $\hat{H}_0$ is equivalent to Eq.\,\eqref{eq:H0MT} with $\hat{c}_{\pm}\to\hat{c}_{\pm}(0)$ and the optomechanical interaction part $\hat{H}_{\rm int}(t)$ reads:
\equ{\label{eq:HintMT}\hat{H}_{\rm int}(x)=2i\omega_s\hbar k_p x\Big(\hat{c}_-^\dagger(0)\hat{c}_+(0)-\textrm{h.c.}\Big).}
The feature of coherent coupling is shown in Eq.\,\eqref{eq:HintMT} explicitly: the mechanical oscillation $x$ induces the coupling between two original optical modes $\hat{c}_\pm(0)$ which have non-degenerate frequencies.

In the derivation until now, $x$ merely works as a parameter. Alternatively, one can start from the total Lagrangian including the mechanical degree of freedom and follow the canonical formulation\,\cite{law1995interaction,khorasani2017higher,pang2018theoretical}, $x$ can be upgraded to be a dynamical variable and further becomes a quantum operator $\hat x$ after quantization. 

To describe the system in the general framework of Eq.\,\eqref{eq:generalH}, we need to express the total Hamiltonian in terms of the new resonant modes $\hat c_{\pm}(\hat x)$. Applying similar transformation as in Eq.\,\eqref{eq:cmpxparaMT}, we can express the input modes $\hat{a}_{1,2}$ into antisymmetric and symmetric ones:
\equ{\label{eq:cpmina12MT}\hat{c}_{\pm\rm in}(\hat x)=\frac{1}{\sqrt{2}}e^{ik_pL/2}(e^{-ik_p\hat x}\hat{a}_1\pm e^{ik_p\hat x}\hat{a}_2).}
Up to linear order in $\hat x$, the cavity-environment interaction Hamiltonian in Eq.\,\eqref{eq:HacMT} can be transformed to (see Eq.\,\eqref{eq:cinx}):
\begin{equation}\label{eq:ringHdissMT}
\hat H_{\gamma}=i\hbar\sqrt{2\gamma} \Big(\hat c_-^{\dagger}(\hat x)\hat c_{-\rm in}(\hat x)+c_+^{\dagger}(\hat 0)\hat c_{+\rm in}(\hat x)-{\rm h.c.}\Big).
\end{equation}
It is clear from Eq.\,\eqref{eq:ringHdissMT} that the coupling rate with the environment doesn't depend on $\hat x$. Thus, the optomechanical coupling has no dissipative feature. 

To sum up, the total Hamiltonian reads:
\equ{\label{eq:ringHtotal}\hat H(x)=\hat H_{\rm opt}(x)+\hat H_{\gamma}+\hat H_m,}
where the cavity optomechanical part $\hat H_{\rm opt}(x)$ is given in Eq.\,\eqref{eq:ringHoptMT}, the cavity-environment interaction part $\hat H_{\gamma}$ is given in Eq.\,\eqref{eq:ringHdissMT} and the free mechanical part is:
\equ{\hat H_m=\frac{\hat p^2}{2m}+\frac{1}{2}m \Omega_m^2\hat x^2-G\hat x}
with $G$ representing any external force exerted on the mechanical oscillator. 
If we write it in the canonical form of Eq.\,\eqref{eq:generalH}, we obtain:
\equ{
\mathbb{\Omega}=\left(
\begin{matrix}
\omega_- & 0 \\
0 & \omega_+
\end{matrix}
\right),\ \
\mathbb{\Gamma}=
\left(
\begin{matrix}
i\sqrt{2\gamma} & 0 \\
0 & i\sqrt{2\gamma}
\end{matrix}
\right),}
and the mode operators are $\hat {\bf{a}}(\hat x)=(\hat c_-(\hat x),\hat c_+(\hat x))^{\rm T}$, $\hat {\bf{a}}_{\rm in}(\hat x)=(\hat c_{-\rm in}(\hat x),\hat c_{+\rm in}(\hat x))^{\rm T}$. There is neither dispersive nor dissipative feature in the Hamiltonian above, and thus the coherent coupling is verified.

In the next section, we will discuss the advantage of coherent coupling in enhanced optomechanical cooling.

\subsection{Application: enhanced cooling}
Optomechanical cooling provides a zero temperature bath through the laser light to remove the thermal noise and cool down the mechanical oscillation to its ground state\,\cite{scully_zubairy_1997}. It contributes to fundamental physics in studying the quantum effects of macroscopic objects\,\cite{chen2013macroscopic,purdy2013observation}. It is also beneficial in the application aspect of frequency conversion\,\cite{tian2010optical,hill2012coherent,ockeloen2016low,lecocq2016mechanically} and quantum information processing\,\cite{huang2018dissipative}.

In the Hamiltonian linearized with respect to $x$, the coherent coupling starts with two non-degenerate optical modes and then couples them by mechanical oscillation. This coupling doesn't change the resonance frequency up to linear order in $\hat x$. Thus, the double resonance structure of coherent coupling systems can potentially provide a more efficient cooling compared with the standard dispersive-coupling-based cooling\,\cite{yong2013review}, because of the additional resonant enhancement of the pumping field: 
When the mechanical frequency matches the frequency distance between the two resonance peaks and the lower frequency is pumped, both the pumping field and the upper mechanical sideband are resonant inside the cavity.

The optoacoustic interaction\,\cite{miao2009quantum} in Sec.\,\ref{sec:coherent} has similar physics properties with the ring cavity system. Contrary to the cooling described above, when the cavity mode with upper frequency is pumped, both the pumping field and the lower mechanical sideband are resonant and the enhanced heating occurs. That explains the principle of parametric instability\,\cite{braginsky2001parametric,evans2015observation}. Different from the optoacoustic interaction which influences the transverse mechanical oscillation, the ring cavity system interact with the longitudinal mechanical oscillation. In this section, we will focus on the cooling of the ring cavity system and compare it with the single cavity dispersive coupling case.


The detailed derivation of optomechanical cooling and mechanical occupation number limit in the ring cavity system can be found in Appendix\,\ref{sec:coolinglim}. Here we only list the main results. Under the resolved sideband condition $\Omega_m \gg \gamma$, we obtain the optical damping rate (see Eq.\,\eqref{eq:gammaopt}): 
\equ{\label{eq:gammaoptMT}\gamma_{\rm opt}=\frac{2|A_{\rm in}|^2k_p^2\hbar \omega_s}{m\gamma^2},}
such that the equation of motion for mechanical operator $\hat x$ becomes:
\equ{\label{eq:xEOMeffMT}m\ddot{\hat x}=\hat F_{\rm bafl}-m\Omega_m^2\hat x-m(\gamma+\gamma_{\rm opt})\dot{\hat x},}
where $\hat F_{\rm bafl}$ is the $x$-independent part of fluctuating backaction force (see Eq.\,\eqref{eq:EOMeff} and the contents below it). The mechanical occupation number can be expressed as:
\equ{\left<\hat n\right>=\frac{\gamma_{\rm opt}}{\gamma_m+\gamma _{\rm opt}}\frac{1}{2}\left[\frac{\gamma ^2}{4 \Omega_m^2}-\frac{\gamma_m}{\gamma_{\rm opt}}\right]+\frac{\gamma_m}{\gamma_m+\gamma_{\rm opt}}\frac{k_B T}{\Omega_m \hbar}.}
Under further condition $\gamma_{\rm opt}\gg\gamma_{\text{m}}$, we can obtain the ultimate cooling limit:
\equ{\left<\hat n\right>=\frac{\gamma_m}{\gamma_m+\gamma_{\rm opt}}n_{\rm th}+\frac{\gamma_{\rm opt}}{\gamma_m+\gamma _{\rm opt}}n_{\rm ba}\approx n_{\rm ba},}
where $n_{\rm th}=k_B T/\Omega_m \hbar$ is the thermal occupation number and $n_{\rm ba}=\gamma ^2/8 \Omega_m^2$ is the back-action limited occupation number.

\begin{figure}
	\begin{center}
		\includegraphics[scale=0.4]{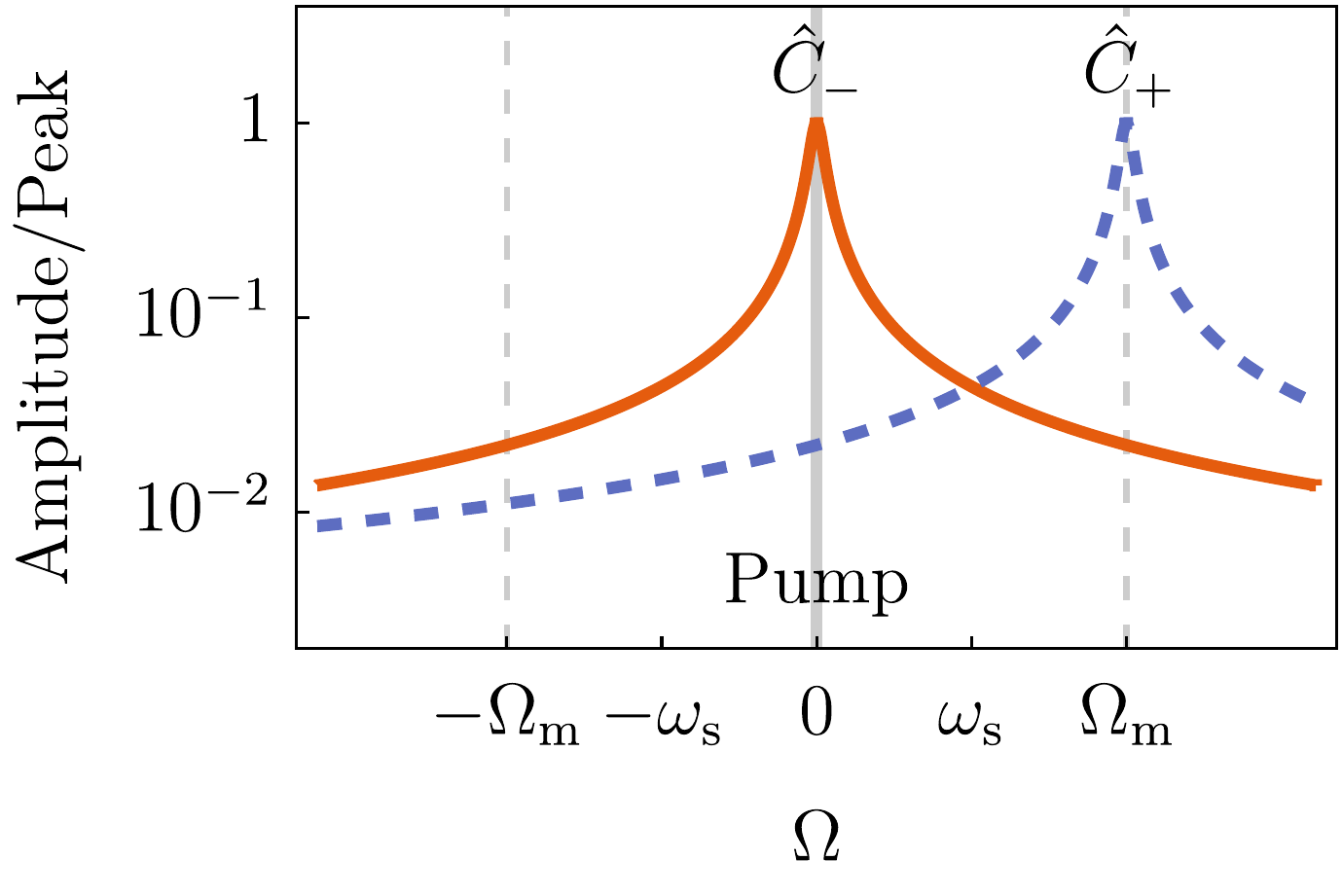}
		\caption{\label{fig:cooling}Pumping regime of sideband cooling. Coherent coupling has a potential advantage over dispersive coupling in sideband cooling because the pumping frequency is also resonant inside the cavity. Dispersive coupling only has the right resonance peak shown by the blue dashed line.}
	\end{center}
\end{figure}

We then compare the cooling rate in ring cavity (with coherent coupling) and the one in a single cavity (with dispersive coupling). We assume the two systems have the same optical bandwidth $\gamma$ and similar \emph{round trip length} $L,L_{\rm sc}$, and are used to cool a mechanical oscillator with the same resonant frequency $\Omega_m$. Both of the two systems are pumped with frequency $\omega_p$. For the ring cavity case, $\omega_p=\omega_-$ and the pumping is injected from the left port as analyzed above. For the single cavity case, $\omega_p$ is red detuned by $\Omega_m$ from its resonance. In both cases, $\Omega_m\ll \Delta\omega_{\rm FSR}$ and thus the two-mode or single mode approximation is feasible. The ultimate occupation number $n_{\rm ba}$ is determined by the ratio between $\gamma$ and $\Omega_m$ and is the same in the two cases. The advantage of coherent coupling is the simultaneous resonant enhancement of pumping field and the upper sideband, which can support higher intracavity field and thus provide larger optical damping $\gamma_{\rm opt}$. The intracavity field amplitudes in two cases are:
\begin{subequations}\label{eq:RSintracavityfields}\begin{align}
C_-&=\frac{A_{\rm in}}{\sqrt{\gamma}} &&\textrm{in the ring cavity}, \\
A_{\rm sc}&=\frac{\sqrt{2\gamma}A_{\rm in}}{\gamma+i\Omega_m}\approx\frac{\sqrt{2\gamma}A_{\rm in}}{i\Omega_m} &&\textrm{in the single cavity},
\end{align}\end{subequations}
which are related as $|C_-|\gg|A_{\rm sc}|$ for the same input amplitude $A_{\rm in}$. The optical damping rates of the ring cavity and the single cavity are:
\begin{subequations}\begin{align}\label{eq:optrcMT}
\gamma_{\rm opt,rc}=\frac{2\omega_sk_p^2 \hbar}{m \gamma}|C_-|^2,\\
\label{eq:optscMT}\gamma_{\rm opt,sc}=\frac{g_{\rm sc}^2\hbar }{m \gamma \Omega_m}|A_{\rm sc}|^2,
\end{align}\end{subequations}
where $2\omega_s=\Omega_{m}$ is the setting of ring cavity resonance and 
$g_{\rm sc}=2 \omega_p/L_{\rm sc}$ is the dispersive coupling strength (see Eq.\,\eqref{eq:gdisper}) expressed in cavity round trip length $L_{\rm sc}$. According to Eq.\,\eqref{eq:HintMT} with coherent coupling strength defined as $g_{\rm rc}=2 i \omega_s k_p$, the damping rate in Eq.\,\eqref{eq:optrcMT} takes the form:
\equ{\gamma_{\rm opt,rc}=\frac{|g_{\rm rc}|^2 \hbar}{m \gamma\Omega_m}|C_-|^2,}
which has the same as Eq.\,\eqref{eq:optscMT}. Assuming the same single photon coupling rates, 
the advantage of intracavity resonance in the coherent coupling case described in Eq.\,\eqref{eq:RSintracavityfields} shows up. Substituting Eq.\,\eqref{eq:RSintracavityfields} in, we obtain the damping rates under the same input amplitude:
\begin{subequations}\begin{align}
\gamma_{\rm opt,rc}&=\frac{\Omega_m k_p^2 \hbar  }{m \gamma^2}|A_{\rm in}| ^2,\\
\gamma_{\rm opt,sc}&=\frac{8 \omega_p^2 \hbar }{m L_{\rm sc}^2 \Omega_m^3}|A_{\rm in}|^2.
\end{align}\end{subequations}
The ratio between the damping rates of the two cases is:
\equ{R_{\rm rc/sc}=\frac{\Omega_m^4 L_{\rm sc}^2}{8c^2 \gamma^2}.}
The ring cavity has the advantage in cooling efficiency over a single cavity so long as $\Omega_m$ is larger than the geometric mean of $\Delta\omega_{\rm FSR}$ and $\gamma$, \textit{i.e.} $\Omega_m\gg\sqrt{\Delta\omega_{\rm FSR} \gamma}$. Because of Eqs.\,\eqref{eq:omegasMT}\eqref{eq:gammaMT}, the ratio above can also be expressed as:
\equ{R_{\rm rc/sc}=\frac{8 L_{\rm sc}^2 (\arcsin r)^4}{L^2 t_0^4} = \frac{L_s^2 L^2 \Omega_m^4}{2t_o^4 c^4}.}

These equations demonstrates that the ring cavity can provide benefit in a larger scale optomechanical setup with long cavities or with high-frequency mechanical oscillators. For example, if we compare a single and a ring cavity with a mechanical membrane as an oscillator of frequency of $2.5$\,MHz \cite{nielsen2017multimode}, front mirror transmission of $0.01$\%, and an equal length of $\sim 40$cm, we find that the cooling rate in the ring cavity is 2.4 times higher than in a single cavity. 
The ring cavity thus could be beneficial for long cavities, used, \textit{e.g.} as optomechanical filters for gravitational-wave detectors \cite{Miao2015}.

\section{\label{sec:discussion}Discussion}
In this paper, we build an unambiguous framework for classifying the optomechanical interaction in a unique way. This framework prescribes to express each Hamiltonian in the canonical form and examine the dependence of its terms on the mechanical oscillation. The canonical form of the Hamiltonian is unique for each system, hence the classification based on that is mutually exclusive. No ambiguity in classifying similar systems, as it was illustrated in the Introduction, can occur.
There are some limitations in our classification framework: we only consider Hamiltonians linear in $x$ and linear in optical modes. Quadratic optomechanical coupling and nonlinear optical effects, for example, are not covered in our framework and need further consideration. 

Based on our framework, we analyze several optomechanical systems, including the newly investigated ring cavity system which exhibits purely coherent coupling. We show that coherent coupling is fundamentally different from either dispersive or dissipative coupling and allows to complete our classification framework.
Our \mbox{analysis} reveals a previously underestimated relevance of coherent coupling in optomechanical systems. It will show up whenever the system has two or more optical modes with non-degenerate frequencies get coupled by mechanical oscillation. We show that although coherent coupling occurs even in some well-studied systems, yet has never been identified as such. For instance, in the system of two coupled cavities with a movable central mirror, both dispersive and coherent coupling coexist, as we show in the Sec.\,\ref{sec:Coexisting}.

The nature of the optomechanical coupling defines the strengths and weaknesses of the system in one desired application. Our classification strategy will allow to approach the experimental design in a systematic way and choose the system that would perform optimally. As a concrete result, we show that coherent coupling allows for more effective laser cooling of the mechanical oscillation due to the simultaneous resonant of pumping field and the upper mechanical sideband.

We anticipate our classification framework to serve as a methodological and practical guide in the growing field of optomechanics. We believe the recognition of the highlighted coherent coupling will lead to the development of novel quantum optomechanical systems and new parameter regimes in the existing ones.


\emph{Acknowledgements} We thank Farid Khalili for many useful discussions. 
The work of M.K. and R.S. was supported by the Deutsche Forschungsgemeinschaft (DFG) (SCHN 757/6-1). X.L., Y.M. and Y.C. are supported by the National Science Foundation through Grants PHY-1612816, PHY-1708212 and PHY-1708213, the Brinson Foundation, and the Simons Foundation (Award Number 568762).


\appendix
\section{\label{sec:HamDisper}Dispersive Hamiltonian}
The resonance condition gives wave vector $k_n(x)=n\pi /(L+x)$ and the corresponding resonant frequency $\omega_n(x)=ck_n(x)$. The electric field inside a single cavity is the superposition of modes with multiple resonant frequencies:
\equ{\hat{E}^{+}(z)=\sum_n\sqrt{\frac{\hbar\omega_n(x)}{4 \mathcal{A} \epsilon_{0} (L+x)}} 2\cos(k_n(x)z)\hat{a}_n,}
with $z$ ranging in $(0,L+x)$. The counterintuitive node antinode distribution is because we choose zero reflection phase in two cavity mirrors. 

The cavity Hamiltonian can be obtained from the total optical energy\,\cite{cheung2011nonadiabatic}:
\equ{\label{eq:HLx}\hat{H}_{\rm cav}(x)=\int_{0}^{L+x} 2 \mathcal{A} \epsilon_{0} \hat{E}^{-}(z) \hat{E}^{+}(z)dz,}
where the factor $2$ accounts for both electric and magnetic energy. The integration result of Eq.\,\eqref{eq:HLx} is:
\equ{\hat{H}_{\rm cav}(x)=\sum_n \hbar\omega_n(x)\hat{a}_n^{\dagger}\hat{a}_n.}
Under single mode approximation, only one specific mode $n$ is considered and we obtain the cavity Hamiltonian:
\equ{\label{eq:singlecavityH}\hat H_{\rm cav}(x)=\hbar(\omega_a-g_{\omega}x)\hat a^\dagger\hat a,}
by defining $\omega_a=\omega_n(0)$ for mode $n$. The dispersive coupling strength $g_{\omega}$ can be extracted from the expressions above as:
\equ{\label{eq:gdisper}g_{\omega}=\frac{\omega_a}{L}.}

\section{\label{sec:HamParaInst}Optoacoustic Hamiltonian}
The Hamiltonian of three-modes optoacoustic interaction\,\cite{miao2009quantum} in Eq.\,\eqref{eq:Hthreemode} is originated from the following cavity integral:
\be
H\propto\int dr_\perp (L+xu_z)\bigg[\epsilon(E_0+E_1)^2+\frac{1}{\mu}(B_0+B_1)^2\bigg],
\ee
where $u_z=u_z(\vec{r}_{\perp})$ is the transverse spatial profile of mechanical oscillation and $(E_0,H_0),(E_1,H_1)$ are two optical modes $\mathrm{TEM}_{00}, \mathrm{TEM}_{01}$ with orthogonal transverse profile $f_0(\vec{r}_{\perp}), f_1(\vec{r}_{\perp})$. The dispersive coupling and the three-modes coupling in this system appear as:
\be
\begin{split}
&H_{\rm disp}\propto\sum_{0,1}\int d\vec{r}_{\perp} xu_z \left[\epsilon E_{0,1}^2+\frac{1}{\mu}B_{0,1}^2\right],\\
&H_{\rm 3-mode}\propto\int d\vec{r}_{\perp} xu_z \left[\epsilon E_0E_1+\frac{1}{\mu}B_0B_1\right].
\end{split}
\ee
Apparently, they contain overlapping function $\Lambda_{0,1}=\int d\vec{r}_{\perp} u_zf_{0,1}^2$ and $\Lambda_{01}=\int d\vec{r}_{\perp} u_z f_0f_1$, respectively. In general, both terms should exist. The reason for the vanishing of dispersive coupling here is simply because of the vanishing of the overlapping function $\Lambda_{0,1}$. In Eq.\,\eqref{eq:Hthreemode} the coupling constant is defined as $G_{0} \equiv \sqrt{\Lambda \hbar \omega_{0} \omega_{1} /\left(m \Omega_{m} L^{2}\right)}$ with the geometrical overlapping factor $\Lambda \equiv\left(L \Lambda_{01} / V\right)^{2}$.


\section{\label{sec:ringH}Hamiltonian derivation of ring cavity system}
\subsection{Input-output relation}
We start the rigorous derivation by writing down the input-output relations\,\cite{chen2013macroscopic} for the coupling of incoming electromagnetic fields $\hat a_{1,2}(k)$ of wavenumber $k$ to all intracavity fields shown in Fig.\,\ref{fig:ringconfig}. Unless claimed otherwise, we will view $x$ as a parameter in the following contents as the two counterpropagating fields $\hat e_{1,2}(k)$ are defined at the instantaneous position of the membrane. The input-output relations can be derived from the frequency space field transfer matrices:
\begin{subequations}
\begin{align}
    \hat {\bf e}(k)&=\mathbb{T}_{ec}(k;x)\hat {\bf c}(k),\\
    \hat {\bf c}(k)&=\mathbb{T}_{cf}(k;x)\hat {\bf f}(k)+\mathbb{T}_{ca}\hat {\bf a}(k),\\
    \hat {\bf f}(k)&=\mathbb{M}\hat {\bf e}(k),
\end{align}
\end{subequations}
where the field vectors are:
\equ{\begin{split}
\hat {\bf e}(k)&=\begin{pmatrix}
\hat e_{1}(k)\\\hat e_{2}(k)
\end{pmatrix},\ \hat {\bf f}(k)=\begin{pmatrix}
\hat f_{1}(k)\\\hat f_{2}(k),
\end{pmatrix}\\
\hat {\bf c}(k)&=\begin{pmatrix}
\hat c_{1}(k)\\\hat c_{2}(k)
\end{pmatrix},\ \hat {\bf a}(k)=\begin{pmatrix}
\hat a_{1}(k)\\\hat a_{2}(k)
\end{pmatrix},\end{split}}
and the transfer matrices read:
\begin{equation}\label{eq:Tmatrices}\begin{split}
\mathbb{T}_{ec}(k;x)&=\begin{pmatrix}
0 & e^{ik(L/2+x)} \\
e^{ik(L/2-x)} & 0
\end{pmatrix},\\
\mathbb{T}_{cf}(k;x)&=\begin{pmatrix}
r_0 e^{ik(L/2+x)} & 0 \\
0 & r_0 e^{ik(L/2-x)}
\end{pmatrix},\\
\mathbb{T}_{ca}&=\begin{pmatrix}
t_0 & 0 \\
0 & t_0
\end{pmatrix},\ \mathbb{M}=\begin{pmatrix}
r & it \\
it & r
\end{pmatrix}.
\end{split}\end{equation}
The $\hat e_{1,2}(k)$ fields thus take the form:
\equ{\label{eq:einputoutput}\hat {\bf e}(k)=\mathbb{T}_p(k)\hat {\bf e}(k)+\mathbb{T}_{\rm in}(k;x)\hat {\bf a}(k),}
where $\mathbb{T}_p(k)=\mathbb{T}_{ec}(k;x)\mathbb{T}_{cf}(k;x)\mathbb{M}$ is the transfer matrix describing the circulation of $\hat {\bf e}(k)$ inside the cavity and it no longer has $x$-dependence; $\mathbb{T}_{\rm in}(k;x)=\mathbb{T}_{ec}(k;x)\mathbb{T}_{ca}$ is the transfer matrix describing the process of $\hat {\bf a}(k)$ coupling into the cavity and propagating to join $\hat{\bf e}(k)$. Eq.\,\eqref{eq:einputoutput} can further be written as:
\equ{\label{eq:efroma}\hat {\bf e}(k)=\mathbb{T}_c(k)^{-1}\mathbb{T}_{\rm in}(k;x)\hat {\bf a}(k).}
where $\mathbb{T}_c(k)$ is the closed form transfer matrix defined as:
\equ{\label{eq:Tck}\mathbb{T}_c(k)\equiv \mathbb{I}-\mathbb{T}_p(k)=\begin{pmatrix}
1-ite^{ikL} & -re^{ikL} \\
-re^{ikL} & 1-ite^{ikL}
\end{pmatrix},}
and $\mathbb{T}_c(k)^{-1}$ works as the feedback kernel that describes the effect of cavity circulation.

\subsection{Resonance structure}
To derive the Hamiltonian, we first consider resonant cavity modes assuming a perfectly reflective front mirror $M_0$. 
In this case no outside field can couple in, \textit{i.e.} $\mathbb{T}_{\rm in}(k;x)\equiv\mathbb{0}$, and therefore Eq.\,\eqref{eq:einputoutput} becomes:
\equ{\label{eq:eTc}\mathbb{T}_c(k)\hat {\bf e}(k)=\hat 0.}
The resonance condition can be obtained from the nontrivial solutions of Eq.\,\eqref{eq:eTc} which requires $\det \mathbb{T}_c=0$. Within one FSR, the ring cavity can support two resonances with different propagation phases:
\begin{subequations}\label{eq:kpmcpm}
\begin{align}
& e^{ik_+L}=r-it,\\
& e^{ik_-L}=-r-it,
\end{align}
\end{subequations}
where $k_{\pm}$ are the resonant wavenumbers that depend only on the total cavity length $L$ and the optical property $r,t$ of the membrane. 
The distance between the frequencies $\omega_{\pm}=c k_{\pm}$ of the two modes within one FSR is:
\equ{\label{eq:omegas}\omega_s\equiv \frac{\omega_+-\omega_-}{2} =\frac{c\arcsin r}{L}.}
To work in a parameter regime which only involves two optical resonances closely separated within one FSR, as shown approximately by the red line in FIG.\,\ref{fig:modesplit}, we assume the membrane to have low reflectivity $r\ll 1$ such that $\omega_s\ll\Delta\omega_{\rm FSR}$. In this case, we don't need to consider the other optical resonances out of one FSR. We then assign $\hat{c}_{\pm}$ to represent the annihilation operator of the two cavity modes with optical  frequencies $\omega_{\pm}$. The only nonzero commutators between them are:
\equ{[\hat{c}_-, \hat{c}_-^{\dagger}]=1,\ \ \textrm{and}\ \ [\hat{c}_+, \hat{c}_+^{\dagger}]=1.}
The relation between $\hat e_{1,2}$ in the two resonant modes $\hat{c}_{\pm}$ can be obtained by plugging $k_\pm$ into Eq.\,\eqref{eq:Tck} and Eq.\,\eqref{eq:eTc} thus becomes:
\equ{\begin{pmatrix}1&-1\\-1&1\end{pmatrix}\begin{pmatrix}\hat e_1(k_+)\\\hat e_2(k_+)\end{pmatrix}=\hat 0,\ \ \begin{pmatrix}1&1\\1&1\end{pmatrix}\begin{pmatrix}\hat e_1(k_-)\\\hat e_2(k_-)\end{pmatrix}=\hat 0.}
The two solutions $\hat{\bf e}(k_{\pm})$ have the following feature:
\begin{subequations}\label{eq:espatial}\begin{align}
\hat e_{1}(k_+)&=\hat e_{2}(k_+),\\
\hat e_{1}(k_-)&=-\hat e_{2}(k_-).
\end{align}\end{subequations}
The operator vector $\hat{\bf f}(k)=(\hat f_{1}(k),\hat f_{2}(k))^{\rm T}$ is defined similarly as $\hat{\bf e}(k)$, containing two fields $\hat f_{1,2}(k)$ that propagates away from the membrane, as shown in FIG.\,\ref{fig:ringconfig}. $\hat{\bf f}(k)$ satisfies $\mathbb{T}_c(k)\hat{\bf f}(k)=\hat 0$, the same as $\hat{\bf e}(k)$ in Eq.\,\eqref{eq:eTc}, and its two solutions have the same feature as $\hat{\bf e}(k_\pm)$ in Eq.\,\eqref{eq:espatial}. These field operators $\hat e_{1,2}(k_{\pm}),\hat f_{1,2}(k_{\pm})$, represented by $\hat e_{1,2 \pm},\hat f_{1,2 \pm}$ in the following contents for convenience, are related by the propagation phases:
\begin{subequations}\label{eq:efrelation}\begin{align}
\hat e_{1\pm}&=e^{ik_\pm L}\hat f_{2\pm},\\
\hat e_{2\pm}&=e^{ik_\pm L}\hat f_{1\pm}.
\end{align}\end{subequations}
According to the spatial features mentioned above, $\hat c_+$ is named symmetric mode and $\hat c_-$ is named antisymmetric mode. They can be constructed from $\hat e_{1,2\pm}$ fields as follow:
\begin{subequations}\label{eq:cfrome}
\begin{align}
\hat c_+\equiv \frac{\hat e_{2+}+\hat e_{1+}}{\sqrt{2}},\\
\hat c_-\equiv \frac{\hat e_{2-}-\hat e_{1-}}{\sqrt{2}},
\end{align}
\end{subequations}
such that $\hat e_{2+}=\hat c_+/\sqrt{2}$ and $e_{2-}=\hat c_-/\sqrt{2}$ because of Eq.\,\eqref{eq:espatial}. We will use these expressions above to construct the electric field in the following contents.

\subsection{Electric field standing wave distribution}

To quantitatively describe the electric field distribution, we construct a coordinate system inside the ring cavity, as shown in FIG.\,\ref{fig:ringconfig}. The origin of this $z$-coordinate is the front mirror $M_0$ and it increases clockwise along the optical axis of the ring cavity. It becomes $z_x= L/2+x$ at the instantaneous position of the membrane and finally becomes $z=L$ when reaching the front mirror again. The coordinate system here is folded and thus $z=L$ represents the same position as $z=0$.

For a beam with cross-sectional area $\mathcal{A}$ inside the ring cavity, the frequency-dependent normalization factor is: \equ{\label{eq:Norm}\mathcal{N}(\omega)=\sqrt{\frac{\hbar \omega}{2 \mathcal{A} \epsilon_0L}},}
such that the positive frequency part of the electric field at any spatial coordinate $z$ can be written as\,\cite{xuereb2011dissipative}:
\equ{\label{eq:Eintegral}\hat E^+(z,t;x)=\int_0^{+\infty} d\omega\mathcal{N}(\omega) \tilde c (z,\omega/c;x) e^{-i\omega t},}
where the mode function $\tilde c (z,\omega/c;x)\equiv\hat c(z,k;x)e^{i\omega t}$ with $k=\omega/c$ can be constructed from $\hat e_{1,2}(k),\hat f_{1,2}(k)$ fields taking the membrane position $z_x$ as the phase reference point:
\equ{\label{eq:cfromef}\begin{split}&\hat c(z,k;x)=\\
&\left\{\begin{array}{lr}
e^{ik(z-z_x)}\hat e_1(k) + e^{-ik(z-z_x)}\hat f_1(k) &z\in(0,z_x),\\
e^{ik(z-z_x)}\hat e_2(k) + e^{-ik(z-z_x)}\hat f_2(k) & z\in (z_x,L),\\
\end{array}\right.\end{split}}
because the optical relaxation time is much less than the mechanical one and thus the field distribution can adjust itself simultaneously when $z_x$ changes. The field inside a perfect cavity is rigorously restricted by the resonance structure and thus has a discretized frequency space distribution, as shown in Eq.\,\eqref{eq:kpmcpm}. Therefore, instead of an integral over the whole spectrum as in Eq.\,\eqref{eq:Eintegral}, the electric field takes the summation of components with discretized frequencies $\omega_{\pm}$:
\equ{\label{eq:EfieldCom}\begin{split}&\hat{E}^+(z;x)=\sqrt{\frac{\hbar}{2\mathcal{A}\epsilon_0 L}}\times\\
&\left\{\begin{array}{lr}
\sum_{\pm} \sqrt{\omega_\pm} e^{ik_{\pm}(z-z_x)}\hat e_{1\pm} &\\
\ +\sum_{\pm} \sqrt{\omega_\pm} e^{-ik_{\pm}(z-z_x)}\hat f_{1\pm} &z\in(0,z_x),\\
\\
\sum_{\pm} \sqrt{\omega_\pm} e^{ik_{\pm}(z-z_x)}\hat f_{2\pm} &\\
\ +\sum_{\pm} \sqrt{\omega_\pm} e^{-ik_{\pm}(z-z_x)}\hat e_{2\pm} & z\in (z_x,L).\\
\end{array}\right.\end{split}}
Considering all Eqs.\,\eqref{eq:kpmcpm}\eqref{eq:espatial}\eqref{eq:efrelation}\eqref{eq:cfrome} and plugging them into Eq.\,\eqref{eq:EfieldCom}, where the process is actually a unitary transformation from $(\hat e_{\pm},\hat f_{\pm})$ basis to $(\hat c_+,\hat c_-)$ basis, the optical standing wave inside the ring cavity can be derived as:
\equ{\label{eq:Efieldx}\begin{split}&\hat{E}^+(z;x)=\sqrt{\frac{\hbar}{4\mathcal{A}\epsilon_0 L}}\times\\
&\left\{\begin{array}{lr}
2i\sin(k_-(z-x))\sqrt{\omega_-}\hat{c}_- &\\
\ +2\cos (k_+(z-x))\sqrt{\omega_+}\hat{c}_+ &z\in(0,z_x),\\
\\
2i\sin(k_-(z-L-x))\sqrt{\omega_-}\hat{c}_- &\\
\ +2\cos(k_+(z-L-x))\sqrt{\omega_+}\hat{c}_+ & z\in (z_x,L).\\
\end{array}
\right.\end{split}}
Or equivalently,
\equ{\label{eq:EfromcP}\hat{E}^+(z;x)=\mathcal{N}(\omega_-)P_-(z;x)\hat c_-+\mathcal{N}(\omega_+)P_+(z;x)\hat c_+,}
where $P_-(z;x)$ and $P_+(z;x)$ are the wavefunctions of the two modes $\hat c_-$ and $\hat c_+$ along $z$ axis: 
\begin{subequations}\label{eq:PmPp}
\begin{align}
&P_-(z;x)=\left\{\begin{array}{lr}
2i\sin(k_-(z-x)) &z\in(0,z_x),\\
2i\sin(k_-(z-L-x)) &z\in (z_x,L),\\
\end{array}\right.\\
&P_+(z;x)=\left\{\begin{array}{lr}
2\cos (k_+(z-x)) &z\in(0,z_x),\\
2\cos(k_+(z-L-x)) &z\in (z_x,L).\\
\end{array}\right.
\end{align}\end{subequations}
$P_{\pm}(z;x)$ repersent the electric field standing wave distribution and are qualitatively plotted in FIG.\,\ref{fig:modeprofile}. The position of the nodes for both symmetric and antisymmetric modes are shifted with the membrane position $z_x$ and $P_{\pm}(z;x)$ has the following features:
\begin{subequations}\label{eq:PpPmfeature}\begin{align}\label{eq:PmPpsymmetry}
P_{\pm}(z_x-\zeta\ {\rm mod}\ L;x)&=\pm P_{\pm}(z_x+\zeta\ {\rm mod}\ L;x),\\
|P_-(z=x;x)|&=0,\\
|P_+(z=x;x)|&=\max\limits_{z\in(0,L)}|P_+(z;x)|,
\end{align}\end{subequations}
where $\zeta$ represents the distance from an arbitrary point to the membrane. That is, starting from $z_x$ and going in two directions, the standing wave amplitude of $\hat c_{+(-)}$ mode remains the same (opposite sign), until it reach maximum (zero) at $z=x$, which is $L/2$ away from $z_x$ both clockwise and counterclockwise. The standing wave feature of $\hat c_{+(-)}$ mode agrees with the naming of (anti)symmetric mode.

\subsection{Conservative cavity Hamiltonian}

The cavity Hamiltonian can be obtained from the total optical energy inside the ring cavity\,\cite{cheung2011nonadiabatic}: 
\equ{\label{eq:Hintegral}\hat{H}_{\rm cav}=2\mathcal{A}\epsilon_0\int_0^{L}\hat{E}^-(z;x)\hat{E}^+(z;x)dz.}
Substituting Eq.\,\eqref{eq:Efieldx} in, we can obtain:
\equ{\label{eq:H0}\hat{H}_{\rm cav}=\hbar \omega_-\hat{c}_-^\dagger\hat{c}_-+\hbar \omega_+\hat{c}_+^\dagger\hat{c}_+.}
The cavity Hamiltonian doesn't have $x$-dependence because the ring cavity is a closed quantum system until now, as shown by the $x$-independent equation Eq.\,\eqref{eq:eTc} that we start from. Explicitly illustrated in Eq.\,\eqref{eq:PmPpsymmetry}, the field distribution relative to the position of the membrane remains the same regardless of the value of $x$. As the coordinate system is a folded one, the integration in Eq.\,\eqref{eq:Hintegral} doesn't contain $x$.

\subsection{Interaction with the environment}

To complete the total Hamiltonian derivation and reveal the $x$-dependence, we consider the coupling of the cavity modes to the outside continuum by assuming the front mirror to have low transmittance $t_0\ll 1$. The cavity linewidth $\gamma$ can be obtained from the imaginary part of the pole of Eq.\,\eqref{eq:efroma} and it only depends on $L$ and front mirror transmittance $t_0$:
\equ{\label{eq:gamma}\gamma=\frac{ct_0^2}{2L}.}
Also according to Eq.\,\eqref{eq:efroma}, as an open passive system when $\mathbb{T}_{\rm in}(k;x)\neq\mathbb{0}$, the ring cavity actually only carries the pumping frequency $\omega_p=k_p c$. The extent to which $\hat c_{\pm}$ modes are excited depends on the detuning of the pumping frequency to the resonant ones: $\omega_p-\omega_{\pm}$. 
In the following contents we will use the wavevector $k_p$ of the pumping field instead of the resonant wavevectors $k_{\pm}$. We use $\hat{c}_{1,2}$ to represent the counterclockwise and clockwise propagating fields that the environment fields $\hat{a}_{1,2}$ directly couples to. According to $\mathbb{T}_{ec}(k;x)$ in Eq.\,\eqref{eq:Tmatrices}, the field operators $\hat{c}_{1,2}$ and $\hat{e}_{1,2}$ are related by the propagation phases:
\begin{subequations}\label{eq:ecrelation}\begin{align}
\hat e_1&=e^{ik_p z_x}\hat c_2,\\
\hat e_2&=e^{ik_p (L-z_x)}\hat c_1.
\end{align}\end{subequations}The cavity-environment interaction Hamiltonian can be expressed as:
\begin{equation}\label{eq:Hac}
\hat{H}_{\gamma}=i\hbar\sqrt{2\gamma}(\hat c_1^{\dagger}\hat a_1-{\rm h.c.})+i\hbar\sqrt{2\gamma}(\hat c_2^{\dagger}\hat a_2-{\rm h.c.})
\end{equation}
For $\hat{c}_{1,2}$ as well as the fields leaking out from the ring cavity towards the detector, the front mirror position is a natural choice of the phase reference point. However, the resonant modes $\hat c_{\pm}$ take the membrane position $z_x$ as the phase reference point, as shown in Eqs.\,\eqref{eq:cfromef}\eqref{eq:EfieldCom}\eqref{eq:Efieldx}. 
The $x$-dependent way that $\hat{c}_{1,2}$ get superimposed to form $\hat c_{\pm}$ is revealed by Eqs.\,\eqref{eq:cfrome}\eqref{eq:ecrelation}. The transformation between $\hat{c}_{1,2}$ and $\hat c_{\pm}$ thus depends on $x$:
\equ{\label{eq:cmpxpara}\hat{c}_{\pm}(x)=\frac{1}{\sqrt{2}}e^{ik_pL/2}(e^{-ik_px}\hat{c}_1\pm e^{ik_px}\hat{c}_2).}
Note that $\hat{c}_{\pm}(0)$ can be seen as the \emph{original} optical modes that the outside modes directly couple to and they have the specific distribution when the membrane stays on its equilibrium $x=0$. For the outside modes, $\hat{c}_{\pm}(x)$ are the new resonant modes when the membrane is displaced by $x$. It is equivalent to say that the mechanical oscillation $x$ changes the way of interference between $\hat c_{1,2}$ that leads to the formation of different resonant modes $\hat{c}_{\pm}(x)$.  

To show the feature of optomechanical coupling, we linearize Eq.\,\eqref{eq:cmpxpara} with respect to $x$:
\begin{subequations}\label{eq:modemixing}\begin{align}
\hat{c}_-(x)&=\hat{c}_-(0)-ik_px\hat{c}_+(0),\\
\hat{c}_+(x)&=\hat{c}_+(0)-ik_px\hat{c}_-(0).
\end{align}\end{subequations}
Thus, the optomechanical Hamiltonian can be expressed as:
\equ{\label{eq:ringHopt}\begin{split}\hat{H}_{\rm opt}(x)&=\hbar \omega_-\hat{c}_-^\dagger(x)\hat{c}_-(x)+\hbar \omega_+\hat{c}_+^\dagger(x)\hat{c}_+(x)\\
&\equiv\hat{H}_0+ \hat{H}_{\rm int}(x),\end{split}}
where the free part $\hat{H}_0$ is equivalent to Eq.\,\eqref{eq:H0} and the optomechanical interaction part $\hat{H}_{\rm int}(t)$ reads:
\equ{\label{eq:Hint}\hat{H}_{\rm int}(x)=2i\omega_s\hbar k_p x\Big(\hat{c}_-^\dagger(0)\hat{c}_+(0)-\textrm{h.c.}\Big).}
The interaction Hamiltonian in Eq.\,\eqref{eq:Hint} explicitly shows the feature of coherent coupling: the mechanical oscillation $x$ induces the coupling between two original optical modes $\hat{c}_\pm(0)$ which have non-degenerate frequencies. 
Note that in the previous derivation $x$ merely works as a parameter. Alternatively, one can start with the total Lagrangian including the mechanical degree of freedom and follow the canonical formulation\,\cite{law1995interaction,khorasani2017higher,pang2018theoretical}, $x$ can thus be upgraded to be a dynamical variable and further becomes a quantum operator $\hat x$ after quantization. To describe the system under the general framework of Eq.\,\eqref{eq:generalH}, we need to express the total Hamiltonian with the new resonant modes $\hat c_{\pm}(\hat x)$. Applying the same transformation as in Eq.\,\eqref{eq:cmpxpara}, we can express the input modes $\hat{a}_{1,2}$ into antisymmetric and symmetric ones:
\equ{\label{eq:cpmina12}\hat{c}_{\pm\rm in}(\hat x)=\frac{1}{\sqrt{2}}e^{ik_pL/2}(e^{-ik_p\hat x}\hat{a}_1\pm e^{ik_p\hat x}\hat{a}_2).}
Thus, the cavity-environment interaction Hamiltonian in Eq.\,\eqref{eq:Hac} can be transformed to:
\begin{equation}\label{eq:ringHdiss}
\hat H_{\gamma}=i\hbar\sqrt{2\gamma} \Big(\hat c_-^{\dagger}(0)\hat c_{-\rm in}(0)+c_+^{\dagger}(0)\hat c_{+\rm in}(0)-{\rm h.c.}\Big),
\end{equation}
which is equivalent to the generalized expression up to linear order in $\hat x$: 
\equ{\label{eq:cinx}\begin{split}&\hat H_{\gamma}(x)=\sum_{\pm}i\hbar\sqrt{2\gamma}\Big(\hat c_\pm^{\dagger}(\hat x)\hat c_{\pm\rm in}(\hat x)-{\rm h.c.}\Big)\\
\approx&\sum_{\pm}i\hbar\sqrt{2\gamma}\Big( (1-2k_p^2\hat x^2)\hat c_\pm^{\dagger}(0)\hat c_{\pm\rm in}(0)-{\rm h.c.}\Big)\\
=&\hat H_\gamma+\mathcal{O}(\hat x^2).\end{split}}
It is clear from Eq.\,\eqref{eq:cinx} that, even expressed in $\hat c_{\pm \rm in}(\hat x)$ for consistency under the framework, the coupling rate with the environment doesn't depend on $\hat x$. Thus, the optomechanical coupling has no dissipative feature.

\section{\label{sec:coolinglim}Optomechanical cooling limit in the ring cavity system}

\subsection{\label{sec:coupledOMeq}Coupled optical and mechanical equations of motion}
For notational convenience, in the following contents, all expressions without explicit arguments are by default in time domain with temporal argument $t$; all derivatives represented by dot are with respect to $t$, \textit{i.e.} $\dot{\hat c}_{\pm}\equiv \partial \hat c_{\pm} /\partial t$; ${\hat{c}}_{\pm ({\rm in})}$ are used to represent ${\hat{c}}_{\pm ({\rm in})}(0)$; $C_{\pm}\equiv\left<\hat c_{\pm}\right>$ are used to represent the expectation value of optical modes, \textit{i.e.} the classical amplitude. We assume single port pumping from $\hat a_1$ with frequency $\omega_-$ and amplitude $A_1$. According to Eq.\,\eqref{eq:cpmina12}, the pumping amplitude of cavity modes are $C_{\pm\rm in}=A_1/\sqrt{2}$. We work in the rotating frame with pumping frequency $\omega_-$. Based on the Hamiltonian in Eq.\,\eqref{eq:ringHtotal}, the equations of motion for ${\hat{c}}_{\pm}$ modes read:
\begin{subequations}\label{eq:cmo}
\begin{align}
&\dot{\hat{c}}_-=2\omega_s k_p\hat x\hat{c}_+-\gamma {\hat c}_-+\sqrt{2\gamma}\hat{c}_{-\rm in},\\
&\dot{\hat{c}}_+=-2i\omega_s {\hat c}_+-2\omega_sk_p\hat x\hat{c}_--\gamma {\hat c}_++\sqrt{2\gamma}\hat{c}_{+\rm in}.
\end{align}
\end{subequations}
The intracavity amplitudes of the two modes are given by the static solutions of Eq.\,\eqref{eq:cmo} with $\hat x=0$:
\begin{subequations}\label{eq:Cstatic}
\begin{align}
&C_-=\frac{C_{-\rm in}\sqrt{2\gamma}}{\gamma}=\frac{A_1}{\sqrt{\gamma}},\\
&C_+=\frac{C_{+\rm in}\sqrt{2\gamma}}{\gamma+2i\omega_s}=\frac{A_1\sqrt{\gamma}}{\gamma+2i\omega_s}.
\end{align}\end{subequations}
It can be seen from Eq.\,\eqref{eq:Cstatic} that $\hat c_-$ mode is on resonance while $\hat c_+$ mode is off resonance. 

Each optical field can be divided into static amplitude and quantum fluctuation $\hat{c}_{\pm}\to C_{\pm} +\hat{c}_{\pm}$ and Eq.\,\eqref{eq:cmo} can thus be linearized as:
\begin{subequations}\label{eq:cmolinear}\begin{align}
& \dot{\hat{c}}_-=2\omega_sk_p\hat x C_+-\gamma {\hat c}_- +\sqrt{2\gamma}\hat{c}_{-\rm in},\\
& \dot{\hat{c}}_+=-2i\omega_s {\hat c}_+-2\omega_sk_p\hat xC_--\gamma {\hat c}_++\sqrt{2\gamma}\hat{c}_{+\rm in}.
\end{align}\end{subequations} 
The mechanical equations of motion are:
\begin{subequations}\label{eq:xpmo}
\begin{align}\label{xmo}
\dot{\hat x}=&\frac{\hat p}{m}\\
\label{eq:pmoinc}\begin{split}
\dot{\hat p}=&2i\omega_sk_p\hbar(\hat c_+^\dagger \hat c_- - \hat c_-^\dagger \hat c_+)\\
&-4\omega_sk_p^2\hbar\hat x(\hat c_-^\dagger \hat c_- - \hat c_+^\dagger \hat c_+)\\
&+G-m\Omega_m^2\hat x-m\gamma_m\dot{\hat x}
\end{split}
\end{align}
\end{subequations}
The linearization of Eq.\,\eqref{eq:pmoinc} gives:
\begin{equation}\label{eq:pmoc}\begin{split}
\dot{\hat p}=&2i\omega_sk_p\hbar(C_+^* C_- - C_-^* C_+),\\
&+2i\omega_sk_p\hbar(\hat c_+^\dagger C_--\hat c_-^\dagger C_+-\hat c_+C_-^*+\hat c_-C_+^*)\\
&-4\omega_sk_p^2\hbar\hat x(C_-^* C_- - C_+^* C_+)\\
&+G-m\Omega_m^2\hat x-m\gamma_m\dot{\hat x}.
\end{split}\end{equation}
where the first line represents the static radiation pressure, the second line is fluctuating radiation pressure and the third line represents optical trapping due to standing wave energy distribution. 

\subsection{\label{sec:sidebandOPT}Sideband feature and optical damping}
To solve the coupled optical and mechanical EOMs in Eqs.\,\eqref{eq:cmolinear}\eqref{xmo}\eqref{eq:pmoc}, we transfer them into frequency domain\,\cite{aspelmeyer2014cavity}. The $\hat x$-dependence in sidebands of each mode $\hat{c}_{\pm}$ is obtained by scattering from the other mode amplitude $C_{\mp}$ and can be expressed as:
\begin{subequations}\label{SignalInEigenmodes}
\begin{align}
\hat c_-^{(\hat x)}(+\Omega)&=C_+\times\frac{2\omega_s k_p\hat x}{\gamma-i\Omega},  \\
\hat c_-^{\dagger (\hat x)}(-\Omega)&=C_+^*\times\frac{2\omega_s k_p\hat x}{\gamma-i\Omega},  \\
\hat c_+^{(\hat x)}(+\Omega)&=C_-\times\frac{-2\omega_s k_p\hat x}{\gamma-i(\Omega-2\omega_s)},\\
\hat c_+^{\dagger(\hat x)}(-\Omega)&=C_-^*\times\frac{-2\omega_sk_p\hat x}{\gamma-i(\Omega+2\omega_s)}.
\end{align}
\end{subequations}
Note that $\hat c_-^{(\hat x)}$ is gained by scattering from $C_+$ which is off-resonance in cavity and the two sidebands $\hat c_-^{(\hat x)}(+\Omega)$ and $\hat c_-^{\dagger (\hat x)}(-\Omega)$ are symmetric. Similarly, $\hat c_+^{(\hat x)}$ is gained by scattering from the resonant mode $C_-$. However, as $\hat c_+$ is peaked at a higher frequency from pumping, the two sidebands $\hat c_+^{(\hat x)}(+\Omega)$ and $\hat c_+^{\dagger (\hat x)}(-\Omega)$ are extremely unbalanced. See Fig.\ref{fig:cooling} for illustration. Based on sideband expressions in Eq.\,\eqref{SignalInEigenmodes}, we derive the $\hat x$-dependent part in $(\hat c_+^\dagger C_--\hat c_-^\dagger C_+-\hat c_+C_-^*+\hat c_-C_+^*)$ as below:
\begin{equation}
\begin{split}
&(\hat c_+^\dagger C_--\hat c_-^\dagger C_+-\hat c_+C_-^*+\hat c_-C_+^*)^{(\hat x)}\\
=&2\omega_sk_p\hat x|C_-|^2\left(\frac{1}{\gamma-i(\Omega-2\omega_s)}-\frac{1}{\gamma-i(\Omega+2\omega_s)}\right)\\
&+2\omega_sk_p\hat x|C_+|^2\left(\frac{1}{\gamma-i\Omega}-\frac{1}{\gamma-i\Omega}\right)\\
\approx &2\omega_sk_p\hat x\frac{|A_{\rm in}|^2}{\gamma} \frac{-4i\omega_s}{(\gamma-i\Omega)^2+4\omega_s^2}
\end{split}
\end{equation}
The beating between $C_-$ and two highly unbalanced sidebands in $\hat c_+$ mode provides strong optical rigidity and damping to the mechanical oscillator. Note that strong average field in each mode only beats with quantum fluctuation in the other mode. This scattering-like-interaction between non-degenerate optical modes is the essential feature of coherent optomechanical couplings. Because the two optical resonance is splitted by $2 \omega_s$, the two sidebands will have maximum difference when $\Omega\approx 2 \omega_s$. As a result, optical cooling happens when we pump the antisymmetric mode with frequency $\omega_-$ and will be the strongest when $\Omega_m\approx 2 \omega_s$. 

We will focus on parameter regime $\Omega\approx\Omega_m\approx 2 \omega_s$ in the following contents of this section. The momentum equation of motion near that frequency is:
\begin{equation}\label{eq:EOMeff}
\mathcal{F}[\dot{\hat p}]={\hat F}_{\rm bafl}(\Omega)+G_{\rm eff}-m\Omega_{\rm eff}^2 \hat x(\Omega)+im\gamma_{\rm eff}\Omega \hat x(\Omega)
\end{equation}
where ${\hat F}_{\rm bafl}$ is the $x$-independent part of radiation pressure force $2i\omega_sk\hbar(\hat c_+^\dagger C_--\hat c_-^\dagger C_+-\hat c_+C_-^*+\hat c_-C_+^*)$, $G_{\rm eff}=G-2|A_{\rm in}|^2k_p\hbar$ is the offsetted external force, $\Omega_{\rm eff}^2=\Omega_m^2+3|A_{\rm in}|^2k_p^2\hbar \omega_s/m\gamma$ is the mechanical resonance together with optical rigidity, $\gamma_{\rm eff}=\gamma_m+2|A_{\rm in}|^2k_p^2\hbar \omega_s/m\gamma^2$ is effective mechanical damping rate including optical cooling. All expressions above are obtained under approximation condition $\Omega _m\gg\gamma$. We can extract the optical spring and optical damping terms from the approximated formulas above:
\begin{subequations}
\begin{align}
&\textrm{Optical spring:}&&\Omega_{\rm opt}^2=\frac{3|A_{\rm in}|^2k_p^2\hbar \omega_s}{m\gamma}\\\label{eq:gammaopt}
&\textrm{Optical damping:}&&\gamma_{\rm opt}=\frac{2|A_{\rm in}|^2k_p^2\hbar \omega_s}{m\gamma^2}
\end{align}
\end{subequations}
Compared with original mechanical properties, optical spring is always negligible within reachable input power ($P_{\rm in}<1 \rm W$) while optical damping is comparable with mechanical damping when $P_{\rm in}\approx 0.04 \rm W$ and is much bigger with higher input power.

\subsection{\label{sec:numberlimit}Quantum limit of mechanical occupation number}
We then calculate the occupation number limit. In the case where the mechanical object is a high-Q-oscillator, we represent the mechanical oscillation in terms of mechanical creation and annihilation operators $\tilde{m}^{\dagger}$ and $\tilde{m}$ in the rotating frame of mechanical resonant frequency $\Omega_m$:
\equ{\label{eq:xtotildem}\hat x=x_{\rm ZPF}(\tilde m e^{-i\Omega_m t}+\tilde m^\dag e^{i\Omega_m t}),}
where $x_{\rm ZPF}=\sqrt{\hbar/2m\Omega_m}$ is the zero-point fluctuation of the mechanical oscillator. Using $\omega$ to represent the sideband frequency of mechanical oscillator around $\Omega_m$, the Fourier components of $\tilde m$ and $\tilde m^{\dagger}$ read: 
\begin{subequations}\label{eq:tildemFT}
\begin{align}
    \tilde m(t)&=\int_{-\Omega_m}^{+\infty}\frac{d\omega}{2\pi}\tilde m(\omega)e^{-i\omega t},\\
    \tilde m^{\dagger}(t)&=\int_{-\infty}^{\Omega_m}\frac{d\omega}{2\pi}\tilde m^{\dagger}(-\omega)e^{-i\omega t}.
\end{align}
\end{subequations}
Although the upper and lower limits for both integrals in Eq.\,\eqref{eq:tildemFT} can be extended to infinity under condition $\omega\ll\Omega_m$, we keep this rigorous form for clearer future reference. 
The average mechanical occupation number is defined as\,\cite{Sakurai:1341875}: 
\equ{\label{eq:occunumDEF}\left<n(t)\right>\equiv\frac{\left<\tilde{m}(t)\tilde{m}^{\dagger}(t)+\tilde{m}^{\dagger}(t)\tilde{m}(t)\right>-1}{2}.}
According to Eq.\,\eqref{eq:tildemFT} $\tilde{m}(t)\tilde{m}^{\dagger}(t)$ and $\tilde{m}^{\dagger}(t)\tilde{m}(t)$ can be expressed as:
\begin{subequations}\label{eq:mtexp}
\begin{align}
    \tilde{m}(t)\tilde{m}^{\dagger}(t)=\iint_{-\Omega_m}^{+\infty}\frac{d\omega d\omega'}{(2\pi)^2}\tilde{m}(\omega)\tilde{m}^{\dagger}(\omega')e^{i(\omega'-\omega) t},\\
    \tilde{m}^{\dagger}(t)\tilde{m}(t)=\iint_{-\Omega_m}^{+\infty}\frac{d\omega d\omega'}{(2\pi)^2}\tilde{m}^{\dagger}(\omega')\tilde{m}(\omega)e^{i(\omega'-\omega) t},
\end{align}
\end{subequations}
To obtain the mechanical occupation number, we need to calculate the second order correlation function of mechanical operators $\left<\tilde{m}(\omega)\tilde{m}^{\dagger}(\omega')\right>$ and $\left<\tilde{m}^{\dagger}(\omega')\tilde{m}(\omega)\right>$. Therefore, we need to obtain equations of motion of $\tilde m,\tilde m^{\dagger}$ by rephrasing those of $\hat x,\hat p$ in Sec.\,\ref{sec:coupledOMeq} and Sec.\,\ref{sec:sidebandOPT}. 

Quoting Eq.\,\eqref{eq:EOMeff} and ignoring the static force by letting $G_{\rm eff}=0$, we obtain the second order equation of motion of $\hat x$:
\equ{\label{eq:xEOMeff}m\ddot{\hat x}=\hat F_{\rm bafl}-m\Omega_m^2\hat x-m\gamma_{\rm eff}\dot{\hat x}.}
Transferred into frequency domain, Eq.\,\eqref{eq:xEOMeff} becomes:
\equ{m(\Omega_m^2-\Omega^2-i\Omega\gamma_{\rm eff})\hat x(\Omega)=\hat F_{\rm bafl}(\Omega),}
which can be factorized under condition $\gamma_{\rm eff}\ll \Omega_m$ as:
\equ{\label{eq:xEOMfactorized}\left[\frac{\gamma_{\rm eff}}{2}-i(\Omega-\Omega_m)\right]\left[\frac{\gamma_{\rm eff}}{2}-i(\Omega+\Omega_m)\right]\hat x(\Omega)=\frac{\hat F_{\rm bafl}(\Omega)}{m}.}
According to Eqs.\,\eqref{eq:xtotildem}\eqref{eq:tildemFT}, the Fourier Transformation of $\hat x$ reads:
\begin{widetext}
\equ{\label{eq:xFT}\begin{split}
\hat x(\Omega)&\equiv
\int_{-\infty}^{+\infty}dt\hat x(t)e^{i\Omega t}=x_{\rm ZPF}\int_{-\infty}^{+\infty}dt\left[e^{i(\Omega-\Omega_m) t}\int_{-\Omega_m}^{+\infty}\frac{d\omega}{2\pi}\tilde m(\omega)e^{-i\omega t}+e^{i(\Omega+\Omega_m) t}\int_{-\infty}^{\Omega_m}\frac{d\omega}{2\pi}\tilde m^{\dagger}(-\omega)e^{-i\omega t}\right]\\
&=x_{\rm ZPF}\left[\tilde m(\Omega-\Omega_m)\theta(\Omega)+m^{\dagger}(-\Omega-\Omega_m)\theta(-\Omega)\right]
\end{split}}
\end{widetext}
where $\theta(\Omega)$ is the Heaviside step function and $\delta(\Omega)$ is the Dirac delta function. Plugging Eq.\,\eqref{eq:xFT} into Eq.\,\eqref{eq:xEOMfactorized} and considering the thermal force $\hat F_{\text{th}}$, we obtain the equations of motion for $\tilde{m}^{\dagger},\tilde{m}$ in their frequency domain:
\begin{subequations}
\begin{align}
\begin{split}
\Big[\frac{\gamma _m+\gamma_{\rm opt}}{2}&-i\omega\Big]\tilde{m}(\omega)=\\
\frac{i x_{\rm ZPF}}{\hbar}&\Big[\hat{F}_{\rm bafl}(\Omega_m+\omega)+\hat F_{\rm th}(\Omega_m+\omega)\Big],\end{split}\\
\begin{split}
\Big[\frac{\gamma _m+\gamma _{\rm opt}}{2}&+i\omega\Big]\tilde{m}^{\dagger}(\omega )=\\
-\frac{i x_{\rm ZPF}}{\hbar}&\Big[\hat{F}_{\rm bafl}(-(\Omega_m+\omega))+\hat F_{\rm th}(-(\Omega_m+\omega))\Big].\end{split}
\end{align}
\end{subequations}
The fluctuating backaction force on the mechanical oscillator in the frequency domain reads:
\be\begin{split}\hat{F}_{\text{bafl}}(\Omega )=2 i \omega _s \hbar k_p
\Big(& C_+^* \hat{c}_-(\Omega)-C_+ \hat{c}^{\dagger}_-(-\Omega)\\
&-C_-^* \hat{c}_+(\Omega)+C_- \hat{c}^{\dagger}_+(-\Omega)\Big),\end{split}\ee
and satisfies the relation $\hat{F}_{\rm bafl}^{\dagger}(\Omega )=\hat{F}_{\rm bafl}(-\Omega )$. The spectrum $S_F(\Omega)$ of the backaction force $\hat{F}_{\text{bafl}}(\Omega)$ is defined as:
\be
\left<\hat{F}_{\rm bafl}(\Omega ) \hat{F}_{\rm bafl}\left(\Omega'\right)\right>=2 \pi \delta\left(\Omega '+\Omega\right)S_F(\Omega )\ee
and takes the following expression:
\be
\begin{split}S_F(\Omega )=8 A_{\rm in}^2 \omega _s^2 \hbar ^2 k_p^2 \gamma ^2&\Bigg[
\frac{1}{\gamma^2\left(\gamma^2+\left(\Omega-2\omega_s\right)^2\right)}\\
&+\frac{1}{\left(\gamma^2+\Omega ^2\right)\left(\gamma ^2+4 \omega_s^2\right)}\Bigg].\end{split}
\ee
Also, the thermal force $\hat F_{\rm th}$ has white spectrum:
\equ{\left<\hat{F}_{\rm th}(\Omega ) \hat{F}_{\rm th} (\Omega')\right>=2 \pi \delta\left(\Omega '+\Omega\right)2mk_BT \gamma_m.}
Thus, the second order correlation function of mechanical operators can be calculated by:
\begin{subequations}
\begin{align}
\left<\tilde{m}(\omega)\tilde{m}^{\dagger}(\omega')\right>&=2\pi\delta(\omega-\omega')S_+(\omega),\\
\left<\tilde{m}^{\dagger}(\omega')\tilde{m}(\omega)\right>&=2\pi\delta(\omega-\omega')S_-(\omega),
\end{align}
\end{subequations}
with $S_+(\omega)$ and $S_-(\omega)$ defined as:
\begin{subequations}
\begin{align}
S_+(\omega)&\equiv\frac{x_{\rm ZPF}^2/{\hbar^2}\left[S_{F}(\Omega_m+\omega)+2mk_BT \gamma_m\right]}{ \left(\frac{\gamma_m+\gamma_{\rm opt}}{2}\right)^2+\omega ^2},\\
S_-(\omega)&\equiv\frac{x_{\rm ZPF}^2/{\hbar^2}\left[S_{F}(-(\Omega_m+\omega))+2mk_BT \gamma_m\right]}{ \left(\frac{\gamma_m+\gamma_{\rm opt}}{2}\right)^2+\omega ^2}.
\end{align}
\end{subequations}
According to Eq.\,\eqref{eq:mtexp}, the time domain mechanical correlation functions can be calculated as follows:
\begin{widetext}\begin{subequations}
\begin{align}
\left<\tilde{m}(t)\tilde{m}^{\dagger}(t)\right>=\int_{-\Omega_m}^{+\infty}S_+(\omega)\frac{d\omega}{2\pi}&=\frac{4 A_{\rm in}^2 \gamma^2k_p^2\hbar\omega _s^2}{m\Omega _m \left(\gamma_m+\gamma_{\rm opt}\right)}\left[\frac{1}{\gamma^2 \left(\gamma^2+\left(\Omega_m-2 \omega_s\right)^2\right)}+\frac{1}{4 \Omega _m^2 \omega_s^2}\right]+\frac{\gamma_m }{\gamma_m+\gamma_{\rm opt}}\frac{k_B T}{\Omega_m\hbar},\\
\left<\tilde{m}^{\dagger}(t)\tilde{m}(t)\right>=\int_{-\Omega_m}^{+\infty}S_-(\omega)\frac{d\omega}{2\pi}&=\frac{4 A_{\rm in}^2 \gamma ^2k_p^2 \hbar  \omega _s^2}{m\Omega _m \left(\gamma_m+\gamma_{\rm opt}\right)}\left[\frac{1}{\gamma^2 \left(\Omega _m+2 \omega_s\right){}^2}+\frac{1}{4\Omega _m^2 \omega _s^2}\right]+\frac{\gamma_m }{\gamma_m+\gamma_{\rm opt}}\frac{k_B T}{\Omega_m\hbar}.
\end{align}
\end{subequations}\end{widetext}

Based on all derivation above, under condition $\Omega_m \gg \gamma$, the mechanical occupation number defined in Eq.\,\eqref{eq:occunumDEF} can be expressed as:
\equ{\label{eq:mechnum}\left<\hat n\right>=\frac{\gamma_{\rm opt}}{\gamma_m+\gamma _{\rm opt}}\frac{1}{2}\left[\frac{\gamma ^2}{4 \Omega_m^2}-\frac{\gamma_m}{\gamma_{\rm opt}}\right]+\frac{\gamma_m}{\gamma_m+\gamma_{\rm opt}}\frac{k_B T}{\Omega_m \hbar}.}
Under further condition $\gamma_{\rm opt}\gg\gamma_{\text{m}}$, we can rewrite the expression above to get the ultimate cooling limit:
\equ{\label{eq:mechlimit}\left<\hat n\right>=\frac{\gamma_m}{\gamma_m+\gamma_{\rm opt}}n_{\rm th}+\frac{\gamma_{\rm opt}}{\gamma_m+\gamma _{\rm opt}}n_{\rm ba}\approx n_{\rm ba},}
where $n_{\rm th}=k_B T/\Omega_m \hbar$ is the thermal occupation number and $n_{\rm ba}=\gamma ^2/8 \Omega_m^2$ is the back-action limited occupation number.

\bibliography{ringcavity.bib}

\begin{thebibliography}{79}%
\makeatletter
\providecommand \@ifxundefined [1]{%
 \@ifx{#1\undefined}
}%
\providecommand \@ifnum [1]{%
 \ifnum #1\expandafter \@firstoftwo
 \else \expandafter \@secondoftwo
 \fi
}%
\providecommand \@ifx [1]{%
 \ifx #1\expandafter \@firstoftwo
 \else \expandafter \@secondoftwo
 \fi
}%
\providecommand \natexlab [1]{#1}%
\providecommand \enquote  [1]{``#1''}%
\providecommand \bibnamefont  [1]{#1}%
\providecommand \bibfnamefont [1]{#1}%
\providecommand \citenamefont [1]{#1}%
\providecommand \href@noop [0]{\@secondoftwo}%
\providecommand \href [0]{\begingroup \@sanitize@url \@href}%
\providecommand \@href[1]{\@@startlink{#1}\@@href}%
\providecommand \@@href[1]{\endgroup#1\@@endlink}%
\providecommand \@sanitize@url [0]{\catcode `\\12\catcode `\$12\catcode
  `\&12\catcode `\#12\catcode `\^12\catcode `\_12\catcode `\%12\relax}%
\providecommand \@@startlink[1]{}%
\providecommand \@@endlink[0]{}%
\providecommand \url  [0]{\begingroup\@sanitize@url \@url }%
\providecommand \@url [1]{\endgroup\@href {#1}{\urlprefix }}%
\providecommand \urlprefix  [0]{URL }%
\providecommand \Eprint [0]{\href }%
\providecommand \doibase [0]{http://dx.doi.org/}%
\providecommand \selectlanguage [0]{\@gobble}%
\providecommand \bibinfo  [0]{\@secondoftwo}%
\providecommand \bibfield  [0]{\@secondoftwo}%
\providecommand \translation [1]{[#1]}%
\providecommand \BibitemOpen [0]{}%
\providecommand \bibitemStop [0]{}%
\providecommand \bibitemNoStop [0]{.\EOS\space}%
\providecommand \EOS [0]{\spacefactor3000\relax}%
\providecommand \BibitemShut  [1]{\csname bibitem#1\endcsname}%
\let\auto@bib@innerbib\@empty
\bibitem [{\citenamefont {Kippenberg}\ and\ \citenamefont
  {Vahala}(2008)}]{kippenberg2008cavity}%
  \BibitemOpen
  \bibfield  {author} {\bibinfo {author} {\bibfnamefont {T.~J.}\ \bibnamefont
  {Kippenberg}}\ and\ \bibinfo {author} {\bibfnamefont {K.~J.}\ \bibnamefont
  {Vahala}},\ }\href@noop {} {\bibfield  {journal} {\bibinfo  {journal}
  {Science}\ }\textbf {\bibinfo {volume} {321}},\ \bibinfo {pages} {1172}
  (\bibinfo {year} {2008})}\BibitemShut {NoStop}%
\bibitem [{\citenamefont {Favero}\ and\ \citenamefont
  {Karrai}(2009)}]{favero2009optomechanics}%
  \BibitemOpen
  \bibfield  {author} {\bibinfo {author} {\bibfnamefont {I.}~\bibnamefont
  {Favero}}\ and\ \bibinfo {author} {\bibfnamefont {K.}~\bibnamefont
  {Karrai}},\ }\href@noop {} {\bibfield  {journal} {\bibinfo  {journal} {Nature
  Photonics}\ }\textbf {\bibinfo {volume} {3}},\ \bibinfo {pages} {201}
  (\bibinfo {year} {2009})}\BibitemShut {NoStop}%
\bibitem [{\citenamefont {Aspelmeyer}\ \emph {et~al.}(2014)\citenamefont
  {Aspelmeyer}, \citenamefont {Kippenberg},\ and\ \citenamefont
  {Marquardt}}]{aspelmeyer2014cavity}%
  \BibitemOpen
  \bibfield  {author} {\bibinfo {author} {\bibfnamefont {M.}~\bibnamefont
  {Aspelmeyer}}, \bibinfo {author} {\bibfnamefont {T.~J.}\ \bibnamefont
  {Kippenberg}}, \ and\ \bibinfo {author} {\bibfnamefont {F.}~\bibnamefont
  {Marquardt}},\ }\href@noop {} {\bibfield  {journal} {\bibinfo  {journal}
  {Reviews of Modern Physics}\ }\textbf {\bibinfo {volume} {86}},\ \bibinfo
  {pages} {1391} (\bibinfo {year} {2014})}\BibitemShut {NoStop}%
\bibitem [{\citenamefont {Bowen}\ and\ \citenamefont
  {Milburn}(2015)}]{bowen2015quantum}%
  \BibitemOpen
  \bibfield  {author} {\bibinfo {author} {\bibfnamefont {W.~P.}\ \bibnamefont
  {Bowen}}\ and\ \bibinfo {author} {\bibfnamefont {G.~J.}\ \bibnamefont
  {Milburn}},\ }\href@noop {} {\emph {\bibinfo {title} {Quantum
  optomechanics}}}\ (\bibinfo  {publisher} {CRC press},\ \bibinfo {year}
  {2015})\BibitemShut {NoStop}%
\bibitem [{\citenamefont {Bhattacharya}\ and\ \citenamefont
  {Meystre}(2007)}]{bhattacharya2007trapping}%
  \BibitemOpen
  \bibfield  {author} {\bibinfo {author} {\bibfnamefont {M.}~\bibnamefont
  {Bhattacharya}}\ and\ \bibinfo {author} {\bibfnamefont {P.}~\bibnamefont
  {Meystre}},\ }\href@noop {} {\bibfield  {journal} {\bibinfo  {journal}
  {Physical Review Letters}\ }\textbf {\bibinfo {volume} {99}},\ \bibinfo
  {pages} {073601} (\bibinfo {year} {2007})}\BibitemShut {NoStop}%
\bibitem [{\citenamefont {Yong-Chun}\ \emph {et~al.}(2013)\citenamefont
  {Yong-Chun}, \citenamefont {Yu-Wen}, \citenamefont {Wei},\ and\ \citenamefont
  {Yun-Feng}}]{yong2013review}%
  \BibitemOpen
  \bibfield  {author} {\bibinfo {author} {\bibfnamefont {L.}~\bibnamefont
  {Yong-Chun}}, \bibinfo {author} {\bibfnamefont {H.}~\bibnamefont {Yu-Wen}},
  \bibinfo {author} {\bibfnamefont {W.~C.}\ \bibnamefont {Wei}}, \ and\
  \bibinfo {author} {\bibfnamefont {X.}~\bibnamefont {Yun-Feng}},\ }\href@noop
  {} {\bibfield  {journal} {\bibinfo  {journal} {Chinese Physics B}\ }\textbf
  {\bibinfo {volume} {22}},\ \bibinfo {pages} {114213} (\bibinfo {year}
  {2013})}\BibitemShut {NoStop}%
\bibitem [{\citenamefont {Sawadsky}\ \emph {et~al.}(2015)\citenamefont
  {Sawadsky}, \citenamefont {Kaufer}, \citenamefont {Nia}, \citenamefont
  {Tarabrin}, \citenamefont {Khalili}, \citenamefont {Hammerer},\ and\
  \citenamefont {\mbox{Schnabel}}}]{sawadsky2015observation}%
  \BibitemOpen
  \bibfield  {author} {\bibinfo {author} {\bibfnamefont {A.}~\bibnamefont
  {Sawadsky}}, \bibinfo {author} {\bibfnamefont {H.}~\bibnamefont {Kaufer}},
  \bibinfo {author} {\bibfnamefont {R.~M.}\ \bibnamefont {Nia}}, \bibinfo
  {author} {\bibfnamefont {S.~P.}\ \bibnamefont {Tarabrin}}, \bibinfo {author}
  {\bibfnamefont {F.~Y.}\ \bibnamefont {Khalili}}, \bibinfo {author}
  {\bibfnamefont {K.}~\bibnamefont {Hammerer}}, \ and\ \bibinfo {author}
  {\bibfnamefont {R.}~\bibnamefont {\mbox{Schnabel}}},\ }\href@noop {}
  {\bibfield  {journal} {\bibinfo  {journal} {Physical Review Letters}\
  }\textbf {\bibinfo {volume} {114}},\ \bibinfo {pages} {043601} (\bibinfo
  {year} {2015})}\BibitemShut {NoStop}%
\bibitem [{\citenamefont {Purdy}\ \emph
  {et~al.}(2013{\natexlab{a}})\citenamefont {Purdy}, \citenamefont {Yu},
  \citenamefont {Peterson}, \citenamefont {Kampel},\ and\ \citenamefont
  {Regal}}]{purdy2013strong}%
  \BibitemOpen
  \bibfield  {author} {\bibinfo {author} {\bibfnamefont {T.~P.}\ \bibnamefont
  {Purdy}}, \bibinfo {author} {\bibfnamefont {P.-L.}\ \bibnamefont {Yu}},
  \bibinfo {author} {\bibfnamefont {R.}~\bibnamefont {Peterson}}, \bibinfo
  {author} {\bibfnamefont {N.}~\bibnamefont {Kampel}}, \ and\ \bibinfo {author}
  {\bibfnamefont {C.}~\bibnamefont {Regal}},\ }\href@noop {} {\bibfield
  {journal} {\bibinfo  {journal} {Physical Review X}\ }\textbf {\bibinfo
  {volume} {3}},\ \bibinfo {pages} {031012} (\bibinfo {year}
  {2013}{\natexlab{a}})}\BibitemShut {NoStop}%
\bibitem [{\citenamefont {Kronwald}\ \emph {et~al.}(2014)\citenamefont
  {Kronwald}, \citenamefont {Marquardt},\ and\ \citenamefont
  {Clerk}}]{kronwald2014dissipative}%
  \BibitemOpen
  \bibfield  {author} {\bibinfo {author} {\bibfnamefont {A.}~\bibnamefont
  {Kronwald}}, \bibinfo {author} {\bibfnamefont {F.}~\bibnamefont {Marquardt}},
  \ and\ \bibinfo {author} {\bibfnamefont {A.~A.}\ \bibnamefont {Clerk}},\
  }\href@noop {} {\bibfield  {journal} {\bibinfo  {journal} {New Journal of
  Physics}\ }\textbf {\bibinfo {volume} {16}},\ \bibinfo {pages} {063058}
  (\bibinfo {year} {2014})}\BibitemShut {NoStop}%
\bibitem [{\citenamefont {Aggarwal}\ \emph {et~al.}(2018)\citenamefont
  {Aggarwal}, \citenamefont {Cullen}, \citenamefont {Cripe}, \citenamefont
  {Cole}, \citenamefont {Lanza}, \citenamefont {Libson}, \citenamefont
  {Follman}, \citenamefont {Heu}, \citenamefont {Corbitt},\ and\ \citenamefont
  {Mavalvala}}]{aggarwal2018room}%
  \BibitemOpen
  \bibfield  {author} {\bibinfo {author} {\bibfnamefont {N.}~\bibnamefont
  {Aggarwal}}, \bibinfo {author} {\bibfnamefont {T.}~\bibnamefont {Cullen}},
  \bibinfo {author} {\bibfnamefont {J.}~\bibnamefont {Cripe}}, \bibinfo
  {author} {\bibfnamefont {G.~D.}\ \bibnamefont {Cole}}, \bibinfo {author}
  {\bibfnamefont {R.}~\bibnamefont {Lanza}}, \bibinfo {author} {\bibfnamefont
  {A.}~\bibnamefont {Libson}}, \bibinfo {author} {\bibfnamefont
  {D.}~\bibnamefont {Follman}}, \bibinfo {author} {\bibfnamefont
  {P.}~\bibnamefont {Heu}}, \bibinfo {author} {\bibfnamefont {T.}~\bibnamefont
  {Corbitt}}, \ and\ \bibinfo {author} {\bibfnamefont {N.}~\bibnamefont
  {Mavalvala}},\ }\href@noop {} {\bibfield  {journal} {\bibinfo  {journal}
  {arXiv preprint arXiv:1812.09942}\ } (\bibinfo {year} {2018})}\BibitemShut
  {NoStop}%
\bibitem [{\citenamefont {Schnabel}(2017)}]{schnabel2017squeezedSec56}%
  \BibitemOpen
  \bibfield  {author} {\bibinfo {author} {\bibfnamefont {R.}~\bibnamefont
  {Schnabel}},\ }\href@noop {} {\bibfield  {journal} {\bibinfo  {journal}
  {Physics Reports}\ }\textbf {\bibinfo {volume} {684}},\ \bibinfo {pages} {1}
  (\bibinfo {year} {2017})}\BibitemShut {NoStop}%
\bibitem [{\citenamefont {Bose}\ \emph {et~al.}(1997)\citenamefont {Bose},
  \citenamefont {Jacobs},\ and\ \citenamefont {Knight}}]{bose1997preparation}%
  \BibitemOpen
  \bibfield  {author} {\bibinfo {author} {\bibfnamefont {S.}~\bibnamefont
  {Bose}}, \bibinfo {author} {\bibfnamefont {K.}~\bibnamefont {Jacobs}}, \ and\
  \bibinfo {author} {\bibfnamefont {P.}~\bibnamefont {Knight}},\ }\href@noop {}
  {\bibfield  {journal} {\bibinfo  {journal} {Physical Review A}\ }\textbf
  {\bibinfo {volume} {56}},\ \bibinfo {pages} {4175} (\bibinfo {year}
  {1997})}\BibitemShut {NoStop}%
\bibitem [{\citenamefont {Bose}\ \emph {et~al.}(1999)\citenamefont {Bose},
  \citenamefont {Jacobs},\ and\ \citenamefont {Knight}}]{bose1999scheme}%
  \BibitemOpen
  \bibfield  {author} {\bibinfo {author} {\bibfnamefont {S.}~\bibnamefont
  {Bose}}, \bibinfo {author} {\bibfnamefont {K.}~\bibnamefont {Jacobs}}, \ and\
  \bibinfo {author} {\bibfnamefont {P.~L.}\ \bibnamefont {Knight}},\
  }\href@noop {} {\bibfield  {journal} {\bibinfo  {journal} {Physical Review
  A}\ }\textbf {\bibinfo {volume} {59}},\ \bibinfo {pages} {3204} (\bibinfo
  {year} {1999})}\BibitemShut {NoStop}%
\bibitem [{\citenamefont {Vitali}\ \emph {et~al.}(2007)\citenamefont {Vitali},
  \citenamefont {Gigan}, \citenamefont {Ferreira}, \citenamefont {B{\"o}hm},
  \citenamefont {Tombesi}, \citenamefont {Guerreiro}, \citenamefont {Vedral},
  \citenamefont {Zeilinger},\ and\ \citenamefont
  {Aspelmeyer}}]{vitali2007optomechanical}%
  \BibitemOpen
  \bibfield  {author} {\bibinfo {author} {\bibfnamefont {D.}~\bibnamefont
  {Vitali}}, \bibinfo {author} {\bibfnamefont {S.}~\bibnamefont {Gigan}},
  \bibinfo {author} {\bibfnamefont {A.}~\bibnamefont {Ferreira}}, \bibinfo
  {author} {\bibfnamefont {H.}~\bibnamefont {B{\"o}hm}}, \bibinfo {author}
  {\bibfnamefont {P.}~\bibnamefont {Tombesi}}, \bibinfo {author} {\bibfnamefont
  {A.}~\bibnamefont {Guerreiro}}, \bibinfo {author} {\bibfnamefont
  {V.}~\bibnamefont {Vedral}}, \bibinfo {author} {\bibfnamefont
  {A.}~\bibnamefont {Zeilinger}}, \ and\ \bibinfo {author} {\bibfnamefont
  {M.}~\bibnamefont {Aspelmeyer}},\ }\href@noop {} {\bibfield  {journal}
  {\bibinfo  {journal} {Physical Review Letters}\ }\textbf {\bibinfo {volume}
  {98}},\ \bibinfo {pages} {030405} (\bibinfo {year} {2007})}\BibitemShut
  {NoStop}%
\bibitem [{\citenamefont {Miao}\ \emph {et~al.}(2010)\citenamefont {Miao},
  \citenamefont {Danilishin},\ and\ \citenamefont {Chen}}]{miao2010universal}%
  \BibitemOpen
  \bibfield  {author} {\bibinfo {author} {\bibfnamefont {H.}~\bibnamefont
  {Miao}}, \bibinfo {author} {\bibfnamefont {S.}~\bibnamefont {Danilishin}}, \
  and\ \bibinfo {author} {\bibfnamefont {Y.}~\bibnamefont {Chen}},\ }\href@noop
  {} {\bibfield  {journal} {\bibinfo  {journal} {Physical Review A}\ }\textbf
  {\bibinfo {volume} {81}},\ \bibinfo {pages} {052307} (\bibinfo {year}
  {2010})}\BibitemShut {NoStop}%
\bibitem [{\citenamefont {Mancini}\ \emph {et~al.}(2002)\citenamefont
  {Mancini}, \citenamefont {Giovannetti}, \citenamefont {Vitali},\ and\
  \citenamefont {Tombesi}}]{mancini2002entangling}%
  \BibitemOpen
  \bibfield  {author} {\bibinfo {author} {\bibfnamefont {S.}~\bibnamefont
  {Mancini}}, \bibinfo {author} {\bibfnamefont {V.}~\bibnamefont
  {Giovannetti}}, \bibinfo {author} {\bibfnamefont {D.}~\bibnamefont {Vitali}},
  \ and\ \bibinfo {author} {\bibfnamefont {P.}~\bibnamefont {Tombesi}},\
  }\href@noop {} {\bibfield  {journal} {\bibinfo  {journal} {Physical Review
  Letters}\ }\textbf {\bibinfo {volume} {88}},\ \bibinfo {pages} {120401}
  (\bibinfo {year} {2002})}\BibitemShut {NoStop}%
\bibitem [{\citenamefont {Bhattacharya}\ \emph {et~al.}(2008)\citenamefont
  {Bhattacharya}, \citenamefont {Giscard},\ and\ \citenamefont
  {Meystre}}]{bhattacharya2008entangling}%
  \BibitemOpen
  \bibfield  {author} {\bibinfo {author} {\bibfnamefont {M.}~\bibnamefont
  {Bhattacharya}}, \bibinfo {author} {\bibfnamefont {P.-L.}\ \bibnamefont
  {Giscard}}, \ and\ \bibinfo {author} {\bibfnamefont {P.}~\bibnamefont
  {Meystre}},\ }\href@noop {} {\bibfield  {journal} {\bibinfo  {journal}
  {Physical Review A}\ }\textbf {\bibinfo {volume} {77}},\ \bibinfo {pages}
  {030303} (\bibinfo {year} {2008})}\BibitemShut {NoStop}%
\bibitem [{\citenamefont {Hartmann}\ and\ \citenamefont
  {Plenio}(2008)}]{hartmann2008steady}%
  \BibitemOpen
  \bibfield  {author} {\bibinfo {author} {\bibfnamefont {M.~J.}\ \bibnamefont
  {Hartmann}}\ and\ \bibinfo {author} {\bibfnamefont {M.~B.}\ \bibnamefont
  {Plenio}},\ }\href@noop {} {\bibfield  {journal} {\bibinfo  {journal}
  {Physical Review Letters}\ }\textbf {\bibinfo {volume} {101}},\ \bibinfo
  {pages} {200503} (\bibinfo {year} {2008})}\BibitemShut {NoStop}%
\bibitem [{\citenamefont {Schnabel}(2015)}]{schnabel2015einstein}%
  \BibitemOpen
  \bibfield  {author} {\bibinfo {author} {\bibfnamefont {R.}~\bibnamefont
  {Schnabel}},\ }\href@noop {} {\bibfield  {journal} {\bibinfo  {journal}
  {Physical Review A}\ }\textbf {\bibinfo {volume} {92}},\ \bibinfo {pages}
  {012126} (\bibinfo {year} {2015})}\BibitemShut {NoStop}%
\bibitem [{\citenamefont {Nation}(2013)}]{nation2013nonclassical}%
  \BibitemOpen
  \bibfield  {author} {\bibinfo {author} {\bibfnamefont {P.}~\bibnamefont
  {Nation}},\ }\href@noop {} {\bibfield  {journal} {\bibinfo  {journal}
  {Physical Review A}\ }\textbf {\bibinfo {volume} {88}},\ \bibinfo {pages}
  {053828} (\bibinfo {year} {2013})}\BibitemShut {NoStop}%
\bibitem [{\citenamefont {Brunelli}\ \emph {et~al.}(2018)\citenamefont
  {Brunelli}, \citenamefont {Houhou}, \citenamefont {Moore}, \citenamefont
  {Nunnenkamp}, \citenamefont {Paternostro},\ and\ \citenamefont
  {Ferraro}}]{brunelli2018unconditional}%
  \BibitemOpen
  \bibfield  {author} {\bibinfo {author} {\bibfnamefont {M.}~\bibnamefont
  {Brunelli}}, \bibinfo {author} {\bibfnamefont {O.}~\bibnamefont {Houhou}},
  \bibinfo {author} {\bibfnamefont {D.~W.}\ \bibnamefont {Moore}}, \bibinfo
  {author} {\bibfnamefont {A.}~\bibnamefont {Nunnenkamp}}, \bibinfo {author}
  {\bibfnamefont {M.}~\bibnamefont {Paternostro}}, \ and\ \bibinfo {author}
  {\bibfnamefont {A.}~\bibnamefont {Ferraro}},\ }\href@noop {} {\bibfield
  {journal} {\bibinfo  {journal} {Physical Review A}\ }\textbf {\bibinfo
  {volume} {98}},\ \bibinfo {pages} {063801} (\bibinfo {year}
  {2018})}\BibitemShut {NoStop}%
\bibitem [{\citenamefont {Davis}\ \emph {et~al.}(2018)\citenamefont {Davis},
  \citenamefont {Wang}, \citenamefont {Safavi-Naeini},\ and\ \citenamefont
  {Schleier-Smith}}]{davis2018painting}%
  \BibitemOpen
  \bibfield  {author} {\bibinfo {author} {\bibfnamefont {E.~J.}\ \bibnamefont
  {Davis}}, \bibinfo {author} {\bibfnamefont {Z.}~\bibnamefont {Wang}},
  \bibinfo {author} {\bibfnamefont {A.~H.}\ \bibnamefont {Safavi-Naeini}}, \
  and\ \bibinfo {author} {\bibfnamefont {M.~H.}\ \bibnamefont
  {Schleier-Smith}},\ }\href@noop {} {\bibfield  {journal} {\bibinfo  {journal}
  {Physical Review Letters}\ }\textbf {\bibinfo {volume} {121}},\ \bibinfo
  {pages} {123602} (\bibinfo {year} {2018})}\BibitemShut {NoStop}%
\bibitem [{\citenamefont {Chen}(2013)}]{chen2013macroscopic}%
  \BibitemOpen
  \bibfield  {author} {\bibinfo {author} {\bibfnamefont {Y.}~\bibnamefont
  {Chen}},\ }\href@noop {} {\bibfield  {journal} {\bibinfo  {journal} {Journal
  of Physics B: Atomic, Molecular and Optical Physics}\ }\textbf {\bibinfo
  {volume} {46}},\ \bibinfo {pages} {104001} (\bibinfo {year}
  {2013})}\BibitemShut {NoStop}%
\bibitem [{\citenamefont {Purdy}\ \emph
  {et~al.}(2013{\natexlab{b}})\citenamefont {Purdy}, \citenamefont {Peterson},\
  and\ \citenamefont {Regal}}]{purdy2013observation}%
  \BibitemOpen
  \bibfield  {author} {\bibinfo {author} {\bibfnamefont {T.~P.}\ \bibnamefont
  {Purdy}}, \bibinfo {author} {\bibfnamefont {R.~W.}\ \bibnamefont {Peterson}},
  \ and\ \bibinfo {author} {\bibfnamefont {C.}~\bibnamefont {Regal}},\
  }\href@noop {} {\bibfield  {journal} {\bibinfo  {journal} {Science}\ }\textbf
  {\bibinfo {volume} {339}},\ \bibinfo {pages} {801} (\bibinfo {year}
  {2013}{\natexlab{b}})}\BibitemShut {NoStop}%
\bibitem [{\citenamefont {Bawaj}\ \emph {et~al.}(2015)\citenamefont {Bawaj},
  \citenamefont {Biancofiore}, \citenamefont {Bonaldi}, \citenamefont
  {Bonfigli}, \citenamefont {Borrielli}, \citenamefont {Di~Giuseppe},
  \citenamefont {Marconi}, \citenamefont {Marino}, \citenamefont {Natali},
  \citenamefont {Pontin} \emph {et~al.}}]{bawaj2015probing}%
  \BibitemOpen
  \bibfield  {author} {\bibinfo {author} {\bibfnamefont {M.}~\bibnamefont
  {Bawaj}}, \bibinfo {author} {\bibfnamefont {C.}~\bibnamefont {Biancofiore}},
  \bibinfo {author} {\bibfnamefont {M.}~\bibnamefont {Bonaldi}}, \bibinfo
  {author} {\bibfnamefont {F.}~\bibnamefont {Bonfigli}}, \bibinfo {author}
  {\bibfnamefont {A.}~\bibnamefont {Borrielli}}, \bibinfo {author}
  {\bibfnamefont {G.}~\bibnamefont {Di~Giuseppe}}, \bibinfo {author}
  {\bibfnamefont {L.}~\bibnamefont {Marconi}}, \bibinfo {author} {\bibfnamefont
  {F.}~\bibnamefont {Marino}}, \bibinfo {author} {\bibfnamefont
  {R.}~\bibnamefont {Natali}}, \bibinfo {author} {\bibfnamefont
  {A.}~\bibnamefont {Pontin}},  \emph {et~al.},\ }\href@noop {} {\bibfield
  {journal} {\bibinfo  {journal} {Nature Communications}\ }\textbf {\bibinfo
  {volume} {6}},\ \bibinfo {pages} {7503} (\bibinfo {year} {2015})}\BibitemShut
  {NoStop}%
\bibitem [{\citenamefont {Li}\ \emph {et~al.}(2016)\citenamefont {Li},
  \citenamefont {Zippilli}, \citenamefont {Zhang},\ and\ \citenamefont
  {Vitali}}]{li2016discriminating}%
  \BibitemOpen
  \bibfield  {author} {\bibinfo {author} {\bibfnamefont {J.}~\bibnamefont
  {Li}}, \bibinfo {author} {\bibfnamefont {S.}~\bibnamefont {Zippilli}},
  \bibinfo {author} {\bibfnamefont {J.}~\bibnamefont {Zhang}}, \ and\ \bibinfo
  {author} {\bibfnamefont {D.}~\bibnamefont {Vitali}},\ }\href@noop {}
  {\bibfield  {journal} {\bibinfo  {journal} {Physical Review A}\ }\textbf
  {\bibinfo {volume} {93}},\ \bibinfo {pages} {050102} (\bibinfo {year}
  {2016})}\BibitemShut {NoStop}%
\bibitem [{\citenamefont {Belenchia}\ \emph {et~al.}(2017)\citenamefont
  {Belenchia}, \citenamefont {Benincasa}, \citenamefont {Liberati},
  \citenamefont {Marin}, \citenamefont {Marino},\ and\ \citenamefont
  {Ortolan}}]{belenchia2017tests}%
  \BibitemOpen
  \bibfield  {author} {\bibinfo {author} {\bibfnamefont {A.}~\bibnamefont
  {Belenchia}}, \bibinfo {author} {\bibfnamefont {D.~M.}\ \bibnamefont
  {Benincasa}}, \bibinfo {author} {\bibfnamefont {S.}~\bibnamefont {Liberati}},
  \bibinfo {author} {\bibfnamefont {F.}~\bibnamefont {Marin}}, \bibinfo
  {author} {\bibfnamefont {F.}~\bibnamefont {Marino}}, \ and\ \bibinfo {author}
  {\bibfnamefont {A.}~\bibnamefont {Ortolan}},\ }\href@noop {} {\bibfield
  {journal} {\bibinfo  {journal} {Physical Review D}\ }\textbf {\bibinfo
  {volume} {95}},\ \bibinfo {pages} {026012} (\bibinfo {year}
  {2017})}\BibitemShut {NoStop}%
\bibitem [{\citenamefont {Tian}\ and\ \citenamefont
  {Wang}(2010)}]{tian2010optical}%
  \BibitemOpen
  \bibfield  {author} {\bibinfo {author} {\bibfnamefont {L.}~\bibnamefont
  {Tian}}\ and\ \bibinfo {author} {\bibfnamefont {H.}~\bibnamefont {Wang}},\
  }\href@noop {} {\bibfield  {journal} {\bibinfo  {journal} {Physical Review
  A}\ }\textbf {\bibinfo {volume} {82}},\ \bibinfo {pages} {053806} (\bibinfo
  {year} {2010})}\BibitemShut {NoStop}%
\bibitem [{\citenamefont {Hill}\ \emph {et~al.}(2012)\citenamefont {Hill},
  \citenamefont {Safavi-Naeini}, \citenamefont {Chan},\ and\ \citenamefont
  {Painter}}]{hill2012coherent}%
  \BibitemOpen
  \bibfield  {author} {\bibinfo {author} {\bibfnamefont {J.~T.}\ \bibnamefont
  {Hill}}, \bibinfo {author} {\bibfnamefont {A.~H.}\ \bibnamefont
  {Safavi-Naeini}}, \bibinfo {author} {\bibfnamefont {J.}~\bibnamefont {Chan}},
  \ and\ \bibinfo {author} {\bibfnamefont {O.}~\bibnamefont {Painter}},\
  }\href@noop {} {\bibfield  {journal} {\bibinfo  {journal} {Nature
  Communications}\ }\textbf {\bibinfo {volume} {3}},\ \bibinfo {pages} {1196}
  (\bibinfo {year} {2012})}\BibitemShut {NoStop}%
\bibitem [{\citenamefont {Ockeloen-Korppi}\ \emph {et~al.}(2016)\citenamefont
  {Ockeloen-Korppi}, \citenamefont {Damsk{\"a}gg}, \citenamefont
  {Pirkkalainen}, \citenamefont {Heikkil{\"a}}, \citenamefont {Massel},\ and\
  \citenamefont {Sillanp{\"a}{\"a}}}]{ockeloen2016low}%
  \BibitemOpen
  \bibfield  {author} {\bibinfo {author} {\bibfnamefont {C.}~\bibnamefont
  {Ockeloen-Korppi}}, \bibinfo {author} {\bibfnamefont {E.}~\bibnamefont
  {Damsk{\"a}gg}}, \bibinfo {author} {\bibfnamefont {J.-M.}\ \bibnamefont
  {Pirkkalainen}}, \bibinfo {author} {\bibfnamefont {T.}~\bibnamefont
  {Heikkil{\"a}}}, \bibinfo {author} {\bibfnamefont {F.}~\bibnamefont
  {Massel}}, \ and\ \bibinfo {author} {\bibfnamefont {M.}~\bibnamefont
  {Sillanp{\"a}{\"a}}},\ }\href@noop {} {\bibfield  {journal} {\bibinfo
  {journal} {Physical Review X}\ }\textbf {\bibinfo {volume} {6}},\ \bibinfo
  {pages} {041024} (\bibinfo {year} {2016})}\BibitemShut {NoStop}%
\bibitem [{\citenamefont {Lecocq}\ \emph {et~al.}(2016)\citenamefont {Lecocq},
  \citenamefont {Clark}, \citenamefont {Simmonds}, \citenamefont {Aumentado},\
  and\ \citenamefont {Teufel}}]{lecocq2016mechanically}%
  \BibitemOpen
  \bibfield  {author} {\bibinfo {author} {\bibfnamefont {F.}~\bibnamefont
  {Lecocq}}, \bibinfo {author} {\bibfnamefont {J.}~\bibnamefont {Clark}},
  \bibinfo {author} {\bibfnamefont {R.}~\bibnamefont {Simmonds}}, \bibinfo
  {author} {\bibfnamefont {J.}~\bibnamefont {Aumentado}}, \ and\ \bibinfo
  {author} {\bibfnamefont {J.}~\bibnamefont {Teufel}},\ }\href@noop {}
  {\bibfield  {journal} {\bibinfo  {journal} {Physical Review Letters}\
  }\textbf {\bibinfo {volume} {116}},\ \bibinfo {pages} {043601} (\bibinfo
  {year} {2016})}\BibitemShut {NoStop}%
\bibitem [{\citenamefont {Huang}\ \emph {et~al.}(2018)\citenamefont {Huang},
  \citenamefont {Li}, \citenamefont {Chin}, \citenamefont {Cai}, \citenamefont
  {Gu}, \citenamefont {Karim}, \citenamefont {Wu}, \citenamefont {Chen},
  \citenamefont {Yang}, \citenamefont {Hao} \emph
  {et~al.}}]{huang2018dissipative}%
  \BibitemOpen
  \bibfield  {author} {\bibinfo {author} {\bibfnamefont {J.}~\bibnamefont
  {Huang}}, \bibinfo {author} {\bibfnamefont {Y.}~\bibnamefont {Li}}, \bibinfo
  {author} {\bibfnamefont {L.~K.}\ \bibnamefont {Chin}}, \bibinfo {author}
  {\bibfnamefont {H.}~\bibnamefont {Cai}}, \bibinfo {author} {\bibfnamefont
  {Y.}~\bibnamefont {Gu}}, \bibinfo {author} {\bibfnamefont {M.~F.}\
  \bibnamefont {Karim}}, \bibinfo {author} {\bibfnamefont {J.}~\bibnamefont
  {Wu}}, \bibinfo {author} {\bibfnamefont {T.}~\bibnamefont {Chen}}, \bibinfo
  {author} {\bibfnamefont {Z.}~\bibnamefont {Yang}}, \bibinfo {author}
  {\bibfnamefont {Y.}~\bibnamefont {Hao}},  \emph {et~al.},\ }\href@noop {}
  {\bibfield  {journal} {\bibinfo  {journal} {Applied Physics Letters}\
  }\textbf {\bibinfo {volume} {112}},\ \bibinfo {pages} {051104} (\bibinfo
  {year} {2018})}\BibitemShut {NoStop}%
\bibitem [{\citenamefont {Wu}\ \emph {et~al.}(2014)\citenamefont {Wu},
  \citenamefont {Hryciw}, \citenamefont {Healey}, \citenamefont {Lake},
  \citenamefont {Jayakumar}, \citenamefont {Freeman}, \citenamefont {Davis},\
  and\ \citenamefont {Barclay}}]{wu2014dissipative}%
  \BibitemOpen
  \bibfield  {author} {\bibinfo {author} {\bibfnamefont {M.}~\bibnamefont
  {Wu}}, \bibinfo {author} {\bibfnamefont {A.~C.}\ \bibnamefont {Hryciw}},
  \bibinfo {author} {\bibfnamefont {C.}~\bibnamefont {Healey}}, \bibinfo
  {author} {\bibfnamefont {D.~P.}\ \bibnamefont {Lake}}, \bibinfo {author}
  {\bibfnamefont {H.}~\bibnamefont {Jayakumar}}, \bibinfo {author}
  {\bibfnamefont {M.~R.}\ \bibnamefont {Freeman}}, \bibinfo {author}
  {\bibfnamefont {J.~P.}\ \bibnamefont {Davis}}, \ and\ \bibinfo {author}
  {\bibfnamefont {P.~E.}\ \bibnamefont {Barclay}},\ }\href@noop {} {\bibfield
  {journal} {\bibinfo  {journal} {Physical Review X}\ }\textbf {\bibinfo
  {volume} {4}},\ \bibinfo {pages} {021052} (\bibinfo {year}
  {2014})}\BibitemShut {NoStop}%
\bibitem [{\citenamefont {McClelland}\ \emph {et~al.}(2011)\citenamefont
  {McClelland}, \citenamefont {Mavalvala}, \citenamefont {Chen},\ and\
  \citenamefont {\mbox{Schnabel}}}]{mcclelland2011advanced}%
  \BibitemOpen
  \bibfield  {author} {\bibinfo {author} {\bibfnamefont {D.~E.}\ \bibnamefont
  {McClelland}}, \bibinfo {author} {\bibfnamefont {N.}~\bibnamefont
  {Mavalvala}}, \bibinfo {author} {\bibfnamefont {Y.}~\bibnamefont {Chen}}, \
  and\ \bibinfo {author} {\bibfnamefont {R.}~\bibnamefont {\mbox{Schnabel}}},\
  }\href@noop {} {\bibfield  {journal} {\bibinfo  {journal} {Laser \& Photonics
  Reviews}\ }\textbf {\bibinfo {volume} {5}},\ \bibinfo {pages} {677} (\bibinfo
  {year} {2011})}\BibitemShut {NoStop}%
\bibitem [{\citenamefont {Abbott}\ \emph {et~al.}(2016)\citenamefont {Abbott}
  \emph {et~al.}}]{abbott2016observation}%
  \BibitemOpen
  \bibfield  {author} {\bibinfo {author} {\bibfnamefont {B.~P.}\ \bibnamefont
  {Abbott}} \emph {et~al.} (\bibinfo {collaboration} {LIGO Scientific
  Collaboration and Virgo Collaboration}),\ }\href@noop {} {\bibfield
  {journal} {\bibinfo  {journal} {Physical Review Letters}\ }\textbf {\bibinfo
  {volume} {116}},\ \bibinfo {pages} {061102} (\bibinfo {year}
  {2016})}\BibitemShut {NoStop}%
\bibitem [{\citenamefont {Abbott}\ \emph {et~al.}(2018)\citenamefont {Abbott}
  \emph {et~al.}}]{abbott2018prospects}%
  \BibitemOpen
  \bibfield  {author} {\bibinfo {author} {\bibfnamefont {B.~P.}\ \bibnamefont
  {Abbott}} \emph {et~al.} (\bibinfo {collaboration} {KAGRA Collaboration, LIGO
  Scientific Collaboration and Virgo Collaboration}),\ }\href@noop {}
  {\bibfield  {journal} {\bibinfo  {journal} {Living Reviews in Relativity}\
  }\textbf {\bibinfo {volume} {21}},\ \bibinfo {pages} {3} (\bibinfo {year}
  {2018})}\BibitemShut {NoStop}%
\bibitem [{\citenamefont {Harry}\ and\ \citenamefont {the LIGO
  Scientific~Collaboration}(2010)}]{harry2010advanced}%
  \BibitemOpen
  \bibfield  {author} {\bibinfo {author} {\bibfnamefont {G.~M.}\ \bibnamefont
  {Harry}}\ and\ \bibinfo {author} {\bibnamefont {the LIGO
  Scientific~Collaboration}},\ }\href@noop {} {\bibfield  {journal} {\bibinfo
  {journal} {Classical and Quantum Gravity}\ }\textbf {\bibinfo {volume}
  {27}},\ \bibinfo {pages} {084006} (\bibinfo {year} {2010})}\BibitemShut
  {NoStop}%
\bibitem [{\citenamefont {Aasi}\ \emph {et~al.}(2015)\citenamefont {Aasi} \emph
  {et~al.}}]{aasi2015advanced}%
  \BibitemOpen
  \bibfield  {author} {\bibinfo {author} {\bibfnamefont {J.}~\bibnamefont
  {Aasi}} \emph {et~al.} (\bibinfo {collaboration} {The LIGO scientific
  Collaboration}),\ }\href@noop {} {\bibfield  {journal} {\bibinfo  {journal}
  {Classical and quantum gravity}\ }\textbf {\bibinfo {volume} {32}},\ \bibinfo
  {pages} {074001} (\bibinfo {year} {2015})}\BibitemShut {NoStop}%
\bibitem [{\citenamefont {Acernese}\ \emph {et~al.}(2014)\citenamefont
  {Acernese} \emph {et~al.}}]{acernese2014advanced}%
  \BibitemOpen
  \bibfield  {author} {\bibinfo {author} {\bibfnamefont {F.}~\bibnamefont
  {Acernese}} \emph {et~al.} (\bibinfo {collaboration} {The Virgo
  Collaboration}),\ }\href@noop {} {\bibfield  {journal} {\bibinfo  {journal}
  {Classical and Quantum Gravity}\ }\textbf {\bibinfo {volume} {32}},\ \bibinfo
  {pages} {024001} (\bibinfo {year} {2014})}\BibitemShut {NoStop}%
\bibitem [{\citenamefont {Acernese}\ \emph {et~al.}(2015)\citenamefont
  {Acernese} \emph {et~al.}}]{acernese2015advanced}%
  \BibitemOpen
  \bibfield  {author} {\bibinfo {author} {\bibfnamefont {F.}~\bibnamefont
  {Acernese}} \emph {et~al.} (\bibinfo {collaboration} {The Virgo
  Collaboration}),\ }in\ \href@noop {} {\emph {\bibinfo {booktitle} {Journal of
  Physics: Conference Series}}},\ Vol.\ \bibinfo {volume} {610}\ (\bibinfo
  {organization} {IOP Publishing},\ \bibinfo {year} {2015})\ p.\ \bibinfo
  {pages} {012014}\BibitemShut {NoStop}%
\bibitem [{\citenamefont {L{\"u}ck}\ and\ \citenamefont {the
  GEO600~Team}(1997)}]{luck1997geo600}%
  \BibitemOpen
  \bibfield  {author} {\bibinfo {author} {\bibfnamefont {H.}~\bibnamefont
  {L{\"u}ck}}\ and\ \bibinfo {author} {\bibnamefont {the GEO600~Team}},\
  }\href@noop {} {\bibfield  {journal} {\bibinfo  {journal} {Classical and
  quantum gravity}\ }\textbf {\bibinfo {volume} {14}},\ \bibinfo {pages} {1471}
  (\bibinfo {year} {1997})}\BibitemShut {NoStop}%
\bibitem [{\citenamefont {Affeldt}\ \emph {et~al.}(2014)\citenamefont
  {Affeldt}, \citenamefont {Danzmann}, \citenamefont {Dooley}, \citenamefont
  {Grote}, \citenamefont {Hewitson}, \citenamefont {Hild}, \citenamefont
  {Hough}, \citenamefont {Leong}, \citenamefont {L{\"u}ck}, \citenamefont
  {Prijatelj} \emph {et~al.}}]{affeldt2014advanced}%
  \BibitemOpen
  \bibfield  {author} {\bibinfo {author} {\bibfnamefont {C.}~\bibnamefont
  {Affeldt}}, \bibinfo {author} {\bibfnamefont {K.}~\bibnamefont {Danzmann}},
  \bibinfo {author} {\bibfnamefont {K.}~\bibnamefont {Dooley}}, \bibinfo
  {author} {\bibfnamefont {H.}~\bibnamefont {Grote}}, \bibinfo {author}
  {\bibfnamefont {M.}~\bibnamefont {Hewitson}}, \bibinfo {author}
  {\bibfnamefont {S.}~\bibnamefont {Hild}}, \bibinfo {author} {\bibfnamefont
  {J.}~\bibnamefont {Hough}}, \bibinfo {author} {\bibfnamefont
  {J.}~\bibnamefont {Leong}}, \bibinfo {author} {\bibfnamefont
  {H.}~\bibnamefont {L{\"u}ck}}, \bibinfo {author} {\bibfnamefont
  {M.}~\bibnamefont {Prijatelj}},  \emph {et~al.},\ }\href@noop {} {\bibfield
  {journal} {\bibinfo  {journal} {Classical and quantum gravity}\ }\textbf
  {\bibinfo {volume} {31}},\ \bibinfo {pages} {224002} (\bibinfo {year}
  {2014})}\BibitemShut {NoStop}%
\bibitem [{\citenamefont {Aso}\ \emph {et~al.}(2013)\citenamefont {Aso},
  \citenamefont {Michimura}, \citenamefont {Somiya}, \citenamefont {Ando},
  \citenamefont {Miyakawa}, \citenamefont {Sekiguchi}, \citenamefont
  {Tatsumi},\ and\ \citenamefont {Yamamoto}}]{aso2013interferometer}%
  \BibitemOpen
  \bibfield  {author} {\bibinfo {author} {\bibfnamefont {Y.}~\bibnamefont
  {Aso}}, \bibinfo {author} {\bibfnamefont {Y.}~\bibnamefont {Michimura}},
  \bibinfo {author} {\bibfnamefont {K.}~\bibnamefont {Somiya}}, \bibinfo
  {author} {\bibfnamefont {M.}~\bibnamefont {Ando}}, \bibinfo {author}
  {\bibfnamefont {O.}~\bibnamefont {Miyakawa}}, \bibinfo {author}
  {\bibfnamefont {T.}~\bibnamefont {Sekiguchi}}, \bibinfo {author}
  {\bibfnamefont {D.}~\bibnamefont {Tatsumi}}, \ and\ \bibinfo {author}
  {\bibfnamefont {H.}~\bibnamefont {Yamamoto}} (\bibinfo {collaboration} {The
  KAGRA Collaboration}),\ }\href@noop {} {\bibfield  {journal} {\bibinfo
  {journal} {Physical Review D}\ }\textbf {\bibinfo {volume} {88}},\ \bibinfo
  {pages} {043007} (\bibinfo {year} {2013})}\BibitemShut {NoStop}%
\bibitem [{\citenamefont {Somiya}(2012)}]{somiya2012detector}%
  \BibitemOpen
  \bibfield  {author} {\bibinfo {author} {\bibfnamefont {K.}~\bibnamefont
  {Somiya}},\ }\href@noop {} {\bibfield  {journal} {\bibinfo  {journal}
  {Classical and Quantum Gravity}\ }\textbf {\bibinfo {volume} {29}},\ \bibinfo
  {pages} {124007} (\bibinfo {year} {2012})}\BibitemShut {NoStop}%
\bibitem [{\citenamefont {Kleckner}\ \emph {et~al.}(2006)\citenamefont
  {Kleckner}, \citenamefont {Marshall}, \citenamefont {de~Dood}, \citenamefont
  {Dinyari}, \citenamefont {Pors}, \citenamefont {Irvine},\ and\ \citenamefont
  {Bouwmeester}}]{kleckner2006high}%
  \BibitemOpen
  \bibfield  {author} {\bibinfo {author} {\bibfnamefont {D.}~\bibnamefont
  {Kleckner}}, \bibinfo {author} {\bibfnamefont {W.}~\bibnamefont {Marshall}},
  \bibinfo {author} {\bibfnamefont {M.~J.}\ \bibnamefont {de~Dood}}, \bibinfo
  {author} {\bibfnamefont {K.~N.}\ \bibnamefont {Dinyari}}, \bibinfo {author}
  {\bibfnamefont {B.-J.}\ \bibnamefont {Pors}}, \bibinfo {author}
  {\bibfnamefont {W.~T.}\ \bibnamefont {Irvine}}, \ and\ \bibinfo {author}
  {\bibfnamefont {D.}~\bibnamefont {Bouwmeester}},\ }\href@noop {} {\bibfield
  {journal} {\bibinfo  {journal} {Physical Review Letters}\ }\textbf {\bibinfo
  {volume} {96}},\ \bibinfo {pages} {173901} (\bibinfo {year}
  {2006})}\BibitemShut {NoStop}%
\bibitem [{\citenamefont {Xuereb}\ \emph
  {et~al.}(2011{\natexlab{a}})\citenamefont {Xuereb}, \citenamefont
  {\mbox{Schnabel}},\ and\ \citenamefont {Hammerer}}]{xuereb2011dissipative}%
  \BibitemOpen
  \bibfield  {author} {\bibinfo {author} {\bibfnamefont {A.}~\bibnamefont
  {Xuereb}}, \bibinfo {author} {\bibfnamefont {R.}~\bibnamefont
  {\mbox{Schnabel}}}, \ and\ \bibinfo {author} {\bibfnamefont {K.}~\bibnamefont
  {Hammerer}},\ }\href@noop {} {\bibfield  {journal} {\bibinfo  {journal}
  {Physical Review Letters}\ }\textbf {\bibinfo {volume} {107}},\ \bibinfo
  {pages} {213604} (\bibinfo {year} {2011}{\natexlab{a}})}\BibitemShut
  {NoStop}%
\bibitem [{\citenamefont {Vyatchanin}\ and\ \citenamefont
  {Matsko}(2016)}]{vyatchanin2016quantum}%
  \BibitemOpen
  \bibfield  {author} {\bibinfo {author} {\bibfnamefont {S.~P.}\ \bibnamefont
  {Vyatchanin}}\ and\ \bibinfo {author} {\bibfnamefont {A.~B.}\ \bibnamefont
  {Matsko}},\ }\href@noop {} {\bibfield  {journal} {\bibinfo  {journal}
  {Physical Review A}\ }\textbf {\bibinfo {volume} {93}},\ \bibinfo {pages}
  {063817} (\bibinfo {year} {2016})}\BibitemShut {NoStop}%
\bibitem [{\citenamefont {Khalili}\ \emph {et~al.}(2016)\citenamefont
  {Khalili}, \citenamefont {Tarabrin}, \citenamefont {Hammerer},\ and\
  \citenamefont {\mbox{Schnabel}}}]{khalili2016generalized}%
  \BibitemOpen
  \bibfield  {author} {\bibinfo {author} {\bibfnamefont {F.~Y.}\ \bibnamefont
  {Khalili}}, \bibinfo {author} {\bibfnamefont {S.~P.}\ \bibnamefont
  {Tarabrin}}, \bibinfo {author} {\bibfnamefont {K.}~\bibnamefont {Hammerer}},
  \ and\ \bibinfo {author} {\bibfnamefont {R.}~\bibnamefont
  {\mbox{Schnabel}}},\ }\href@noop {} {\bibfield  {journal} {\bibinfo
  {journal} {Physical Review A}\ }\textbf {\bibinfo {volume} {94}},\ \bibinfo
  {pages} {013844} (\bibinfo {year} {2016})}\BibitemShut {NoStop}%
\bibitem [{\citenamefont {Xuereb}\ \emph
  {et~al.}(2011{\natexlab{b}})\citenamefont {Xuereb}, \citenamefont {Horak},\
  and\ \citenamefont {Freegarde}}]{Xuereb2011ring}%
  \BibitemOpen
  \bibfield  {author} {\bibinfo {author} {\bibfnamefont {A.}~\bibnamefont
  {Xuereb}}, \bibinfo {author} {\bibfnamefont {P.}~\bibnamefont {Horak}}, \
  and\ \bibinfo {author} {\bibfnamefont {T.}~\bibnamefont {Freegarde}},\ }\href
  {\doibase 10.1080/09500340.2011.559316} {\bibfield  {journal} {\bibinfo
  {journal} {Journal of Modern Optics}\ }\textbf {\bibinfo {volume} {58}},\
  \bibinfo {pages} {1342} (\bibinfo {year} {2011}{\natexlab{b}})},\ \Eprint
  {http://arxiv.org/abs/https://doi.org/10.1080/09500340.2011.559316}
  {https://doi.org/10.1080/09500340.2011.559316} \BibitemShut {NoStop}%
\bibitem [{\citenamefont {Chesi}\ \emph
  {et~al.}(2015{\natexlab{a}})\citenamefont {Chesi}, \citenamefont {Wang},\
  and\ \citenamefont {Twamley}}]{Chesi2015ring}%
  \BibitemOpen
  \bibfield  {author} {\bibinfo {author} {\bibfnamefont {S.}~\bibnamefont
  {Chesi}}, \bibinfo {author} {\bibfnamefont {Y.-D.}\ \bibnamefont {Wang}}, \
  and\ \bibinfo {author} {\bibfnamefont {J.}~\bibnamefont {Twamley}},\ }\href
  {https://doi.org/10.1038/srep07816} {\bibfield  {journal} {\bibinfo
  {journal} {Scientific Reports}\ }\textbf {\bibinfo {volume} {5}},\ \bibinfo
  {pages} {7816 EP } (\bibinfo {year} {2015}{\natexlab{a}})}\BibitemShut
  {NoStop}%
\bibitem [{\citenamefont {Yilmaz}\ \emph
  {et~al.}(2017{\natexlab{a}})\citenamefont {Yilmaz}, \citenamefont {Schuster},
  \citenamefont {Wolf}, \citenamefont {Schmidt}, \citenamefont {Eisele},
  \citenamefont {Zimmermann},\ and\ \citenamefont {Slama}}]{Yilmaz2017ring}%
  \BibitemOpen
  \bibfield  {author} {\bibinfo {author} {\bibfnamefont {A.}~\bibnamefont
  {Yilmaz}}, \bibinfo {author} {\bibfnamefont {S.}~\bibnamefont {Schuster}},
  \bibinfo {author} {\bibfnamefont {P.}~\bibnamefont {Wolf}}, \bibinfo {author}
  {\bibfnamefont {D.}~\bibnamefont {Schmidt}}, \bibinfo {author} {\bibfnamefont
  {M.}~\bibnamefont {Eisele}}, \bibinfo {author} {\bibfnamefont
  {C.}~\bibnamefont {Zimmermann}}, \ and\ \bibinfo {author} {\bibfnamefont
  {S.}~\bibnamefont {Slama}},\ }\href {\doibase 10.1088/1367-2630/aa55ee}
  {\bibfield  {journal} {\bibinfo  {journal} {New Journal of Physics}\ }\textbf
  {\bibinfo {volume} {19}},\ \bibinfo {pages} {013038} (\bibinfo {year}
  {2017}{\natexlab{a}})}\BibitemShut {NoStop}%
\bibitem [{\citenamefont {Law}(1995)}]{law1995interaction}%
  \BibitemOpen
  \bibfield  {author} {\bibinfo {author} {\bibfnamefont {C.}~\bibnamefont
  {Law}},\ }\href@noop {} {\bibfield  {journal} {\bibinfo  {journal} {Physical
  Review A}\ }\textbf {\bibinfo {volume} {51}},\ \bibinfo {pages} {2537}
  (\bibinfo {year} {1995})}\BibitemShut {NoStop}%
\bibitem [{\citenamefont {Khorasani}(2017)}]{khorasani2017higher}%
  \BibitemOpen
  \bibfield  {author} {\bibinfo {author} {\bibfnamefont {S.}~\bibnamefont
  {Khorasani}},\ }\href@noop {} {\bibfield  {journal} {\bibinfo  {journal}
  {Applied Sciences}\ }\textbf {\bibinfo {volume} {7}},\ \bibinfo {pages} {656}
  (\bibinfo {year} {2017})}\BibitemShut {NoStop}%
\bibitem [{\citenamefont {Pang}(2018)}]{pang2018theoretical}%
  \BibitemOpen
  \bibfield  {author} {\bibinfo {author} {\bibfnamefont {B.~H.}\ \bibnamefont
  {Pang}},\ }\emph {\bibinfo {title} {Theoretical Foundations for Quantum
  Measurement in a General Relativistic Framework}},\ \href@noop {} {Ph.D.
  thesis},\ \bibinfo  {school} {California Institute of Technology} (\bibinfo
  {year} {2018})\BibitemShut {NoStop}%
\bibitem [{\citenamefont {Thompson}\ \emph {et~al.}(2008)\citenamefont
  {Thompson}, \citenamefont {Zwickl}, \citenamefont {Jayich}, \citenamefont
  {Marquardt}, \citenamefont {Girvin},\ and\ \citenamefont
  {Harris}}]{thompson2008strong}%
  \BibitemOpen
  \bibfield  {author} {\bibinfo {author} {\bibfnamefont {J.}~\bibnamefont
  {Thompson}}, \bibinfo {author} {\bibfnamefont {B.}~\bibnamefont {Zwickl}},
  \bibinfo {author} {\bibfnamefont {A.}~\bibnamefont {Jayich}}, \bibinfo
  {author} {\bibfnamefont {F.}~\bibnamefont {Marquardt}}, \bibinfo {author}
  {\bibfnamefont {S.}~\bibnamefont {Girvin}}, \ and\ \bibinfo {author}
  {\bibfnamefont {J.}~\bibnamefont {Harris}},\ }\href@noop {} {\bibfield
  {journal} {\bibinfo  {journal} {Nature}\ }\textbf {\bibinfo {volume} {452}},\
  \bibinfo {pages} {72} (\bibinfo {year} {2008})}\BibitemShut {NoStop}%
\bibitem [{\citenamefont {Karuza}\ \emph {et~al.}(2012)\citenamefont {Karuza},
  \citenamefont {Galassi}, \citenamefont {Biancofiore}, \citenamefont
  {Molinelli}, \citenamefont {Natali}, \citenamefont {Tombesi}, \citenamefont
  {Di~Giuseppe},\ and\ \citenamefont {Vitali}}]{karuza2012tunable}%
  \BibitemOpen
  \bibfield  {author} {\bibinfo {author} {\bibfnamefont {M.}~\bibnamefont
  {Karuza}}, \bibinfo {author} {\bibfnamefont {M.}~\bibnamefont {Galassi}},
  \bibinfo {author} {\bibfnamefont {C.}~\bibnamefont {Biancofiore}}, \bibinfo
  {author} {\bibfnamefont {C.}~\bibnamefont {Molinelli}}, \bibinfo {author}
  {\bibfnamefont {R.}~\bibnamefont {Natali}}, \bibinfo {author} {\bibfnamefont
  {P.}~\bibnamefont {Tombesi}}, \bibinfo {author} {\bibfnamefont
  {G.}~\bibnamefont {Di~Giuseppe}}, \ and\ \bibinfo {author} {\bibfnamefont
  {D.}~\bibnamefont {Vitali}},\ }\href@noop {} {\bibfield  {journal} {\bibinfo
  {journal} {Journal of Optics}\ }\textbf {\bibinfo {volume} {15}},\ \bibinfo
  {pages} {025704} (\bibinfo {year} {2012})}\BibitemShut {NoStop}%
\bibitem [{\citenamefont {Xie}\ \emph {et~al.}(2017)\citenamefont {Xie},
  \citenamefont {Liao}, \citenamefont {Shang}, \citenamefont {Ye},\ and\
  \citenamefont {Lin}}]{xie2017phonon}%
  \BibitemOpen
  \bibfield  {author} {\bibinfo {author} {\bibfnamefont {H.}~\bibnamefont
  {Xie}}, \bibinfo {author} {\bibfnamefont {C.-G.}\ \bibnamefont {Liao}},
  \bibinfo {author} {\bibfnamefont {X.}~\bibnamefont {Shang}}, \bibinfo
  {author} {\bibfnamefont {M.-Y.}\ \bibnamefont {Ye}}, \ and\ \bibinfo {author}
  {\bibfnamefont {X.-M.}\ \bibnamefont {Lin}},\ }\href@noop {} {\bibfield
  {journal} {\bibinfo  {journal} {Physical Review A}\ }\textbf {\bibinfo
  {volume} {96}},\ \bibinfo {pages} {013861} (\bibinfo {year}
  {2017})}\BibitemShut {NoStop}%
\bibitem [{\citenamefont {Machado}\ \emph {et~al.}(2019)\citenamefont
  {Machado}, \citenamefont {Slooter},\ and\ \citenamefont
  {Blanter}}]{machado2019quantum}%
  \BibitemOpen
  \bibfield  {author} {\bibinfo {author} {\bibfnamefont {J.}~\bibnamefont
  {Machado}}, \bibinfo {author} {\bibfnamefont {R.}~\bibnamefont {Slooter}}, \
  and\ \bibinfo {author} {\bibfnamefont {Y.~M.}\ \bibnamefont {Blanter}},\
  }\href@noop {} {\bibfield  {journal} {\bibinfo  {journal} {Physical Review
  A}\ }\textbf {\bibinfo {volume} {99}},\ \bibinfo {pages} {053801} (\bibinfo
  {year} {2019})}\BibitemShut {NoStop}%
\bibitem [{\citenamefont {Miao}\ \emph
  {et~al.}(2009{\natexlab{a}})\citenamefont {Miao}, \citenamefont {Zhao},
  \citenamefont {Ju},\ and\ \citenamefont {Blair}}]{miao2009quantum}%
  \BibitemOpen
  \bibfield  {author} {\bibinfo {author} {\bibfnamefont {H.}~\bibnamefont
  {Miao}}, \bibinfo {author} {\bibfnamefont {C.}~\bibnamefont {Zhao}}, \bibinfo
  {author} {\bibfnamefont {L.}~\bibnamefont {Ju}}, \ and\ \bibinfo {author}
  {\bibfnamefont {D.~G.}\ \bibnamefont {Blair}},\ }\href@noop {} {\bibfield
  {journal} {\bibinfo  {journal} {Physical Review A}\ }\textbf {\bibinfo
  {volume} {79}},\ \bibinfo {pages} {063801} (\bibinfo {year}
  {2009}{\natexlab{a}})}\BibitemShut {NoStop}%
\bibitem [{\citenamefont {Braginsky}\ \emph {et~al.}(2001)\citenamefont
  {Braginsky}, \citenamefont {Strigin},\ and\ \citenamefont
  {Vyatchanin}}]{braginsky2001parametric}%
  \BibitemOpen
  \bibfield  {author} {\bibinfo {author} {\bibfnamefont {V.}~\bibnamefont
  {Braginsky}}, \bibinfo {author} {\bibfnamefont {S.}~\bibnamefont {Strigin}},
  \ and\ \bibinfo {author} {\bibfnamefont {S.~P.}\ \bibnamefont {Vyatchanin}},\
  }\href@noop {} {\bibfield  {journal} {\bibinfo  {journal} {Physics Letters
  A}\ }\textbf {\bibinfo {volume} {287}},\ \bibinfo {pages} {331} (\bibinfo
  {year} {2001})}\BibitemShut {NoStop}%
\bibitem [{\citenamefont {Evans}\ \emph {et~al.}(2015)\citenamefont {Evans},
  \citenamefont {Gras}, \citenamefont {Fritschel}, \citenamefont {Miller},
  \citenamefont {Barsotti}, \citenamefont {Martynov}, \citenamefont {Brooks},
  \citenamefont {Coyne}, \citenamefont {Abbott}, \citenamefont {Adhikari} \emph
  {et~al.}}]{evans2015observation}%
  \BibitemOpen
  \bibfield  {author} {\bibinfo {author} {\bibfnamefont {M.}~\bibnamefont
  {Evans}}, \bibinfo {author} {\bibfnamefont {S.}~\bibnamefont {Gras}},
  \bibinfo {author} {\bibfnamefont {P.}~\bibnamefont {Fritschel}}, \bibinfo
  {author} {\bibfnamefont {J.}~\bibnamefont {Miller}}, \bibinfo {author}
  {\bibfnamefont {L.}~\bibnamefont {Barsotti}}, \bibinfo {author}
  {\bibfnamefont {D.}~\bibnamefont {Martynov}}, \bibinfo {author}
  {\bibfnamefont {A.}~\bibnamefont {Brooks}}, \bibinfo {author} {\bibfnamefont
  {D.}~\bibnamefont {Coyne}}, \bibinfo {author} {\bibfnamefont
  {R.}~\bibnamefont {Abbott}}, \bibinfo {author} {\bibfnamefont {R.~X.}\
  \bibnamefont {Adhikari}},  \emph {et~al.},\ }\href@noop {} {\bibfield
  {journal} {\bibinfo  {journal} {Physical Review Letters}\ }\textbf {\bibinfo
  {volume} {114}},\ \bibinfo {pages} {161102} (\bibinfo {year}
  {2015})}\BibitemShut {NoStop}%
\bibitem [{\citenamefont {Miao}\ \emph
  {et~al.}(2009{\natexlab{b}})\citenamefont {Miao}, \citenamefont {Danilishin},
  \citenamefont {Corbitt},\ and\ \citenamefont {Chen}}]{miao2009standard}%
  \BibitemOpen
  \bibfield  {author} {\bibinfo {author} {\bibfnamefont {H.}~\bibnamefont
  {Miao}}, \bibinfo {author} {\bibfnamefont {S.}~\bibnamefont {Danilishin}},
  \bibinfo {author} {\bibfnamefont {T.}~\bibnamefont {Corbitt}}, \ and\
  \bibinfo {author} {\bibfnamefont {Y.}~\bibnamefont {Chen}},\ }\href@noop {}
  {\bibfield  {journal} {\bibinfo  {journal} {Physical Review Letters}\
  }\textbf {\bibinfo {volume} {103}},\ \bibinfo {pages} {100402} (\bibinfo
  {year} {2009}{\natexlab{b}})}\BibitemShut {NoStop}%
\bibitem [{\citenamefont {Ma}\ \emph {et~al.}(2014)\citenamefont {Ma},
  \citenamefont {Danilishin}, \citenamefont {Zhao}, \citenamefont {Miao},
  \citenamefont {Korth}, \citenamefont {Chen}, \citenamefont {Ward},\ and\
  \citenamefont {Blair}}]{ma2014narrowing}%
  \BibitemOpen
  \bibfield  {author} {\bibinfo {author} {\bibfnamefont {Y.}~\bibnamefont
  {Ma}}, \bibinfo {author} {\bibfnamefont {S.~L.}\ \bibnamefont {Danilishin}},
  \bibinfo {author} {\bibfnamefont {C.}~\bibnamefont {Zhao}}, \bibinfo {author}
  {\bibfnamefont {H.}~\bibnamefont {Miao}}, \bibinfo {author} {\bibfnamefont
  {W.~Z.}\ \bibnamefont {Korth}}, \bibinfo {author} {\bibfnamefont
  {Y.}~\bibnamefont {Chen}}, \bibinfo {author} {\bibfnamefont {R.~L.}\
  \bibnamefont {Ward}}, \ and\ \bibinfo {author} {\bibfnamefont {D.~G.}\
  \bibnamefont {Blair}},\ }\href@noop {} {\bibfield  {journal} {\bibinfo
  {journal} {Physical Review Letters}\ }\textbf {\bibinfo {volume} {113}},\
  \bibinfo {pages} {151102} (\bibinfo {year} {2014})}\BibitemShut {NoStop}%
\bibitem [{\citenamefont {Xuereb}\ \emph
  {et~al.}(2011{\natexlab{c}})\citenamefont {Xuereb}, \citenamefont {Horak},\
  and\ \citenamefont {Freegarde}}]{xuereb2011amplified}%
  \BibitemOpen
  \bibfield  {author} {\bibinfo {author} {\bibfnamefont {A.}~\bibnamefont
  {Xuereb}}, \bibinfo {author} {\bibfnamefont {P.}~\bibnamefont {Horak}}, \
  and\ \bibinfo {author} {\bibfnamefont {T.}~\bibnamefont {Freegarde}},\
  }\href@noop {} {\bibfield  {journal} {\bibinfo  {journal} {Journal of Modern
  Optics}\ }\textbf {\bibinfo {volume} {58}},\ \bibinfo {pages} {1342}
  (\bibinfo {year} {2011}{\natexlab{c}})}\BibitemShut {NoStop}%
\bibitem [{\citenamefont {Chesi}\ \emph
  {et~al.}(2015{\natexlab{b}})\citenamefont {Chesi}, \citenamefont {Wang},\
  and\ \citenamefont {Twamley}}]{chesi2015diabolical}%
  \BibitemOpen
  \bibfield  {author} {\bibinfo {author} {\bibfnamefont {S.}~\bibnamefont
  {Chesi}}, \bibinfo {author} {\bibfnamefont {Y.-D.}\ \bibnamefont {Wang}}, \
  and\ \bibinfo {author} {\bibfnamefont {J.}~\bibnamefont {Twamley}},\
  }\href@noop {} {\bibfield  {journal} {\bibinfo  {journal} {Scientific
  reports}\ }\textbf {\bibinfo {volume} {5}},\ \bibinfo {pages} {7816}
  (\bibinfo {year} {2015}{\natexlab{b}})}\BibitemShut {NoStop}%
\bibitem [{\citenamefont {Yilmaz}\ \emph
  {et~al.}(2017{\natexlab{b}})\citenamefont {Yilmaz}, \citenamefont {Schuster},
  \citenamefont {Wolf}, \citenamefont {Schmidt}, \citenamefont {Eisele},
  \citenamefont {Zimmermann},\ and\ \citenamefont
  {Slama}}]{yilmaz2017optomechanical}%
  \BibitemOpen
  \bibfield  {author} {\bibinfo {author} {\bibfnamefont {A.}~\bibnamefont
  {Yilmaz}}, \bibinfo {author} {\bibfnamefont {S.}~\bibnamefont {Schuster}},
  \bibinfo {author} {\bibfnamefont {P.}~\bibnamefont {Wolf}}, \bibinfo {author}
  {\bibfnamefont {D.}~\bibnamefont {Schmidt}}, \bibinfo {author} {\bibfnamefont
  {M.}~\bibnamefont {Eisele}}, \bibinfo {author} {\bibfnamefont
  {C.}~\bibnamefont {Zimmermann}}, \ and\ \bibinfo {author} {\bibfnamefont
  {S.}~\bibnamefont {Slama}},\ }\href@noop {} {\bibfield  {journal} {\bibinfo
  {journal} {New Journal of Physics}\ }\textbf {\bibinfo {volume} {19}},\
  \bibinfo {pages} {013038} (\bibinfo {year} {2017}{\natexlab{b}})}\BibitemShut
  {NoStop}%
\bibitem [{\citenamefont {Nagorny}\ \emph {et~al.}(2003)\citenamefont
  {Nagorny}, \citenamefont {Els{\"a}sser},\ and\ \citenamefont
  {Hemmerich}}]{nagorny2003collective}%
  \BibitemOpen
  \bibfield  {author} {\bibinfo {author} {\bibfnamefont {B.}~\bibnamefont
  {Nagorny}}, \bibinfo {author} {\bibfnamefont {T.}~\bibnamefont
  {Els{\"a}sser}}, \ and\ \bibinfo {author} {\bibfnamefont {A.}~\bibnamefont
  {Hemmerich}},\ }\href@noop {} {\bibfield  {journal} {\bibinfo  {journal}
  {Physical Review Letters}\ }\textbf {\bibinfo {volume} {91}},\ \bibinfo
  {pages} {153003} (\bibinfo {year} {2003})}\BibitemShut {NoStop}%
\bibitem [{\citenamefont {Kruse}\ \emph {et~al.}(2003)\citenamefont {Kruse},
  \citenamefont {von Cube}, \citenamefont {Zimmermann},\ and\ \citenamefont
  {Courteille}}]{kruse2003observation}%
  \BibitemOpen
  \bibfield  {author} {\bibinfo {author} {\bibfnamefont {D.}~\bibnamefont
  {Kruse}}, \bibinfo {author} {\bibfnamefont {C.}~\bibnamefont {von Cube}},
  \bibinfo {author} {\bibfnamefont {C.}~\bibnamefont {Zimmermann}}, \ and\
  \bibinfo {author} {\bibfnamefont {P.~W.}\ \bibnamefont {Courteille}},\
  }\href@noop {} {\bibfield  {journal} {\bibinfo  {journal} {Physical Review
  Letters}\ }\textbf {\bibinfo {volume} {91}},\ \bibinfo {pages} {183601}
  (\bibinfo {year} {2003})}\BibitemShut {NoStop}%
\bibitem [{\citenamefont {Els{\"a}sser}\ \emph {et~al.}(2004)\citenamefont
  {Els{\"a}sser}, \citenamefont {Nagorny},\ and\ \citenamefont
  {Hemmerich}}]{elsasser2004optical}%
  \BibitemOpen
  \bibfield  {author} {\bibinfo {author} {\bibfnamefont {T.}~\bibnamefont
  {Els{\"a}sser}}, \bibinfo {author} {\bibfnamefont {B.}~\bibnamefont
  {Nagorny}}, \ and\ \bibinfo {author} {\bibfnamefont {A.}~\bibnamefont
  {Hemmerich}},\ }\href@noop {} {\bibfield  {journal} {\bibinfo  {journal}
  {Physical Review A}\ }\textbf {\bibinfo {volume} {69}},\ \bibinfo {pages}
  {033403} (\bibinfo {year} {2004})}\BibitemShut {NoStop}%
\bibitem [{\citenamefont {Klinner}\ \emph {et~al.}(2006)\citenamefont
  {Klinner}, \citenamefont {Lindholdt}, \citenamefont {Nagorny},\ and\
  \citenamefont {Hemmerich}}]{klinner2006normal}%
  \BibitemOpen
  \bibfield  {author} {\bibinfo {author} {\bibfnamefont {J.}~\bibnamefont
  {Klinner}}, \bibinfo {author} {\bibfnamefont {M.}~\bibnamefont {Lindholdt}},
  \bibinfo {author} {\bibfnamefont {B.}~\bibnamefont {Nagorny}}, \ and\
  \bibinfo {author} {\bibfnamefont {A.}~\bibnamefont {Hemmerich}},\ }\href@noop
  {} {\bibfield  {journal} {\bibinfo  {journal} {Physical Review Letters}\
  }\textbf {\bibinfo {volume} {96}},\ \bibinfo {pages} {023002} (\bibinfo
  {year} {2006})}\BibitemShut {NoStop}%
\bibitem [{\citenamefont {Slama}\ \emph {et~al.}(2007)\citenamefont {Slama},
  \citenamefont {Bux}, \citenamefont {Krenz}, \citenamefont {Zimmermann},\ and\
  \citenamefont {Courteille}}]{slama2007superradiant}%
  \BibitemOpen
  \bibfield  {author} {\bibinfo {author} {\bibfnamefont {S.}~\bibnamefont
  {Slama}}, \bibinfo {author} {\bibfnamefont {S.}~\bibnamefont {Bux}}, \bibinfo
  {author} {\bibfnamefont {G.}~\bibnamefont {Krenz}}, \bibinfo {author}
  {\bibfnamefont {C.}~\bibnamefont {Zimmermann}}, \ and\ \bibinfo {author}
  {\bibfnamefont {P.~W.}\ \bibnamefont {Courteille}},\ }\href@noop {}
  {\bibfield  {journal} {\bibinfo  {journal} {Physical Review Letters}\
  }\textbf {\bibinfo {volume} {98}},\ \bibinfo {pages} {053603} (\bibinfo
  {year} {2007})}\BibitemShut {NoStop}%
\bibitem [{\citenamefont {Ritsch}\ \emph {et~al.}(2013)\citenamefont {Ritsch},
  \citenamefont {Domokos}, \citenamefont {Brennecke},\ and\ \citenamefont
  {Esslinger}}]{ritsch2013cold}%
  \BibitemOpen
  \bibfield  {author} {\bibinfo {author} {\bibfnamefont {H.}~\bibnamefont
  {Ritsch}}, \bibinfo {author} {\bibfnamefont {P.}~\bibnamefont {Domokos}},
  \bibinfo {author} {\bibfnamefont {F.}~\bibnamefont {Brennecke}}, \ and\
  \bibinfo {author} {\bibfnamefont {T.}~\bibnamefont {Esslinger}},\ }\href@noop
  {} {\bibfield  {journal} {\bibinfo  {journal} {Reviews of Modern Physics}\
  }\textbf {\bibinfo {volume} {85}},\ \bibinfo {pages} {553} (\bibinfo {year}
  {2013})}\BibitemShut {NoStop}%
\bibitem [{\citenamefont {Schmidt}\ \emph {et~al.}(2014)\citenamefont
  {Schmidt}, \citenamefont {Tomczyk}, \citenamefont {Slama},\ and\
  \citenamefont {Zimmermann}}]{schmidt2014dynamical}%
  \BibitemOpen
  \bibfield  {author} {\bibinfo {author} {\bibfnamefont {D.}~\bibnamefont
  {Schmidt}}, \bibinfo {author} {\bibfnamefont {H.}~\bibnamefont {Tomczyk}},
  \bibinfo {author} {\bibfnamefont {S.}~\bibnamefont {Slama}}, \ and\ \bibinfo
  {author} {\bibfnamefont {C.}~\bibnamefont {Zimmermann}},\ }\href@noop {}
  {\bibfield  {journal} {\bibinfo  {journal} {Physical Review Letters}\
  }\textbf {\bibinfo {volume} {112}},\ \bibinfo {pages} {115302} (\bibinfo
  {year} {2014})}\BibitemShut {NoStop}%
\bibitem [{\citenamefont {Mivehvar}\ \emph {et~al.}(2018)\citenamefont
  {Mivehvar}, \citenamefont {Ostermann}, \citenamefont {Piazza},\ and\
  \citenamefont {Ritsch}}]{mivehvar2018driven}%
  \BibitemOpen
  \bibfield  {author} {\bibinfo {author} {\bibfnamefont {F.}~\bibnamefont
  {Mivehvar}}, \bibinfo {author} {\bibfnamefont {S.}~\bibnamefont {Ostermann}},
  \bibinfo {author} {\bibfnamefont {F.}~\bibnamefont {Piazza}}, \ and\ \bibinfo
  {author} {\bibfnamefont {H.}~\bibnamefont {Ritsch}},\ }\href@noop {}
  {\bibfield  {journal} {\bibinfo  {journal} {Physical Review Letters}\
  }\textbf {\bibinfo {volume} {120}},\ \bibinfo {pages} {123601} (\bibinfo
  {year} {2018})}\BibitemShut {NoStop}%
\bibitem [{\citenamefont {Cheung}\ and\ \citenamefont
  {Law}(2011)}]{cheung2011nonadiabatic}%
  \BibitemOpen
  \bibfield  {author} {\bibinfo {author} {\bibfnamefont {H.}~\bibnamefont
  {Cheung}}\ and\ \bibinfo {author} {\bibfnamefont {C.}~\bibnamefont {Law}},\
  }\href@noop {} {\bibfield  {journal} {\bibinfo  {journal} {Physical Review
  A}\ }\textbf {\bibinfo {volume} {84}},\ \bibinfo {pages} {023812} (\bibinfo
  {year} {2011})}\BibitemShut {NoStop}%
\bibitem [{\citenamefont {Scully}\ and\ \citenamefont
  {Zubairy}(1997)}]{scully_zubairy_1997}%
  \BibitemOpen
  \bibfield  {author} {\bibinfo {author} {\bibfnamefont {M.~O.}\ \bibnamefont
  {Scully}}\ and\ \bibinfo {author} {\bibfnamefont {M.~S.}\ \bibnamefont
  {Zubairy}},\ }\href {\doibase 10.1017/CBO9780511813993} {\emph {\bibinfo
  {title} {Quantum Optics}}}\ (\bibinfo  {publisher} {Cambridge University
  Press},\ \bibinfo {year} {1997})\BibitemShut {NoStop}%
\bibitem [{\citenamefont {Nielsen}\ \emph {et~al.}(2017)\citenamefont
  {Nielsen}, \citenamefont {Tsaturyan}, \citenamefont {M{\o}ller},
  \citenamefont {Polzik},\ and\ \citenamefont
  {Schliesser}}]{nielsen2017multimode}%
  \BibitemOpen
  \bibfield  {author} {\bibinfo {author} {\bibfnamefont {W.~H.~P.}\
  \bibnamefont {Nielsen}}, \bibinfo {author} {\bibfnamefont {Y.}~\bibnamefont
  {Tsaturyan}}, \bibinfo {author} {\bibfnamefont {C.~B.}\ \bibnamefont
  {M{\o}ller}}, \bibinfo {author} {\bibfnamefont {E.~S.}\ \bibnamefont
  {Polzik}}, \ and\ \bibinfo {author} {\bibfnamefont {A.}~\bibnamefont
  {Schliesser}},\ }\href@noop {} {\bibfield  {journal} {\bibinfo  {journal}
  {Proceedings of the National Academy of Sciences}\ }\textbf {\bibinfo
  {volume} {114}},\ \bibinfo {pages} {62} (\bibinfo {year} {2017})}\BibitemShut
  {NoStop}%
\bibitem [{\citenamefont {Miao}\ \emph {et~al.}(2015)\citenamefont {Miao},
  \citenamefont {Ma}, \citenamefont {Zhao},\ and\ \citenamefont
  {Chen}}]{Miao2015}%
  \BibitemOpen
  \bibfield  {author} {\bibinfo {author} {\bibfnamefont {H.}~\bibnamefont
  {Miao}}, \bibinfo {author} {\bibfnamefont {Y.}~\bibnamefont {Ma}}, \bibinfo
  {author} {\bibfnamefont {C.}~\bibnamefont {Zhao}}, \ and\ \bibinfo {author}
  {\bibfnamefont {Y.}~\bibnamefont {Chen}},\ }\href {\doibase
  10.1103/PhysRevLett.115.211104} {\bibfield  {journal} {\bibinfo  {journal}
  {Physical Review Letters}\ }\textbf {\bibinfo {volume} {115}},\ \bibinfo
  {pages} {1} (\bibinfo {year} {2015})},\ \Eprint
  {http://arxiv.org/abs/1506.00117v1} {arXiv:1506.00117v1} \BibitemShut
  {NoStop}%
\bibitem [{\citenamefont {Sakurai}\ and\ \citenamefont
  {Napolitano}(2011)}]{Sakurai:1341875}%
  \BibitemOpen
  \bibfield  {author} {\bibinfo {author} {\bibfnamefont {J.~J.}\ \bibnamefont
  {Sakurai}}\ and\ \bibinfo {author} {\bibfnamefont {J.}~\bibnamefont
  {Napolitano}},\ }\href {https://cds.cern.ch/record/1341875} {\emph {\bibinfo
  {title} {{Modern quantum mechanics; 2nd ed.}}}}\ (\bibinfo  {publisher}
  {Addison-Wesley},\ \bibinfo {address} {San Francisco, CA},\ \bibinfo {year}
  {2011})\BibitemShut {NoStop}%
\end{thebibliography}%

\end{document}